\documentclass[lettersize,twoside,journal]{IEEEtran}
\usepackage{graphicx}
\usepackage{amssymb}
\usepackage{amsmath,cases}
\usepackage{cite}
\usepackage{xcolor}
\usepackage{subfigure}
\usepackage{booktabs}
\usepackage{balance}
\usepackage{acro}
\usepackage{stfloats}
\usepackage{multirow}
\usepackage{mathtools}

\usepackage[mathscr]{euscript}
\PassOptionsToPackage{table}{xcolor}
\usepackage{colortbl}

\usepackage{xcolor}
\usepackage{braket}
\usepackage{bbm}
\usepackage{pifont}

\DeclareMathAlphabet{\mathpzc}{OT1}{pzc}{m}{it}

\DeclareMathAlphabet{\mathitsf}{OML}{cmbr}{m}{it}
\DeclareMathAlphabet{\mathsf}{OT1}{cmbr}{m}{n}

\definecolor{darkgreen}{RGB}{0,128,0}

\newtheorem{definition}{Definition}
\newtheorem{theorem}{Theorem}
\newtheorem{lemma}{Lemma}

\newtheorem{remark}{Remark}

\AtBeginDocument{}

\newcommand{\B}[1]{\mathbf{#1}}
\newcommand{\M}[1]{\boldsymbol{#1}}

\newcommand{\C}{\mathbbmss{C}}

\newcommand{\W}[1]{\M{\mathscr{W}}_{#1}}

\newcommand{\Size}[1]{\ensuremath{\left|#1\right|}}
\newcommand{\Ceil}[1]{\ensuremath{\left\lceil#1\right\rceil}}
\newcommand{\Floor}[1]{\ensuremath{\left\lfloor#1\right\rfloor}}

\newcommand{\lra}[1]{\ensuremath{(#1)}}
\newcommand{\lrb}[1]{\ensuremath{[#1]}}
\newcommand{\lrc}[1]{\ensuremath{\{#1\}}}
\newcommand{\lrd}[1]{\ensuremath{\langle #1\rangle}}

\newcommand{\lrA}[1]{\ensuremath{\left(#1\right)}}
\newcommand{\lrB}[1]{\ensuremath{\left[#1\right]}}
\newcommand{\lrC}[1]{\ensuremath{\left\{#1\right\}}}

\DeclareAcronym{1D}{
short = 1D,
long = one-dimensional
}

\DeclareAcronym{2D}{
short = 2D,
long = two-dimensional
}

\DeclareAcronym{3D}{
short = 3D,
long = three-dimensional
}

\DeclareAcronym{5G}{
short = 5G,
long = fifth generation
}

\DeclareAcronym{6G}{
short = 6G,
long = sixth-generation
}

\DeclareAcronym{AI}{
short = AI,
long = artificial intelligence
}

\DeclareAcronym{ANN}{
short = ANN,
long = artificial neural network
}

\DeclareAcronym{AR}{
short = AR,
long = augmented reality
}

\DeclareAcronym{AO}{
short = AO,
long = absorptive object
}

\DeclareAcronym{AWGN}{
short = AWGN,
long = additive white Gaussian noise
}

\DeclareAcronym{BB84}{
short = BB84,
long = Bennett and Brassard 1984
}

\DeclareAcronym{BEP}{
short = BEP,
long = bit error probability
}

\DeclareAcronym{BLIP}{
short = BLIP,
long = bootstrapping language-image pretraining
}

\DeclareAcronym{BPSK}{
short = BPSK,
long = binary phase-shift keying
}

\DeclareAcronym{BS}{
short = BS,
long = base station
}

\DeclareAcronym{BSM}{
short = BSM,
long = Bell-state measurement
}

\DeclareAcronym{CKA}{
short = CKA,
long = conference key agreement
}

\DeclareAcronym{CNN}{
short = CNN,
long = convolutional neural network
}

\DeclareAcronym{CNOT}{
short = CNOT,
long = controlled-NOT
}

\DeclareAcronym{CPS}{
short = CPS,
long = cyber-physical system
}

\DeclareAcronym{CPTP}{
short = CPTP,
long = completely positive trace-preserving
}

\DeclareAcronym{CPU}{
short = CPU,
long = central processing unit
}

\DeclareAcronym{CSI}{
short = CSI,
long = channel state information
}

\DeclareAcronym{CV}{
short = CV,
long = continuous-variable 
}

\DeclareAcronym{DNN}{
short = DNN,
long = deep neural network
}

\DeclareAcronym{DL}{
short = DL,
long = deep learning
}

\DeclareAcronym{DQL}{
short = DQL,
long = deep Q-learning
}

\DeclareAcronym{DRL}{
short = DRL,
alt = deep RL,
long = deep reinforcement learning
}

\DeclareAcronym{DT}{
short = DT,
long = digital twin
}

\DeclareAcronym{DV}{
short = DV,
long = discrete-variable 
}

\DeclareAcronym{FD}{
short = FD,
long = full-duplex
}

\DeclareAcronym{GAN}{
short = GAN,
long = generative adversarial network
}

\DeclareAcronym{GEO}{
short = GEO,
long = geostationary Earth orbit
}

\DeclareAcronym{GHZ}{
short = GHZ,
long = Greenberger--Horne--Zeilinger
}

\DeclareAcronym{GPS}{
short = GPS,
long = global positioning system
}

\DeclareAcronym{GPT}{
short = GPT,
long = generative pretrained transformer
}

\DeclareAcronym{GPU}{
short = GPU,
long = graphics processing unit 
}

\DeclareAcronym{GUS}{
short = GUS,
long = geometrically uniform symmetry
}

\DeclareAcronym{HQC}{
short = HQC,
long = hybrid quantum-classical
}

\DeclareAcronym{IIoT}{
short = IIoT,
alt = industrial IoT,
long = industrial Internet of Things
}

\DeclareAcronym{IoIV}{
short = IoIV,
long = Internet of intelligent Vehicles
}

\DeclareAcronym{IoT}{
short = IoT,
long = Internet of Things
}

\DeclareAcronym{ISAC}{
short = ISAC,
long = integrated sensing and communication
}

\DeclareAcronym{KPI}{
short = KPI,
long = key performance indicator
}

\DeclareAcronym{LEO}{
short = LEO,
long = low Earth orbit
}

\DeclareAcronym{LLaMA}{
short = LLaMA,
alt = large language model Meta AI,
long = LLM Meta AI
}

\DeclareAcronym{LLM}{
short = LLM,
long = large language model
}

\DeclareAcronym{LPIPS}{
short = LPIPS,
long = learned perceptual image patch similarity
}

\DeclareAcronym{LOCC}{
short = LOCC,
long = local operations and classical communication
}

\DeclareAcronym{LSTM}{
short = LSTM,
long = long short-term memory
}

\DeclareAcronym{MEO}{
short = MEO,
long = medium Earth orbit
}

\DeclareAcronym{MIMO}{
short = MIMO,
long = multiple-input multiple-output
}

\DeclareAcronym{mmWave}{
short = mmWave,
long = millimeter-wave
}

\DeclareAcronym{ML}{
short = ML,
long = machine learning
}

\DeclareAcronym{MLP}{
short = MLP,
long = multi-layer perceptron
}

\DeclareAcronym{MPA}{
short = MPA,
long = mean pixel accuracy
}

\DeclareAcronym{MR}{
short = MR,
long = mixed reality
}

\DeclareAcronym{MSE}{
short = MSE,
long = mean squared error
}

\DeclareAcronym{NISQ}{
short = NISQ,
long = noisy intermediate-scale quantum
}

\DeclareAcronym{NMSE}{
short = NMSE,
alt = normalized MSE,
long = normalized mean squared error
}

\DeclareAcronym{NLP}{
short = NLP,
long = natural language processing
}

\DeclareAcronym{NTN}{
short = NTN,
long = non-terrestrial network
}

\DeclareAcronym{PISQ}{
short = PISQ,
long = perfect intermediate-scale quantum
}

\DeclareAcronym{POVM}{
short = POVM,
long = positive operator-valued measure
}

\DeclareAcronym{PQC}{
short = PQC,
long = parametrized quantum circuit
}

\DeclareAcronym{QAA}{
short = QAA,
long = quantum anonymous authentication
}

\DeclareAcronym{QAB}{
short = QAB,
long = quantum anonymous broadcast
}

\DeclareAcronym{QAC}{
short = QAC,
long = quantum anonymous communication
}

\DeclareAcronym{QACKA}{
short = QA-CKA,
alt = quantum anonymous CKA,
long = quantum anonymous conference key agreement
}

\DeclareAcronym{QACD}{
short = QACD,
long = quantum anonymous collision detection
}

\DeclareAcronym{QAE}{
short = QAE,
long = quantum anonymous entanglement
}

\DeclareAcronym{QAIA}{
short = QAIA,
long = quantum anonymous identity authentication
}

\DeclareAcronym{QAIR}{
short = QAIR,
long = quantum anonymous information retrieval
}

\DeclareAcronym{QAM}{
short = QAM,
long = quadrature amplitude modulation
}

\DeclareAcronym{QAN}{
short = QAN,
long = quantum anonymous network
}

\DeclareAcronym{qAN}{
short = QAN,
long = quantum anonymous notification
}

\DeclareAcronym{QANO}{
short = QANO,
long = quantum anonymous notification
}

\DeclareAcronym{QAOA}{
short = QAOA,
long = quantum approximate optimization algorithm
}

\DeclareAcronym{QAP}{
short = QAP,
long = quantum anonymous publication
}

\DeclareAcronym{QAR}{
short = QAR,
long = quantum anonymous ranking
}

\DeclareAcronym{QAS}{
short = QAS,
long = quantum anonymous sensing
}

\DeclareAcronym{QAT}{
short = QAT,
long = quantum anonymous teleportation
}

\DeclareAcronym{QAV}{
short = QAV,
long = quantum anonymous voting
}

\DeclareAcronym{QCC}{
short = QCC,
long = quantum covert communication
}

\DeclareAcronym{QCRB}{
short = QCRB,
long = quantum Cram\'er--Rao bound
}

\DeclareAcronym{QCS}{
short = QCS,
long = quantum clock synchronization
}

\DeclareAcronym{QEC}{
short = QEC,
long = quantum error correction
}

\DeclareAcronym{QED}{
short = QED,
long = quantum entanglement distribution
}

\DeclareAcronym{QFI}{
short = QFI,
long = quantum Fisher information
}

\DeclareAcronym{QFIM}{
short = QFIM,
alt = QFI matrix,
long = quantum Fisher information matrix
}

\DeclareAcronym{QFT}{
short = QFT,
long = quantum Fourier transform
}

\DeclareAcronym{QIoT}{
short = QIoT,
alt = quantum IoT,
long = quantum Internet of Things
}

\DeclareAcronym{QKD}{
short = QKD,
long = quantum key distribution
}

\DeclareAcronym{QLSTM}{
short = QLSTM,
alt = quantum LSTM,
long = quantum long short-term memory
}

\DeclareAcronym{QML}{
short = QML,
alt = quantum ML,
long = quantum machine learning
}

\DeclareAcronym{QNN}{
short = QNN,
long = quantum neural network
}

\DeclareAcronym{QOC}{
short = QOC,
long = quantum optimal control
}

\DeclareAcronym{QPU}{
short = QPU,
long = quantum processing unit
}

\DeclareAcronym{QSC}{
short = QSC,
alt = quantum SC,
long = quantum semantic communication
}

\DeclareAcronym{QSN}{
short = QSN,
long = quantum sensing network
}

\DeclareAcronym{QUBO}{
short = QUBO,
long = quadratic unconstrained binary optimization
}

\DeclareAcronym{ReLU}{
short = ReLU,
long = rectified linear unit
}

\DeclareAcronym{ResNet}{
short = ResNet,
long = residual network
}

\DeclareAcronym{RGB}{
short = RGB,
long = red-green-blue
}

\DeclareAcronym{RL}{
short = RL,
long = reinforcement learning
}

\DeclareAcronym{RMSE}{
short = RMSE,
alt = root MSE,
long = root mean squared error
}

\DeclareAcronym{RNN}{
short = RNN,
long = recurrent neural network
}

\DeclareAcronym{RSA}{
short = RSA,
long = Rivest--Shamir--Adleman
}

\DeclareAcronym{SAM}{
short = SAM,
long = segment anything model
}

\DeclareAcronym{SC}{
short = SC,
long = semantic communication
}

\DeclareAcronym{SEP}{
short = SEP,
long = symbol error probability
}

\DeclareAcronym{SNR}{
short = SNR,
long = signal-to-noise ratio
}

\DeclareAcronym{SRM}{
short = SRM,
long = square-root measurement
}

\DeclareAcronym{SQL}{
short = SQL,
long = standard quantum limit
}

\DeclareAcronym{SVM}{
short = SVM,
long = support vector machine
}

\DeclareAcronym{SwinT}{
short = SwinT,
long = shifted window transformer
}

\DeclareAcronym{THz}{
short = THz,
long = terahertz
}

\DeclareAcronym{TN}{
short = TN,
long = terrestrial network
}

\DeclareAcronym{UAV}{
short = UAV,
long = unmanned aerial vehicle
}

\DeclareAcronym{U-Net}{
short = U-Net,
long = U-shaped network
}

\DeclareAcronym{VAE}{
short = VAE,
long = variational autoencoder
}

\DeclareAcronym{V2I}{
short = V2I,
long = vehicle-to-infrastructure
}

\DeclareAcronym{V2X}{
short = V2X,
long = vehicle-to-everything
}

\DeclareAcronym{VFM}{
short = VFM,
long = vision foundation model
}

\DeclareAcronym{ViT}{
short = ViT,
long = vision transformer
}

\DeclareAcronym{VQA}{
short = VQA,
long = variational quantum algorithm
}

\DeclareAcronym{VQC}{
short = VQC,
long = variational quantum circuit
}

\DeclareAcronym{VQS}{
short = VQS,
long = variational quantum sensing
}

\DeclareAcronym{VR}{
short = VR,
long = virtual reality
}

\DeclareAcronym{YOLO}{
short = YOLO,
long = you only look once
}

\usepackage{tikz}
\usetikzlibrary{calc,positioning}

\tikzset{
	timeline axis/.style={line width=1pt},
	timeline tick/.style={line width=1pt},
	timeline guide/.style={densely dotted, line width=0.9pt},
}

\newcounter{tlflip}

\newcommand{\entry}[4][]{%
	\pgfmathsetmacro{\stage}{#2}%
	\pgfmathsetmacro{\x}{#3-\tlstart}%
	\pgfmathsetmacro{\y}{0.9*\stage}%
	
	\draw[timeline tick] (\x,0.2) -- (\x,-0.2);
	
	\def\side{#1}%
	\def\sideup{up}%
	\def\sidedown{down}%
	\def\sideupleft{upleft}%
	\def\sideupright{upright}%
	\def\sidedownleft{downleft}%
	\def\sidedownright{downright}%
	
	\ifx\side\empty
	\ifodd\value{tlflip}
	\def\side{up}%
	\else
	\def\side{down}%
	\fi
	\fi
	
	\ifx\side\sideup
	\node[above,align=center] at (\x,0.35) {#3};
	\draw[timeline guide] (\x,0.90) -- (\x,\y);
	\node[above,align=center] at (\x,\y) {#4};
	
	\else\ifx\side\sidedown
	\node[below,align=center] at (\x,-0.35) {#3};
	\draw[timeline guide] (\x,-0.90) -- (\x,-\y);
	\node[below,align=center] at (\x,-\y) {#4};
	
	\else\ifx\side\sideupleft
	\node[above,align=center] at (\x,0.35) {#3};
	\pgfmathsetmacro{\xt}{\x-3.0} % distance to the left
	\draw[timeline guide] (\x,0.9) -- (\xt+0.2,\y-0.2); % angled connector
	\node[above,align=center] at (\xt,\y) {#4};
	
	\else\ifx\side\sideupright
	\node[above,align=center] at (\x,0.35) {#3};
	\pgfmathsetmacro{\xt}{\x+3.0}
	\draw[timeline guide] (\x,0.9) -- (\xt-0.2,\y-0.2);
	\node[above,align=center] at (\xt,\y) {#4};
	
	\else\ifx\side\sidedownleft
	\node[below,align=center] at (\x,-0.35) {#3};
	\pgfmathsetmacro{\xt}{\x-3.0}
	\draw[timeline guide] (\x,-0.9) -- (\xt+0.2,-\y+0.2);
	\node[below,align=center] at (\xt,-\y) {#4};
	
	\else\ifx\side\sidedownright
	\node[below,align=center] at (\x,-0.35) {#3};
	\pgfmathsetmacro{\xt}{\x+3.0}
	\draw[timeline guide] (\x,-0.9) -- (\xt-0.2,-\y+0.2);
	\node[below,align=center] at (\xt,-\y) {#4};
	
	\else
	\node[above,align=center] at (\x,0.35) {#3};
	\draw[timeline guide] (\x,0.90) -- (\x,\y);
	\node[above,align=center] at (\x,\y) {#4};
	\fi\fi\fi\fi\fi\fi
	
	\addtocounter{tlflip}{1}%
}

\tikzset{
	vtaxis/.style={line width=1.0pt},
	vtmark/.style={line width=1.0pt},
	vtlink/.style={line width=0.7pt},
}

\newcounter{vtLine}        % counts printed lines (not entries)
\newcounter{vtTotalLines}  % total printed lines (for centering)

\def\vtRowSep{0.45}   % spacing between EVERY line (uniform!)
\def\vtStartY{0}       % computed dynamically
\def\vtAllCount{}
\def\vtAllDraw{}
\def\vtlastyear{*}

\makeatletter
\newcount\vt@linecount

\def\vtCountLines#1{%
	\vt@linecount=0\relax
	\expandafter\vtCountLinesAux#1\\\vt@nil
}
\def\vtCountLinesAux#1\\#2\vt@nil{%
	\advance\vt@linecount by 1\relax
	\def\vt@next{#2}%
	\ifx\vt@next\empty
	\else
	\expandafter\vtCountLinesAux#2\vt@nil
	\fi
}

\def\vtDrawLines#1{%
	\expandafter\vtDrawLinesAux#1\\\vt@nil
}
\def\vtDrawLinesAux#1\\#2\vt@nil{%
	\stepcounter{vtLine}%
	\pgfmathsetmacro{\idx}{\value{vtLine}-1}%
	\pgfmathsetmacro{\y}{\vtStartY + \idx*\vtRowSep}%
\node[anchor=west, align=left] at (\vtTextX,\y) {#1};

	\def\vt@next{#2}%
	\ifx\vt@next\empty
	\else
	\expandafter\vtDrawLinesAux#2\vt@nil
	\fi
}

\newcommand{\vtDoEntry}[2]{%
	\edef\thisyear{#1}%
	\pgfmathsetmacro{\ybase}{\vtyearscale*(#1-\vtstartyear)}%
	
	\ifx\thisyear\vtlastyear
	\else
	\node[anchor=east] at (\vtYearX,\ybase) {#1};
	\draw[vtmark] (\vtAxisX,\ybase) -- (\vtAxisX+0.01,\ybase);
	\edef\vtlastyear{#1}%
	\fi
	
	\pgfmathsetmacro{\idxfirst}{\value{vtLine}}%
	\pgfmathsetmacro{\yfirst}{\vtStartY + \idxfirst*\vtRowSep}%
	\draw[vtlink] (\vtAxisX+0.01,\ybase) -- (\vtDiagX,\yfirst) -- (\vtTextX,\yfirst);
	
	\vtDrawLines{#2}%
}

\newcommand{\vtCountEntry}[2]{%
	\vtCountLines{#2}%
	\addtocounter{vtTotalLines}{\the\vt@linecount}%
}

\newcommand{\ventry}[2]{%
	\g@addto@macro\vtAllCount{\vtCountEntry{#1}{#2}}%
	\g@addto@macro\vtAllDraw{\vtDoEntry{#1}{#2}}%
}
\makeatother

\newenvironment{vtimeline}[2]{%
	\def\vtstartyear{#1}%
	\def\vtendyear{#2}%
	
	\pgfmathsetmacro{\vtyearscale}{0.6}%
	\pgfmathsetmacro{\vtaxisend}{\vtyearscale*(#2-#1)+0}%
	
	\def\vtAxisX{0.04}
	\def\vtYearX{0.03}
	\def\vtTextX{0.08}
	\def\vtDiagX{0.07}
	
	\def\vtAllCount{}%
	\def\vtAllDraw{}%
	\def\vtlastyear{*}%
	\setcounter{vtLine}{0}%
	\setcounter{vtTotalLines}{0}%
	\def\vtStartY{0}%
	
	\begin{tikzpicture}[x=\linewidth,y=-1cm]
		\draw[vtaxis,-stealth] (\vtAxisX,-0.5) -- (\vtAxisX,\vtaxisend+0.5);
	}{%
		\vtAllCount
		
		\ifnum\value{vtTotalLines}>0
		\pgfmathsetmacro{\vtstackH}{(\value{vtTotalLines}-1)*\vtRowSep}%
		\pgfmathsetmacro{\tmp}{(\vtaxisend-\vtstackH)/2}%
		\pgfmathsetmacro{\tmpclipped}{max(0,\tmp)}%
		\xdef\vtStartY{\tmpclipped}%
		\else
		\xdef\vtStartY{0}%
		\fi
		
		\setcounter{vtLine}{0}%
		\def\vtlastyear{*}%
		\vtAllDraw
		
		\path[use as bounding box] (0,0) rectangle (1,\vtaxisend+1);
	\end{tikzpicture}%
}

\usepackage{makecell}
\usepackage{comment}
\usepackage[colorlinks=true,
            linkcolor=black,
            citecolor=black,
            urlcolor=black]{hyperref}

\usepackage{algorithm}
\usepackage{algorithmic}

\usepackage{booktabs}
\usepackage{fixmath}
\usepackage{bm}
\usepackage{amsbsy}
\usepackage{color}
\usepackage{verbatim}
\usepackage{array}
\usepackage{tikz}
\usetikzlibrary{positioning}

\usetikzlibrary{calc}
\usetikzlibrary{decorations.pathmorphing}
\def\B{\mathcal{B}}
\def\C{\mathcal{C}}
\def\E{\mathcal{E}}

\def\G{\mathcal{G}}
\def\HH{\mathcal{H}}

\def\M{\mathcal{M}}

\def\P{\mathcal{P}}

\def\SS{\mathcal{S}}
\def\T{\mathcal{T}}

\def\V{\mathcal{V}}
\def\W{\mathcal{W}}
\def\X{\mathcal{X}}
\def\Y{\mathcal{Y}}

\DeclareMathOperator{\depth}{\mathrm{depth}}
\DeclareMathOperator{\odd}{\mathrm{Odd}}
\def\c{\mathrm{c}}
\def\w{\hat{w}}
\newcommand{\RR}{\mathbbmss{R}}

\DeclareAcronym{BTSP}{
short = BTSP,
alt = biased TSP,
long = biased traveling salesman problem
}

\DeclareAcronym{FPT}{
short = FPT,
long = fixed-parameter tractable
}

\DeclareAcronym{LP}{
short = LP,
long = linear programming
}

\DeclareAcronym{MST}{
short = MST,
long = minimum-cost spanning tree
}

\DeclareAcronym{PCB}{
short = PCB,
long = printed circuit board
}

\DeclareAcronym{SGPP}{
short = SGPP,
long = subgroup path planning problem
}
\DeclareAcronym{SPP}{
short = SPP,
long = the subpath planning problem
}

\DeclareAcronym{TSP}{
short = TSP,
long = traveling salesman problem
}
\DeclareAcronym{TSPP}{
short = TSPP,
long = traveling salesman path problem
}
\DeclareAcronym{TSPPK}{
short = $k$-TSPP,
long = $k$-person traveling salesman path problem
}
\DeclareAcronym{XP}{
short = XP,
long = slice-wise polynomial-time
}

\begin{document}

\title{
\scalebox{0.95}[1]{Polynomial $\lra{2\alpha+\varepsilon}$- and Quadratic $4$-Approximation}
Algorithms for the Biased Traveling Salesman Problem in Bipartitioned Networks
}
%\IEEEspecialpapernotice{(IEEE JournaI Template)}
\author{
Jingyang~Zhao,
Trung~Q.~Duong,~\IEEEmembership{Fellow,~IEEE},
Een-Kee~Hong,~\IEEEmembership{Senior~Member,~IEEE},
and
Hyundong~Shin,~\IEEEmembership{Fellow,~IEEE}
 %\thanks{
 %This work was supported by the National Research Foundation of Korea (NRF) grant funded by the Korean government
 %(MSIT) (No. 2019R1A2C2007037 and 2022R1A4A3033401) and by the MSIT (Ministry of Science and ICT), Korea, under the ITRC
 %(Information Technology Research Center) support program (IITP-2021-0-02046) supervised by the IITP
 %(Institute for Information $\&$ Communications Technology Planning $\&$ Evaluation).
 %\emph{(Corresponding author: Hyundong~Shin.)}
 %}
 \thanks{
 J.~Zhao,
 E.-K.~Hong,
 and H.~Shin (corresponding author)
 are with the Department of Electronics and Information Convergence Engineering,
 Kyung Hee University,
 1732 Deogyeong-daero, Giheung-gu,
 Yongin-si, Gyeonggi-do 17104, Republic of Korea
 (e-mail: hshin@khu.ac.kr). 
 }
\thanks{
T.~Q.~Duong is with
the Faculty of Engineering and Applied Science, 
Memorial University, St. John's, NL A1C 5S7, Canada, 
and also with 
the School of Electronics, Electrical Engineering and Computer Science, 
Queen's University Belfast, 
Belfast BT7 1NN, Belfast, UK
(e-mail: tduong@mun.ca). 
}
 }
 \markboth{}{ 
 Zhao \textit{\MakeLowercase{et al.}}:
%Polynomial $\lra{2\alpha+\varepsilon}$- and Quadratic $4$-
Approximation Algorithms for the Biased Traveling Salesman Problem in Bipartitioned Networks
 }

\maketitle

\begin{abstract}
The \ac{TSP} is one of the fundamental problems in combinatorial optimization that asks for a minimum-cost Hamiltonian cycle in a complete undirected graph with $N$ vertices. For the metric \ac{TSP}, where edge costs satisfy the triangle inequality, a long line of work has produced constant-factor approximation algorithms, culminating in the best-known approximation ratio $\alpha<3/2-10^{-36}$. The biased \ac{TSP} (\acs{BTSP}) generalizes the metric \ac{TSP} to bipartitioned networks: the vertex set is partitioned into two subsets, edge costs satisfy the triangle inequality within each subset, while edges crossing the partition may violate it by a certain bias factor.
This model captures heterogeneous routing scenarios in which distances are metric only locally but may be distorted across different regions or vertex types. Previously, the only known result for the \ac{BTSP} was a $7/2$-approximation algorithm with running time $O\lra{N^3}$.
In this paper, we present two new approximation algorithms that significantly improve the state of the art. The first achieves an approximation ratio of $2\alpha+\varepsilon$ (strictly less than $3$) in polynomial time $N^{O\lrA{1}}$ for any constant $\varepsilon>0$, where the exponent grows at least linearly with $1/\varepsilon$. The second achieves a $4$-approximation in only $O\lra{N^2}$ running time, reducing the previous cubic-time bound by a full polynomial factor. 
These advances directly support emerging networked systems---including multi-tier \acs{6G} architectures and \acl{HQC} networks---where cross-region transitions incur biased and non-metric overheads.
%These advances apply directly to emerging networked systems, including heterogeneous architectures that integrate terrestrial, aerial, and non-terrestrial tiers, as well as \ac{HQC} communication networks where cross-segment transitions through quantum repeaters or \ac{HQC} interfaces incur biased, non-metric overheads.
\end{abstract}

\begin{IEEEkeywords}
Approximation algorithms,
bipartitioned networks,
combinatorial optimization,
\aclp{TSP}.
\end{IEEEkeywords}

%\section*{Acronyms}

%\noindent
%\begin{tabular}{ll}
%\Acs{AI} & 	\Acl{AI}\\
%\end{tabular}

\acresetall		% reset the acronyms 

%%%%%%%%%%%%%%%%%%%%%%%%%%%%%%%%%%%%%%%%%%%%%%%%%%%%%%%%%%%%%%%%%
%%%%%%%%%%%%%%%%%%%%%%%%%%%%%%%%%%%%%%%%%%%%%%%%%%%%%%%%%%%%%%%%%
\section{Introduction}

\IEEEPARstart{N}{etworked} systems encompass a wide range of routing problems in combinatorial optimization, of which the \ac{TSP} is among the most fundamental and extensively investigated~\cite{TV:2024:book}. % ZKPZZS:26:arXiv 
Its enduring prominence is driven by both its profound theoretical significance in computational complexity and its ubiquity in real-world applications~\cite{ZKPZZS:26:IEEE_O_CSTO,TLLY:25:IEEE_J_EVC,GMZTCD:25:IEEE_J_EVC,BMMD:24:IEEE_J_ITS,LIY:24:IEEE_J_ITS,LXHZYI:23:IEEE_J_EVC,JH:22:IEEE_J_EVC,ZZCSTZWD:2022:IEEE_J_ITS,XLZ:21:IEEE_J_ITS,MLZD:21:IEEE_J_ITS,YHHW:19:IEEE_J_EVC,WCLY:16:IEEE_J_ITS,ABC:2006:book}, ranging from logistics and manufacturing to genetics, intelligent transportation networks, and quantum algorithms~\cite{ZVCZ:26:IEEE_J_ITS,JZWMCL:25:IEEE_J_ITS,ZLZBYMZ:25:IEEE_J_ITS,CLXLEC:24:IEEE_J_ITS,LDNXW:24:IEEE_J_ITS,LZC:23:IEEE_J_ITS,KJKJJ:23:IEEE_J_ITS,JMZ:22:IEEE_J_EVC,ZXFZ:22:IEEE_J_ITS,LGN:21:IEEE_J_ITS,ZMZZ:20:IEEE_J_ITS,WST:20:IEEE_J_ITS,XLZ:20:IEEE_J_ITS,FHTGTTO:20:IEEE_J_ITS,ZQL:19:IEEE_J_ITS,AA:18:IEEE_J_ITS,MLDD:17:IEEE_J_ITS,KOHTZ:15:IEEE_J_ITS,CYL:2025:nQI,NRSTD:25:IEEE_J_ITS,SCYB:2023:nQI,GA:23:IEEE_J_ITS,ABAPF:22:IEEE_J_ITS,GKZ:2022:nQI}.
An instance of the \ac{TSP} is defined on a complete undirected graph $\G=\lrA{\V,\E}$ with vertex set $\V=\lrC{v_1, v_2,\ldots, v_N}$, edge set $\E$, and a non-negative edge-cost function $w: \V \times \V \to \RR_{\geq 0}$. The objective is to compute a Hamiltonian cycle 
$\sigma=\lrd{v_{i_1},v_{i_2},\ldots, v_{i_N},v_{i_1}}$ of minimum cost, where the cost of the cycle is
\begin{align}
w\lrA{\sigma}\coloneqq\sum_{j=1}^{N-1} w\lrA{v_{i_j},v_{i_{j+1}}}+w\lrA{v_{i_N},v_{i_1}}
\end{align}
and $\lrC{i_1,i_2,\ldots,i_N}$ is a permutation of $\lrB{N}=\lrC{1,2,\ldots,N}$.

\subsection{Background and Motivation}

In its most general form, the \ac{TSP} is notoriously intractable. It is not only NP-hard~\cite{GJ:1979:book} but also cannot be approximated within any computable function in polynomial time unless $\mathscr{P}=\mathscr{NP}$, where $\mathscr{P}$ denotes the class of decision problems that can be solved in polynomial time, whereas $\mathscr{NP}$ denotes the class of decision problems for which a proposed solution can be verified in polynomial time~\cite{SG:76:JACM}. 
For any minimization problem $\P$, a $\rho$-approximation algorithm is a polynomial-time algorithm that, for every instance of $\P$, computes a solution whose objective value is at most $\rho$ times the value of an optimal solution~\cite{WS:2011:book}.

Fortunately, strong positive results emerge when the edge costs reflect realistic spatial or network distances---specifically, when the cost function $w$ is a \emph{metric}. In the metric \ac{TSP}, the cost function satisfies $w\lrA{a,a}=0$, $w\lrA{a,b}=w\lrA{b,a}$ (symmetry), and $w\lrA{a,b}+w\lrA{b,c}\geq w\lrA{a,c}$ (the triangle inequality) for all $a,b,c\in \V$. 
Under this metric constraint, the problem becomes highly approximable. Classical milestones include the $2$-approximation double-tree algorithm~\cite{WS:2011:book} and the celebrated $3/2$-approximation algorithm discovered in~\cite{C:22:ORF}. 
Only recently was this long-standing $3/2$-approximation barrier broken, with a polynomial-time algorithm achieving an approximation ratio strictly less than $3/2 - 10^{-36}$~\cite{KKG:21:STOC,KKG:23:IPCO}.
In what follows, we denote this state-of-the-art approximation ratio for the metric \ac{TSP} by $\alpha$.

In practice, however, real-world data is rarely perfectly metric. 
Measurement errors, operational overheads, and heterogeneous network topologies often introduce local violations of the triangle inequality. 
To bridge the approximability gap between general and metric graphs, a substantial body of literature has been dedicated to exploring various relaxed models of the \ac{TSP}~\cite{AB:95:SIAMDM,GGMSTZ:10:IEEE_J_EPM,SHES:16:IJCAI,SYKK:17:IJCAI,ZLG:22:ISAAC,ZZG:25:COCOON,ZXPX:25:AAAI,ZSX:26:AAAI}. 
These models typically aim to capture structured deviations from metricity, thereby enabling meaningful algorithmic guarantees while retaining relevance to practical applications. Examples include bounded-metric violations of the triangle inequality and hybrid settings where only a subset of edges violates the metric property.

Prominent among these models, and the main focus of this paper, is the biased \ac{TSP} (\acs{BTSP})~\cite{MRM:17:TCS}, a variant defined on bipartitioned networks in which non-metricity arises from interactions across distinct structural regions. 
In the \ac{BTSP}, the vertex set $\V$ is bipartitioned into two disjoint subsets $\V_1$ and $\V_2$. 
The cost function behaves as a pure metric within the induced subgraphs $\G\lrB{\V_1}$ and $\G\lrB{\V_2}$.
However, the cost of any edge $uv$ connecting $\V_1$ and $\V_2$ may be scaled by an arbitrary bias factor $\beta\lrA{u,v}\geq 1$, potentially violating the triangle inequality. 
Consequently, all non-metric effects are confined to interactions across the partition, yielding a controlled and analyzable form of deviation from the standard metric setting.
This structural restriction provides a natural balance between expressiveness and tractability. On the one hand, it generalizes the classical metric \ac{TSP} by allowing heterogeneous costs across different regions; on the other hand, it preserves enough structure to enable the design of approximation algorithms with provable performance guarantees. Notably, \cite{MRM:17:TCS} showed that classical approximation algorithms for the metric \ac{TSP}~\cite{WS:2011:book,C:22:ORF} fail to achieve any bounded approximation ratio for the \ac{BTSP}.

From an application perspective, the \ac{BTSP} naturally captures scenarios such as heterogeneous fleet routing~\cite{BBV:2008:incollection}, where different vehicle types operate under distinct cost regimes, or \acl{PCB} drilling with tool changes~\cite{MRM:17:TCS}, where movements within the same operational mode are metric, but switching between modes incurs additional, possibly asymmetric overheads. Similar patterns also arise in multimodal transportation systems and manufacturing processes with setup-dependent costs~\cite{AB:95:SIAMDM,GGMSTZ:10:IEEE_J_EPM,SHES:16:IJCAI,SYKK:17:IJCAI}.
More recently, similar bipartitioned cost structures have emerged in \acs{6G} communication networks. In \ac{6G} architectures, the tight integration of terrestrial networks with \aclp{NTN}---including \acl{LEO} satellites, high-altitude platforms, and \acl{UAV} relays---creates a fundamentally bipartitioned routing landscape: propagation within each tier exhibits well-behaved metric structure, while cross-tier handovers and inter-domain handoffs introduce non-metric overheads stemming from protocol translation, beam realignment, and link adaptation \cite{BHS:25:IEEE_J_JSAC,KPN:25:IEEE_M_WC,KBR:25:IEEE_M_NET}. Analogous structures arise in \acl{HQC} communication networks, where intra-segment routing is metric but cross-segment transitions through quantum repeaters or hybrid interfaces incur additional, asymmetric overheads owing to entanglement swapping, decoherence-aware scheduling, and trust-zone re-establishment \cite{KUW:24:IEEE_M_COM,ZPFSW:23:IEEE_J_JSAC,ZFUJSW:23:IEEE_M_WC}. The \ac{BTSP} thus serves as a natural algorithmic abstraction for routing and tour optimization in such heterogeneous networked systems, where local metric behavior coexists with structurally biased cross-region transitions.

Despite its appealing structure, the \ac{BTSP} remains algorithmically challenging. The interaction between metric substructures and non-metric cross edges complicates both algorithm design and analysis. Existing approaches typically rely on decomposing the problem into intra-partition tours and carefully controlling the number and cost of cross edges. For this problem, the current state-of-the-art is a $7/2$-approximation algorithm~\cite{MRM:17:TCS}. Bridging the gap between this bound and the $3/2$ ratio known for the metric \ac{TSP} remains an intriguing open direction, motivating further investigation into improved structural insights and algorithmic techniques.

\subsection{Contributions}
In this work, we significantly advance the state of the art for the \ac{BTSP} by presenting two new approximation algorithms, each employing a distinct algorithmic perspective.

\subsubsection{An Improved $\lrA{2\alpha+\varepsilon}$-Approximation Algorithm in Polynomial Time} 
Our first algorithm achieves an approximation ratio of $2\alpha+\varepsilon$---which is strictly less than $3$, given $\alpha < 3/2$---in polynomial time for any constant $\varepsilon>0$.
This result relies on a new algorithmic framework that structurally connects the \ac{BTSP} to the well-known \ac{TSPPK}. It leverages an approximation algorithm for the metric \ac{TSP} as a subroutine alongside a structural reduction to the \ac{TSPPK}~\cite{TV:2024:book}.
By fixing an arbitrary minimum-cost Hamiltonian cycle, we reveal that the approximability of the \ac{BTSP} is intimately linked to the number of cross edges---edges whose endpoints belong to different partition subsets---present in this fixed Hamiltonian cycle.

Specifically, when the number of cross edges is bounded by $O\lrA{1/\varepsilon}$, we can enumerate these cross edges in polynomial time and reduce the \ac{BTSP} to an equivalent \ac{TSPPK} instance. This instance can then be solved using an existing polynomial-time $\lrA{\alpha+\varepsilon}$-approximation algorithm~\cite{TV:2024:book}.
In the complementary case, when the Hamiltonian cycle contains $\Omega\lrA{1/\varepsilon}$ cross edges, we introduce a tailored patching technique. We enumerate merely two cross edges, say $uv$ and $u'v'$, and compute decoupled Hamiltonian paths: one between $u$ and $u'$ in the subgraph $\G\lrB{\V_1}$, and another between $v$ and $v'$ in $\G\lrB{\V_2}$. We prove that recombining these two Hamiltonian paths with the two cross edges yields a solution to the \ac{BTSP} with an approximation ratio of $2\alpha+O\lrA{\varepsilon}$.

Since the exact number of cross edges in the fixed Hamiltonian cycle is unknown in advance, our algorithm runs both strategies. Specifically, it iterates the \ac{TSPPK} reduction for all cross-edge counts from $0$ up to a threshold of $O\lrA{1/\varepsilon}$, while also applying the patching technique. 
By returning the best solution found across all executions, we obtain an overall approximation ratio of $2\alpha+O\lrA{\varepsilon}$.
This structural insight may also prove useful for designing constant-factor approximation algorithms for broader generalizations of the \ac{BTSP} with more than two partition subsets.

\subsubsection{A Fast $4$-Approximation Algorithm in $O\lra{N^2}$ Time}
Our second algorithm focuses on computational efficiency, achieving an approximation ratio of $4$ with an improved running time of $O\lra{N^2}$. While the previous approaches in~\cite{MRM:17:TCS} similarly relied on the structural properties of a \ac{MST}, they were ultimately bottlenecked by $O\lra{N^3}$-time operations, such as computing minimum-cost matchings. We bypass this cubic-time barrier by introducing efficient $O\lra{N^2}$-time subroutines to compute approximate Hamiltonian paths directly within the metric subgraphs. In addition, we leverage deeper structural insights in our analysis to ensure this speedup does not come at the cost of significant ratio degradation.

In particular, building upon the structural taxonomy established in~\cite{MRM:17:TCS}---which categorizes spanning trees based on the independence of their cross edges into \emph{0-\acp{MST}} (where all cross edges share a common endpoint) and \emph{1-\acp{MST}} (which contain at least two independent cross edges)---we significantly refine the algorithmic treatment of each topology. By redesigning the sub-algorithms for these distinct structures, we not only accelerate the overall running time but also establish strictly tighter approximation ratios for several critical subcases. 
For instance, we improve the approximation ratios in~\cite{MRM:17:TCS} from $3$ to $5/2$ for $0$-\acp{MST} with a single cross edge, from $3$ to $2$ for $0$-\acp{MST} with multiple cross edges, and from $3$ to $8/3$ for $1$-\acp{MST} whose component tree has depth $1$. 
This $8/3$-approximation ratio is achieved through an \emph{iterative pairwise merging} strategy that effectively resolves the topological bottlenecks that arise when merging three mutually-independent sub-structures.
Finally, these refined analytical frameworks and cost-control arguments provide deep structural insights that, we believe, will serve as the foundation for future $O\lra{N^2}$-time algorithms to break the $4$-approximation barrier.

The remainder of the paper is organized as follows. 
Section~\ref{sec:2} provides a detailed overview of the related literature and reviews the existing approximation results for path-based variants of the metric \ac{TSP}, with a particular focus on global relaxations, local scope relaxations, and parameterized approaches to non-metric \acp{TSP}. 
Section~\ref{sec:3} introduces notations and preliminaries of bipartitioned networks, detailing the biased edge-cost function and its key structural properties.
Section~\ref{sec:4} presents our polynomial-time $\lrA{2\alpha+\varepsilon}$-approximation algorithm, including its cross-edge enumeration and structural patching techniques.
Section~\ref{sec:5} provides our $O\lra{N^2}$-time $4$-approximation algorithm, systematically analyzing its application across distinct \ac{MST} topologies, including $0$-\acp{MST} and $1$-\acp{MST}.
Section~\ref{sec:6} concludes the paper.
Table~\ref{tab:acronyms} summarizes the acronyms used in this paper.

\begin{table*}[t]
    \caption{Important acronyms and their expansions}
    \label{tab:acronyms}
    \renewcommand{\arraystretch}{1.4}
    \providecommand {\cellLeft} [1] {\hspace{0.2cm}#1}
    \centering
    \begin{tabular}{ m{1.5cm}m{6.6cm}|m{1.5cm} m{6.6cm} }
        \hline
        \textbf{Acronym} & \textbf{Expansion} &
        \textbf{Acronym} & \textbf{Expansion} \\
        \hline
        \rowcolor[HTML]{C6ECE5}
        \cellLeft{$k$-TSPP} & $k$-Person Traveling Salesman Path Problem &
        \cellLeft{BTSP} & Biased Traveling Salesman Problem \\

        \cellLeft{FPT} & Fixed-Parameter Tractable &
        \cellLeft{LP} & Linear Programming \\

        \rowcolor[HTML]{C6ECE5}
        \cellLeft{MST} & Minimum-Cost Spanning Tree &
        \cellLeft{SGPP} & The Subgroup Path Planning Problem \\

        \cellLeft{SPP} & The Subpath Planning Problem &
        \cellLeft{TSP} & Traveling Salesman Problem \\

        \rowcolor[HTML]{C6ECE5}
        \cellLeft{TSPP} & Traveling Salesman Path Problem &
        \cellLeft{XP} & Slice-Wise Polynomial-Time \\
        \hline
    \end{tabular}
\end{table*}

% \begin{table*}[t]
%     \caption{Expansions of important acronyms}
%     \label{tab:acronyms}
%     \renewcommand{\arraystretch}{1.4}
%     \providecommand {\cellLeft} [1] {\hspace{0.2cm}#1}
%     \centering
%     \begin{tabular}{ m{1.5cm}m{6.6cm}|m{1.5cm} m{6.6cm} }
%         \hline
%         \textbf{Acronym} & \textbf{Expansion} &
%         \textbf{Acronym} & \textbf{Expansion} \\
%         \hline
%         \rowcolor[HTML]{C6ECE5}
%         \cellLeft{$k$-TSPP} & $k$-Person Traveling Salesman Path Problem &
%         \cellLeft{6G} & Sixth-Generation \\

%         \cellLeft{BTSP} & Biased Traveling Salesman Problem &
%         \cellLeft{FPT} & Fixed-Parameter Tractable \\

%         \rowcolor[HTML]{C6ECE5}
%         \cellLeft{HQC} & Hybrid Quantum-Classical &
%         \cellLeft{LEO} & Low-Earth-Orbit \\

%         \cellLeft{LP} & Linear Programming &
%         \cellLeft{MST} & Minimum-Cost Spanning Tree \\

%         \rowcolor[HTML]{C6ECE5}
%         \cellLeft{NTN} & Non-Terrestrial Network &
%         \cellLeft{PCB} & Printed Circuit Board \\

%         \cellLeft{SGPP} & The Subgroup Path Planning Problem &
%         \cellLeft{SPP} & The Subpath Planning Problem \\

%         \rowcolor[HTML]{C6ECE5}
%         \cellLeft{TSP} & Traveling Salesman Problem &
%         \cellLeft{TSPP} & Traveling Salesman Path Problem \\

%         \cellLeft{UAV} & Unmanned Aerial Vehicle &
%         \cellLeft{XP} & Slice-Wise Polynomial-Time \\
%         \hline
%     \end{tabular}
% \end{table*}

\section{Related Work}
\label{sec:2}

In this section, we review the literature most relevant to our study. We first examine approximation results for path-based variants of the metric \ac{TSP}, which serve as foundational subroutines in our proposed algorithms for the \ac{BTSP}. We then explore various theoretical models for non-metric \acp{TSP}, categorized into global relaxations, local scope relaxations, and recent parameterized approaches designed to handle structural non-metricity.
To facilitate comparison, Table~\ref{tab:related_work} summarizes representative path-based and non-metric \ac{TSP} variants related to the \ac{BTSP}, while Figure~\ref{fig:timeline} illustrates the chronological evolution of major algorithmic advances in these directions.

\subsection{Path-Based Variants of the Metric \ac{TSP}}

While the classical metric \ac{TSP} has been long and extensively studied~\cite{WS:2011:book,C:22:ORF,TV:2024:book}, breaking the long-standing $3/2$-approximation barrier proved remarkably difficult, motivating researchers to explore closely related path-based variants. In this subsection, we focus on the (metric) \ac{TSPP} and its multi-path generalization, the \ac{TSPPK}, as both problems play key roles in our algorithms.

\subsubsection{The \ac{TSPP}}
As the most natural path variant of the \ac{TSP}, the \ac{TSPP} seeks a minimum-cost Hamiltonian path between two pre-specified endpoints in a metric graph, instead of a closed Hamiltonian cycle. Despite the structural similarity between the \ac{TSP} and the \ac{TSPP}, their approximability gap was historically significant.
The first approximation algorithm for the \ac{TSPP} was introduced in~\cite{H:91:ORL}, achieving an approximation ratio of $5/3$. A major breakthrough came later in~\cite{AKS:15:JACM}, which proposed an elegant \ac{LP}-based algorithm that improved the approximation ratio to the golden ratio $\lra{\sqrt{5}+1}/2 < 1.619$. Subsequent research steadily refined this bound through advanced matching and patching techniques~\cite{S:13:IPCO,V:16:SIAMDM,GV:18:MP,S:16:FOCS}, eventually yielding approximation ratios of $3/2+\varepsilon$~\cite{TV:19:JACM} and exactly $3/2$~\cite{Z:19:SODA}. Currently, the state-of-the-art approximation ratio for the \ac{TSPP} is $\alpha+\varepsilon$~\cite{TVZ:22:SICOMP}, matching the best-known approximation ratio for the metric \ac{TSP} up to any fixed constant $\varepsilon > 0$.

\subsubsection{The \ac{TSPPK}}

This problem extends the \ac{TSPP} by introducing multiple source-destination pairs, acting as a bridge between single-path routing and complex, multi-segment spanning problems. Given $k$ pairs of vertices, the objective is to compute $k$ paths---one for each specified pair---such that every vertex in the graph is covered. When $k=1$, the problem naturally reduces to the \ac{TSPP}. Furthermore, when the source and destination of each pair coincide (equivalently requiring a cycle that visits each specified vertex), it generalizes the well-known multiple-depot \ac{TSP}~\cite{TV:2024:book}.
A simple $3$-approximation algorithm has long been known for the \ac{TSPPK}. Recently, this ratio was substantially improved to $1 + 2e^{-1/2} < 2.214$ by the framework introduced in~\cite{B:25:SODA}. Moreover, the $\lrA{\alpha+\varepsilon}$-approximation techniques for the \ac{TSPP} can be generalized to a broader class of \ac{TSP} variants~\cite{TV:2024:book}, yielding an $\lrA{\alpha+\varepsilon}$-approximation algorithm for the \ac{TSPPK} with running time $N^{O_{k,\varepsilon}(1)}$. Hence, for any fixed $k$, the best-known approximation ratio for the \ac{TSPPK} effectively matches that of the \ac{TSPP}.

\subsection{Relaxed Models for Non-Metric \acp{TSP}}
While metric variants assume well-behaved spatial environments, real-world data frequently exhibits deviations from the triangle inequality due to physical constraints or operational overhead. To address such deviations, several relaxed non-metric \ac{TSP} models beyond the \ac{BTSP} have been proposed.

\subsubsection{Global Relaxations}
A prominent line of research considers the $\tau$-triangle inequality, which globally relaxes the metric condition by enforcing $\tau \lrA{w\lrA{a,b}+w\lrA{b,c}}\geq w\lrA{a,c}$ across the entire graph for a given parameter $\tau \geq 1$~\cite{AB:95:SIAMDM}. 
Under this global relaxation, a $\lra{3\tau^2+\tau}/2$-approximation algorithm was first established in~\cite{AB:95:SIAMDM} for the \ac{TSP}. This approximation ratio was subsequently improved to $\tau^2+\tau$~\cite{A:01:NETWORKS} and then to $3\tau^2/2$~\cite{B:02:TCS}. 
The current best-known results offer a split guarantee: an approximation ratio of $4\tau$ for $\tau>13/3$~\cite{BC:00:IPL}, and $3\lra{\tau^2+\tau}/4$ for milder deviations where $\tau\leq 13/3$~\cite{M:15:IPL}. Notably, achieving an approximation ratio of $o\lrA{\tau}$ for the \ac{TSP} is impossible unless $\mathscr{P}=\mathscr{NP}$~\cite{BC:00:IPL}.
The $\tau$-triangle inequality has proven highly versatile and has been adapted for other optimization domains, including the \ac{TSPP}~\cite{FHPS:06:ALGOR}, the construction of $k$-vertex/edge-connected spanning subgraphs~\cite{B:08:JDALG}, single-allocation $p$-hub center problems~\cite{CHHK:22:ALGCA}, and densest $k$-subgraph formulations~\cite{CCHKK:20:ICS}. 
A comprehensive overview of these global applications is given in~\cite{KM:2018:incollection}.

\subsubsection{Local Scope Relaxations} 
In contrast to global bounds, other models restrict the scope of metric violations to specific structural interactions. A primary example is the \ac{SGPP}~\cite{SYKK:17:IJCAI}. 
Here, the vertex set is partitioned into disjoint subgroups, and the objective demands that vertices within each subgroup be visited consecutively. Moreover, the triangle inequality $w\lrA{a,b}+w\lrA{b,c}\geq w\lrA{a,c}$ is permitted to be violated \emph{only} if both $a$ and $c$ belong to the same subgroup.
The primary motivation for this model stems from multi-agent systems and heterogeneous manufacturing processes where targets within different subgroups demand distinct operational modes~\cite{SHES:16:IJCAI,SYKK:17:IJCAI,ZXPX:25:AAAI}. Consequently, executing tasks within a specific subgroup may incur additional, localized setup expenses, causing the effective costs of intra-group edges to be disproportionately higher than inter-group edges and inducing a partial metric structure.
% When the maximum subgroup size is restricted to at most two, the \ac{SGPP} reduces to the \ac{SPP}. This specific variant finds direct applications in electronic printing processes~\cite{GGMSTZ:10:IEEE_J_EPM} and automated polishing robotics~\cite{TDZ:04:ICRA}.

When the maximum subgroup size is restricted to two, the \ac{SGPP} reduces to the \ac{SPP}~\cite{TDZ:04:ICRA,GGMSTZ:10:IEEE_J_EPM}. This specific variant has direct applications in automated polishing robots, where the target is to polish nicks on the surface of a workpiece~\cite{TDZ:04:ICRA}, and in electronic printing, where continuous-spray processes are used to deposit electrically functional materials onto flexible substrates~\cite{GGMSTZ:10:IEEE_J_EPM}.
For the \ac{SPP}, the first result was a $2$-approximation algorithm~\cite{SHES:16:IJCAI}, which was later improved to $3/2$~\cite{SYKK:17:IJCAI}. 
For the broader \ac{SGPP}, a polynomial-time $3$-approximation algorithm was proposed in~\cite{SYKK:17:IJCAI}. 
To achieve better approximation ratios, researchers have turned to parameterized complexity.

For a minimization problem with a parameter $k$, a \ac{FPT} $\rho$-approximation algorithm computes a $\rho$-approximate solution in time $O^*\lra{f\lra{k}}$, where $O^*\lra{\cdot}$ hides polynomials in the input size; similarly, a \ac{XP} $\rho$-approximation algorithm runs in time $N^{O\lra{g\lra{k}}}$ for some computable function $g$.
Using this framework, an \ac{FPT} $5/2$-approximation algorithm with running time $O^*\lra{t2^t}$ was proposed for the \ac{SGPP}, parameterized by the maximum subgroup size $t$~\cite{SYKK:17:IJCAI}. 
More recently, a polynomial-time $2.167$-approximation algorithm was introduced in~\cite{ZXPX:25:AAAI}, alongside an \ac{FPT} $2.072$-approximation algorithm with running time $O^*\lra{t^22^t}$ parameterized by $t$ and an \ac{FPT} $\gamma$-approximation algorithm with running time $O^*\lra{2^m}$ parameterized by the total number of subgroups $m$. Here, $\gamma$ is the state-of-the-art approximation ratio for the \ac{TSPP}~\cite{TV:2024:book,KKG:21:STOC}.

\subsubsection{Parameterized Non-Metricity} 
Most recently, structural relaxations of the triangle inequality have been analyzed by quantifying the exact number of violations in a graph. For the \ac{TSP}, two key non-metricity parameters $p$ and $q$ were introduced in~\cite{ZLG:22:ISAAC}, where $p$ denotes the number of vertices involved in triangles that violate the metric property, while $q$ denotes the minimum number of vertices whose removal restores metricity.
In~\cite{ZLG:22:ISAAC}, an \ac{FPT} $3$-approximation algorithm with running time $O\lra{\lra{3p}!8^p N^2+N^3}$ was proposed for the \ac{TSP} parameterized by $p$ (denoted \ac{TSP}-$p$), whereas an \ac{FPT} $\lra{6q+9}$-approximation algorithm with running time $O\lra{q^{O\lra{q}} N^3}$ and an \ac{XP} $3$-approximation algorithm with running time $O\lra{N^{O\lra{q}}}$ were given for the \ac{TSP} parameterized by $q$ (denoted \ac{TSP}-$q$).

Later, an improved \ac{FPT} $5/2$-approximation algorithm with running time $O\lra{\lra{3p}!8^p N^4}$ was proposed in \cite{BEX:24:arXiv} for the \ac{TSP}-$p$. 
An \ac{FPT} $11$-approximation algorithm with running time $O\lra{q^{O\lra{q}} N^3}$ was proposed in \cite{ZZG:25:COCOON} for the \ac{TSP}-$q$.
More recently, improved \ac{FPT} approximation algorithms were given in~\cite{ZSX:26:AAAI}. For the \ac{TSP}-$p$, this work proposed an \ac{FPT} $3/2$-approximation algorithm with running time $2^{O\lra{p\log p}} N^{O\lra{1}}$, as well as an \ac{FPT} $\lra{\alpha+1}$-approximation algorithm with a single-exponential running time of $2^{O\lra{p}}+N^{O\lra{1}}$. 
In addition, for the \ac{TSP}-$q$, the work proposed an improved \ac{FPT} $3$-approximation algorithm with running time $2^{O\lra{q\log q}} N^{O\lra{1}}$.
This parameterization of non-metricity using $p$ and $q$ has also been applied to three location problems~\cite{ZZWG:25:COCOON} to obtain improved algorithms for graphs that are \emph{nearly} metric.

\begin{figure*}[t!]
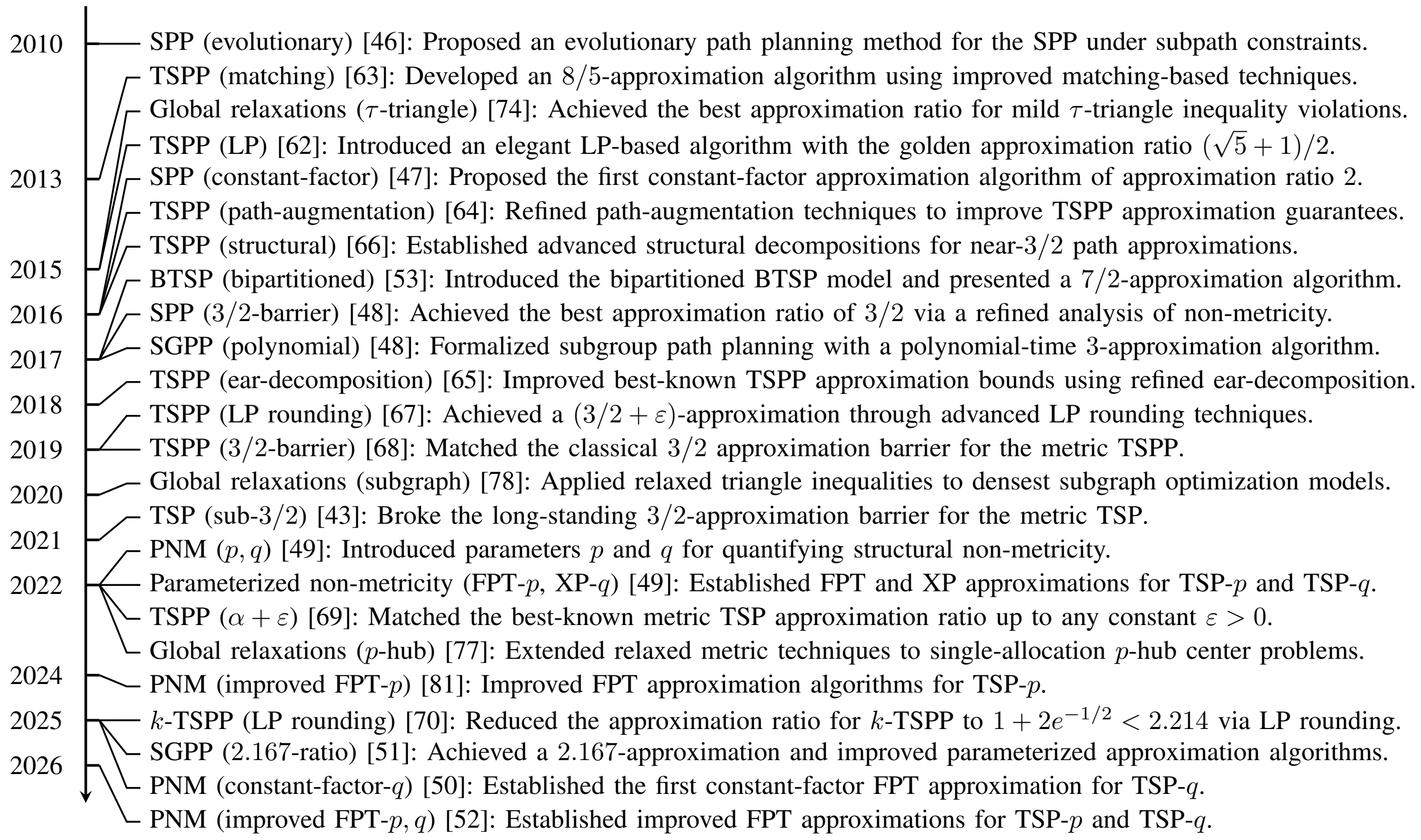

\centering
\resizebox{\textwidth}{!}{%  <-- adjust this number
       
\begin{vtimeline}{2010}{2026}

% \ventry{2006}{Global Relaxations~\cite{FHPS:06:ALGOR}: Extended the $\tau$-triangle inequality framework to the \ac{TSPP} with provable approximation guarantees}

\ventry{2010}{\ac{SPP} (evolutionary)~\cite{GGMSTZ:10:IEEE_J_EPM}: Proposed an evolutionary path planning method for the \ac{SPP} under subpath constraints.}
%Integrated genetic evolution and local graph search for path generation under subpath constraints}

\ventry{2013}{\ac{TSPP} (matching)~\cite{S:13:IPCO}: Developed an $8/5$-approximation algorithm using improved matching-based techniques.}

\ventry{2015}{Global relaxations ($\tau$-triangle)~\cite{M:15:IPL}: Achieved the best approximation ratio for mild $\tau$-triangle inequality violations.}

\ventry{2015}{\ac{TSPP} (\ac{LP})~\cite{AKS:15:JACM}: Introduced an elegant \ac{LP}-based algorithm with the golden approximation ratio $\lra{\sqrt{5}+1}/2$.}

\ventry{2016}{\ac{SPP} (constant-factor)~\cite{SHES:16:IJCAI}: Proposed the first constant-factor approximation algorithm of approximation ratio $2$.}

\ventry{2016}{\ac{TSPP} (path-augmentation)~\cite{V:16:SIAMDM}: Refined path-augmentation techniques to improve \ac{TSPP} approximation guarantees.}

\ventry{2016}{\ac{TSPP} (structural)~\cite{S:16:FOCS}: Established advanced structural decompositions for near-$3/2$ path approximations.}

\ventry{2017}{\ac{BTSP} (bipartitioned)~\cite{MRM:17:TCS}: Introduced the bipartitioned \ac{BTSP} model and presented a $7/2$-approximation algorithm.}

\ventry{2017}{\ac{SPP} ($3/2$-barrier)~\cite{SYKK:17:IJCAI}: Achieved the best approximation ratio of $3/2$ via a refined analysis of non-metricity.}

\ventry{2017}{\ac{SGPP} (polynomial)~\cite{SYKK:17:IJCAI}: Formalized subgroup path planning with a polynomial-time $3$-approximation algorithm.}

\ventry{2018}{\ac{TSPP} (ear-decomposition)~\cite{GV:18:MP}: Improved best-known \ac{TSPP} approximation bounds using refined ear-decomposition.}

\ventry{2019}{\ac{TSPP} (\ac{LP} rounding)~\cite{TV:19:JACM}: Achieved a $(3/2+\varepsilon)$-approximation through advanced \ac{LP} rounding techniques.}

\ventry{2019}{\ac{TSPP} ($3/2$-barrier)~\cite{Z:19:SODA}: Matched the classical $3/2$ approximation barrier for the metric \ac{TSPP}.}

\ventry{2020}{Global relaxations (subgraph)~\cite{CCHKK:20:ICS}: Applied relaxed triangle inequalities to densest subgraph optimization models.}

\ventry{2021}{\ac{TSP} (sub-$3/2$)~\cite{KKG:21:STOC}: Broke the long-standing $3/2$-approximation barrier for the metric \ac{TSP}.}

\ventry{2022}{PNM ($p,q$)~\cite{ZLG:22:ISAAC}: Introduced parameters $p$ and $q$ for quantifying structural non-metricity.}

\ventry{2022}{Parameterized non-metricity (\ac{FPT}-$p$, \ac{XP}-$q$)~\cite{ZLG:22:ISAAC}: Established \ac{FPT} and \ac{XP} approximations for \ac{TSP}-$p$ and \ac{TSP}-$q$.}

\ventry{2022}{\ac{TSPP} ($\alpha+\varepsilon$)~\cite{TVZ:22:SICOMP}: Matched the best-known metric \ac{TSP} approximation ratio up to any constant $\varepsilon>0$.}

\ventry{2022}{Global relaxations ($p$-hub)~\cite{CHHK:22:ALGCA}: Extended relaxed metric techniques to single-allocation $p$-hub center problems.}

\ventry{2024}{PNM (improved \ac{FPT}-$p$)~\cite{BEX:24:arXiv}: Improved \ac{FPT} approximation algorithms for \ac{TSP}-$p$.}

\ventry{2025}{\ac{TSPPK} (\ac{LP} rounding)~\cite{B:25:SODA}: Reduced the approximation ratio for \ac{TSPPK} to $1+2e^{-1/2}<2.214$ 
via \ac{LP} rounding.}

\ventry{2025}{\ac{SGPP} ($2.167$-ratio)~\cite{ZXPX:25:AAAI}: Achieved a $2.167$-approximation and improved parameterized approximation algorithms.}

\ventry{2025}{PNM (constant-factor-$q$)~\cite{ZZG:25:COCOON}: Established the first constant-factor \ac{FPT} approximation for \ac{TSP}-$q$.}

\ventry{2026}{PNM (improved \ac{FPT}-$p,q$)~\cite{ZSX:26:AAAI}: Established improved \ac{FPT} approximations for \ac{TSP}-$p$ and \ac{TSP}-$q$.}

\end{vtimeline}
}

\caption{A timeline illustrating representative algorithms for path-based and non-metric \ac{TSP} variants related to the \ac{BTSP}. Here, PNM stands for parameterized non-metricity.}
\label{fig:timeline}
\end{figure*}

To synthesize the literature reviewed in this section,
Table~\ref{tab:related_work} presents a structured comparison of representative path-based and non-metric \ac{TSP} variants, highlighting their features, advantages, and limitations with respect to structural network heterogeneity. Complementing this taxonomy, Fig.~\ref{fig:timeline} traces the chronological evolution of major algorithmic advances in these domains, contextualizing the recent developments for the \ac{BTSP} addressed in this work.

\begin{table*}[t!] 
\begin{center}
\caption{Representative non-metric and path-based variants related to the \ac{BTSP}}
\label{tab:related_work}

\begin{tabular}{lllll}
\toprule
Class & Model & Feature & Advantage & Limitation \\
\midrule
\midrule

\multirow{4}*{\makecell[l]{Path-based \\ variants}}
& \makecell[l]{\ac{TSPP} \\ \cite{H:91:ORL,AKS:15:JACM,S:13:IPCO,V:16:SIAMDM,GV:18:MP,S:16:FOCS,TV:19:JACM,Z:19:SODA,TVZ:22:SICOMP,KKG:21:STOC}}
& \makecell[l]{Computes a minimum-cost\\ Hamiltonian path between\\ fixed endpoints in a graph.} 
& \makecell[l]{Provides a foundational building\\ block for end-to-end routing\\ in networked systems.} 
& \makecell[l]{Cannot capture structured\\ cross-partition routing\\ cost heterogeneity.} \\

\cmidrule{2-5}
& \makecell[l]{\ac{TSPPK} \\ \cite{TVZ:22:SICOMP,B:25:SODA}} 
& \makecell[l]{Extends \ac{TSPP} to multiple\\ terminal pairs with global\\ vertex coverage constraints.} 
& \makecell[l]{Supports efficient multi-session\\ routing in large-scale\\ heterogeneous networks.} 
& \makecell[l]{Assumes globally metric\\ edge costs and ignores\\ domain-specific link biases.} \\

\midrule 

\makecell[l]{Global \\ relaxations}
& \makecell[l]{$\tau$-triangle inequality \\ \cite{AB:95:SIAMDM,A:01:NETWORKS,B:02:TCS,BC:00:IPL,M:15:IPL}}
& \makecell[l]{Relaxes the triangle\\ inequality uniformly across\\ the entire graph topology.} 
& \makecell[l]{Tolerates globally distorted\\ or noisy network distance\\ measurements.} 
& \makecell[l]{Topology-agnostic; ignores\\ structured non-metricity in\\ partitioned systems.} \\

\midrule

\multirow{4}*{\makecell[l]{Local scope \\ relaxations}} 
& \makecell[l]{\ac{SPP} \\ \cite{TDZ:04:ICRA,GGMSTZ:10:IEEE_J_EPM,SHES:16:IJCAI,SYKK:17:IJCAI}}
& \makecell[l]{Specializes the \ac{SGPP} to\\ strictly consecutive pairwise\\ vertex subgrouping.} 
& \makecell[l]{Admits strong guarantees for\\ precision robotic polishing\\ and circuit board drilling.}
& \makecell[l]{Insufficient expressiveness for\\ complex, multi-tier routing\\ and network systems.} \\

\cmidrule{2-5}
& \makecell[l]{\ac{SGPP} \\ \cite{SYKK:17:IJCAI,ZXPX:25:AAAI}} 
& \makecell[l]{Confines violations within\\ local clusters and enforces\\ consecutive vertex visits.} 
& \makecell[l]{Models coordinated batch-mode\\ operations in localized\\ operational clusters.}
& \makecell[l]{Rigid visit-order constraints\\ limit adaptive cross-domain\\ routing flexibility.} \\

\midrule

\multirow{4}*{\makecell[l]{Parameterized \\ non-metricity}} 
& \makecell[l]{\ac{TSP}-$p$ \\ \cite{ZLG:22:ISAAC,BEX:24:arXiv,ZSX:26:AAAI}} 
& \makecell[l]{Quantifies non-metricity\\ via vertices participating\\ in violated triangles.} 
& \makecell[l]{Characterizes networks with\\ sparse, localized metric\\ anomalies and disruptions.} 
& \makecell[l]{Unsuitable for persistent,\\ structurally-induced cross-\\ partition biases.} \\

\cmidrule{2-5}
& \makecell[l]{\ac{TSP}-$q$ \\ \cite{ZLG:22:ISAAC,ZZG:25:COCOON,ZSX:26:AAAI}}
& \makecell[l]{Measures the minimum\\ vertex set whose removal\\ restores global metricity.} 
& \makecell[l]{Models networks that are\\ nearly metric except for a\\ few non-metric vertices.} 
& \makecell[l]{Does not explicitly model\\ partition-dependent routing\\ costs and transitions.} \\

\midrule

\makecell[l]{Partition-based\\ relaxations\\ (\textbf{this work})}
& \makecell[l]{\ac{BTSP} \\ \cite{MRM:17:TCS}}
& \makecell[l]{Preserves metricity within\\ partitions, allowing biased\\ cross-partition transitions.}
& \makecell[l]{Models partition-dependent\\ routing costs in heterogeneous\\ networked systems.}
% Captures routing in \ac{6G}\\ NTN-integrated and hybrid\\ quantum-classical networks.}
& \makecell[l]{Cross-partition interactions\\ complicate tour construction\\ and cost analysis.} \\

\bottomrule
\end{tabular}
\end{center}
\end{table*}

\section{Bipartitioned Networks}
\label{sec:3}

In this section, we introduce the basic notations and structural definitions used throughout the paper. We first present graph-theoretic preliminaries on general (multi-)graphs, and then specialize to bipartitioned networks under the biased edge-cost function.

\subsection{Basic Graph-Theoretic Notations}

Consider any (multi-)graph $\HH=\lrA{\V_\HH,\E_\HH}$ with an underlying edge-cost function $w_\HH$. 
The \emph{degree} of a vertex $v\in \V_\HH$ in $\HH$ is the number of edges in $\E_\HH$ incident to it.
Let $\odd\lrA{\HH}$ denote the set of odd-degree vertices in $\V_\HH$.
The graph $\HH$ is called \emph{Eulerian} if every vertex in $\V_\HH$ has an even degree, i.e., $\odd\lrA{\HH}=\emptyset$, and $\HH$ is connected.
Let $2\HH$ denote the multigraph obtained by duplicating every edge in $\HH$.
For any edge sets $\E_1$ and $\E_2$, let $\E_1 \uplus \E_2$ denote their multiset union.

For any vertex set $\V'\subseteq \V_\HH$, let $\HH\lrB{\V'}$ denote the subgraph of $\HH$ induced by $\V'$. 
For any subgraph $\HH'$ of $\HH$, let $\V_\HH\lrA{\HH'}$ and $\E_\HH\lrA{\HH'}$ denote the set of vertices and the multiset of edges of $\HH'$, respectively.
For any subset $\E' \subseteq \E_\HH$ of edges, we define its cost as follows:
\begin{align}
w_\HH\lrA{\E'} = \sum_{e \in \E'} w_\HH\lrA{e}.
\end{align}
This definition extends to any subgraph $\HH'$ as $w_\HH\lrA{\HH'}= w_\HH\lrA{\E_\HH\lrA{\HH'}}$, and to any set $\B$ of subgraphs (or trees) as 
\begin{align}
w_\HH\lrA{\B} 
&= 
\sum_{\HH' \in \B} w_\HH\lrA{\HH'}.
\end{align}

% A \emph{matching} in $H$ is a set of pairwise non-adjacent edges. A perfect matching is a matching that covers every vertex of the graph. The cost of a matching $M$ is defined as the sum of the costs of its constituent edges.

In the graph $\HH$, a \emph{walk} $\sigma=\lrd{v_1,v_2,\ldots,v_\ell}$ is a sequence of vertices, where vertices may be repeated. 
Let $\E_\HH\lrA{\sigma}$ denote its edge multiset, i.e., $\E_\HH\lrA{\sigma}=\lrC{v_1v_2,\ldots,v_{\ell-1}v_\ell}$, and the cost of $\sigma$ is defined as the total cost of the edges in $\E_\HH\lrA{\sigma}$, namely, $w_\HH\lrA{\sigma}=\sum_{e \in \E_\HH\lrA{\sigma}} w_\HH\lrA{e}$.
The walk $\sigma$ is called a \emph{path} if it contains no repeated vertices, a \emph{tour}  (or \emph{closed walk}) if $v_1=v_\ell$, and a \emph{cycle} if it is a tour with $\Size{\E_\HH\lrA{\sigma}}=\Size{\V_\HH\lrA{\sigma}}$.
Moreover, $\sigma$ is called a \emph{Hamiltonian path} in $\HH$ if it is a path with $\V_\HH\lrA{\sigma}=\V_\HH$, and a \emph{Hamiltonian cycle} in $\HH$ if it is a cycle with $\V_\HH\lrA{\sigma}=\V_\HH$.
When $\HH$ is an Eulerian graph, $\sigma$ is called an \emph{Eulerian tour} if it is a tour with $\E_\HH\lrA{\sigma}=\E_\HH$.
When $\sigma$ is a path, we call $v_1$ and $v_\ell$ its \emph{terminals}.

For a set $\W$ of walks, we define:
\begin{align}
\E_\HH\lrA{\W}
&=
\biguplus_{\sigma\in\W}
\E_\HH\lrA{\sigma}; \\
w_\HH\lrA{\W} 
&= 
\sum_{\sigma \in \W} 
w_\HH\lrA{\sigma}.
\end{align}
Given a walk $\sigma_1=\lrd{\ldots,x,y,z,\ldots}$, let $\sigma_1'=\lrd{\ldots,x,z,\ldots}$ denote the walk obtained by skipping the vertex $y$.
This operation is called \emph{shortcutting} the edges $xy$ and $yz$ (or, equivalently, the vertex $y$).
If the edge-cost function $w_\HH$ is a metric, then shortcutting never increases the cost of the walk; that is, by the triangle inequality, $w_\HH\lrA{\sigma_1'}\leq w_\HH\lrA{\sigma_1}$.

\subsection{Bipartitioned Networks and the Biased Edge-Cost Function}
\label{sec:3.2}

We now introduce some notations specific to the \ac{BTSP}.
Let $\G=\lrA{\V,\E}$ denote the input complete undirected graph.
Let $w: \E\to\RR_{\geq 0}$ denote an edge-cost metric function that satisfies the triangle inequality.
Let $\P=\lrC{\V_1, \V_2}$ be a partition of the vertex set $\V$ into two (non-empty) subsets, i.e., $\V_1\cup \V_2=\V$ and $\V_1\cap \V_2=\emptyset$.
Moreover, let $\beta: \V_1\times \V_2\to \RR_{\geq 1}$ be a bias-factor function such that $\beta\lrA{u,v} = \beta\lrA{v,u}\geq 1$ for any $u\in \V_1$ and $v\in \V_2$.
We define the biased edge-cost function $\w: \E\to\mathbb{R}_{\geq 0}$ as follows: 
\begin{align}\label{hatcost}
%\w\lrA{u,v}
\w\lrA{uv}
=
\begin{cases}
w\lrA{u,v}, \!\!& u,v\in \V_1 \text{ or } u,v\in \V_2,\\
\beta\lrA{u,v} w\lrA{u,v}, \!\!& u\in \V_1, v\in \V_2.
\end{cases}
\end{align}
Notably, for any $a,b,c\in \V$, the triangle inequality %$\w\lrA{a,b}+\w\lrA{b,c}\geq \w\lrA{a,c}$ 
$\w\lrA{ab}+\w\lrA{bc}\geq \w\lrA{ac}$ still holds in the following cases:
\begin{align}\label{inequalitycases}
\begin{cases}
a,b,c\in \V_1, \text{ or } a,b,c\in \V_2,\\
a,c\in \V_1,\ b\in \V_2, \text{ or } a,c\in \V_2,\ b\in \V_1.\\
\end{cases}
\end{align}

Unless stated otherwise, throughout the remainder of this paper, we consider minimum cost with respect to the biased edge-cost function $\w$ of $\G$. 
For brevity, we extend $\w$ to any edge subset $\SS$ as follows:
\begin{align}
\w\lrA{\SS} \coloneqq \sum_{e \in \SS} \w\lrA{e}
\end{align}
and to any subgraph $\HH$ by $\w\lrA{\HH} \coloneqq \w\lrA{\E\lrA{\HH}}$. 
We further write $\w\lrA{\SS_1,\SS_2,\ldots,\SS_\ell} \coloneqq \sum_{i \in \lrB{\ell}} \w\lrA{\SS_i}$ for individual edge subsets (or simply edges).
%We further write $\w\lrA{e_1,e_2,\ldots,e_\ell} \coloneqq \sum_{i \in \lrB{\ell}} \w\lrA{e_i}$ for individual edges.} 
%The same conventions apply to any edge-cost function used in this paper.}

%Thus, for any subgraph or subset $\HH$ of edges, its cost evaluates to $\w\lrA{\HH}$.

An edge $xy\in \E$ is called a \emph{cross edge} if the vertices $x$ and $y$ do not belong to the same vertex set $\V_1$ or $\V_2$. For any subgraph $\HH$ of $\G$, we let $\E_\c\lrA{\HH}$ denote the set of cross edges in $\HH$.
Two cross edges are called \emph{independent} if they do not share a common vertex.
Consider a tree $\T=\lrA{\V\lrA{\T}, \E\lrA{\T}}$. 
For any two distinct vertices $s,t\in \V\lrA{\T}$, let $P_\T\lrA{s,t}$ denote the unique path between $s$ and $t$ in $\T$, and define the \emph{distance} between $s$ and $t$ in $\T$ as $d_\T\lrA{s,t}=\Size{\E\lrA{P_\T\lrA{s,t}}}$. 
The \emph{depth} of $\T$ is defined as 
\begin{equation}
\depth\lrA{\T} = \min_{v\in \V\lrA{\T}}\max_{u\in \V\lrA{\T}}d_\T\lrA{u,v}.
\end{equation}

Next, we recall several key properties of the biased edge-cost function utilized in our analysis~\cite{MRM:17:TCS}. By definition, the subgraphs $\G\lrB{\V_1}$ and $\G\lrB{\V_2}$ are strictly metric graphs under $\w$. This guarantees that standard metric operations---such as computing approximate Hamiltonian paths and shortcutting vertices entirely within $\V_1$ or entirely within $\V_2$---can be safely performed without exceeding intermediate path bounds. Furthermore, by \eqref{inequalitycases}, shortcutting a vertex $y \in \V_i$ for $i\in[2]$ from a sequence passing through $\lrd{x,y, z}$, where $x, z \in \V_{3-i}$, strictly respects the biased triangle inequality.
More generally, shortcutting all internal vertices in a sequence $\lrd{x,\ldots,z}$, where $x, z \in \V_{3-i}$, also preserves the biased triangle inequality.
In what follows, we let $\sigma^*$ denote a minimum-cost Hamiltonian cycle in $\G$, i.e., an optimal solution to the \ac{BTSP}.

\section{\scalebox{0.95}[1]{A Polynomial $\lrA{2\alpha+\varepsilon}$-Approximation Algorithm}}
\label{sec:4}

In this section, we propose a $\lrA{2\alpha+\varepsilon}$-approximation algorithm for the \ac{BTSP} in polynomial time, where $\varepsilon>0$ denotes any positive constant.
Recall that $\E_{\c}\lrA{\sigma^*}$ denotes the set of cross edges in $\E\lrA{\sigma^*}$. 
Let $N_\c=\Size{\E_\c\lrA{\sigma^*}}$. 
Since $\sigma^*$ is a Hamiltonian cycle in $\G$, we must have $N_\c$ even.

\subsection{Algorithm Overview}

Our algorithm distinguishes two cases based on the magnitude of $N_\c$ relative to $1/\varepsilon$.
The high-level approach is as follows.

\subsubsection{Small $N_\c$ ($N_\c=O\lrA{1/\varepsilon}$)}

We enumerate the cross edge set $\E_\c\lrA{\sigma^*}$ in polynomial time $N^{O\lrA{N_\c}}$.
Fix an arbitrary orientation of $\sigma^*$, either \emph{clockwise} or \emph{counterclockwise}, and select any edge in $\E_\c\lrA{\sigma^*}$ as the first cross edge. 
Then, we enumerate, in constant time $O\lrA{N_\c!}$, the precedence order of all edges in $\E_\c\lrA{\sigma^*}$ along the Hamiltonian cycle $\sigma^*$ with respect to the chosen orientation.
It then suffices to find a minimum-cost Hamiltonian cycle that contains all edges in $\E_\c\lrA{\sigma^*}$ and preserves their precedence order as in $\sigma^*$.
We will show that this problem can be reduced to the \ac{TSPPK} in polynomial time while preserving the approximation ratio.
Since the \ac{TSPPK} admits an $\lrA{\alpha+\varepsilon}$-approximation algorithm for fixed $k$~\cite{TV:2024:book}, we obtain an $\lrA{\alpha+\varepsilon}$-approximation algorithm, denoted as $\textsc{Small}\lrA{N_\c}$, for this case.

\subsubsection{Large $N_\c$ ($N_\c=\Omega\lrA{1/\varepsilon}$)}
\label{sec:LargeOverview}

%Otherwise, $N_\c=\Omega\lrA{1/\varepsilon}$. In this case, 
We show that there exists a pair of independent cross edges $e_1=u_1v_1$ and $e_2=u_2v_2$ in $\E_\c\lrA{\sigma^*}$, where $u_1,u_2\in \V_1$ and $v_1,v_2\in \V_2$, such that:
\begin{align} \label{eq:L:1}
%\w\lrA{e_1}+\w\lrA{e_2}
\w\lrA{e_1,e_2}
=
O\lrA{1/N_\c}\w\lrA{\sigma^*}.
\end{align}
Moreover, the minimum-cost Hamiltonian paths $P^*\lrA{u_1,u_2}$ in $\G\lrB{\V_1}$ and $P^*\lrA{v_1,v_2}$ in $\G\lrB{\V_2}$ satisfy
\begin{align} \label{eq:L:2}
%\w\lrA{P^*\lrA{u_1,u_2}}+\w\lrA{P^*\lrA{v_1,v_2}}
\w\lrA{P^*\lrA{u_1,u_2},P^*\lrA{v_1,v_2}}
\leq 
\lrA{2+O\lrA{1/N_\c}}
\w\lrA{\sigma^*}.
\end{align} 
Consequently, we enumerate $e_1$ and $e_2$, compute Hamiltonian paths $P\lrA{u_1,u_2}$ in $\G\lrB{\V_1}$ and $P\lrA{v_1,v_2}$ in $\G\lrB{\V_2}$ using the $\lrA{\alpha+O\lra{1/N_\c}}$-approximation algorithm for the \ac{TSPP} in \cite{TVZ:22:SICOMP}, and combine these paths with the cross edges to obtain a $\lrA{2\alpha+O\lra{1/N_\c}}$-approximate Hamiltonian cycle.
We denote this algorithm by \textsc{Large}.
Since $N_\c=\Omega\lrA{1/\varepsilon}$, the approximation ratio of \textsc{Large} achieves $2\alpha+O\lrA{\varepsilon}$.

Although $N_\c$ is unknown in advance, we can set a constant threshold $N_0=O\lrA{1/\varepsilon}$ and run $\textsc{Small}\lrA{\ell}$ for each even $\ell\in \lrB{N_0}$ together with \textsc{Large}, obtaining $O\lrA{N_0}$ candidate solutions. Returning the best solution yields a $\lrA{2\alpha+O\lra{\varepsilon}}$-approximation for the \ac{BTSP}.
The full procedure is summarized in Algorithm~\ref{algf}, where $\Ceil{x}$ and $\Floor{x}$ denote the smallest integer greater than or equal to $x$ and the largest integer less than or equal to $x$, respectively.
In what follows, we first introduce the algorithm $\textsc{Small}\lrA{\ell}$.

\begin{algorithm}[t]
\caption{A $\lrA{2\alpha+\varepsilon}$-approximation algorithm}
\label{algf} 
\begin{algorithmic}[1]
\REQUIRE The graph $\G=\lrA{\V,\E}$ and a constant $\varepsilon>0$.
\ENSURE A Hamiltonian cycle in $\G$.

\STATE Set $N_0=\Ceil{32/\varepsilon}$.

\STATE Call \textsc{Large} to obtain a Hamiltonian cycle $\sigma_0$ in $\G$.\label{algsf2}

\FOR{each even $\ell\in \lrB{N_0}$}\label{algsf3}

\STATE Call $\textsc{Small}\lrA{\ell}$ to obtain a Hamiltonian cycle $\sigma_\ell$ in $\G$.\label{algsf4}

\ENDFOR

\RETURN $\arg\min_{\ell\in\lrC{0,2,4,6,\ldots,2\Floor{N_0/2}}} \w\lrA{\sigma_\ell}$.\label{algsf6}
\end{algorithmic}

\end{algorithm}

\subsection{The Algorithm $\textsc{Small}\lrA{\ell}$}
When $N_\c = \ell$ and $\ell=O\lrA{1/\varepsilon}$, the algorithm $\textsc{Small}\lrA{\ell}$ is an $\lrA{\alpha+\varepsilon}$-approximation algorithm for the \ac{BTSP}.
As mentioned earlier, it first enumerates $\ell$ (even) ordered cross edges, denoted by $\lrA{u_0v_0,\ldots,u_{\ell-1}v_{\ell-1}}$, where $u_i\in \V_1$ and $v_i\in \V_2$ for each $0\le i<\ell$.
It then aims to find a Hamiltonian cycle in $\G$ that contains all these edges and preserves their precedence order; that is, a Hamiltonian cycle of the form
\begin{align}
\lrd{u_0,v_0,\ldots,v_1,u_1,%\ldots,u_{\ell-2},v_{\ell-2},
\ldots,v_{\ell-1},u_{\ell-1},\ldots,u_0}.
\end{align}
% For convenience, we may allow $v_{i-1}=v_i$ or $u_{i}=u_{\lrA{i+1}\bmod\ell}$ for some odd $i\in\lrB{\ell-1}$.
% Moreover, to ensure that such a Hamiltonian cycle exists, we require that $v_i\neq v_j$ and $u_{i}\neq u_{j\bmod\ell}$ for any $i<j$ with $\ell-1>j-i>1$.

For convenience, for any $i<j$, we allow $v_{i-1}=v_{j-1}$ or $u_i=u_{j\bmod\ell}$ if and only if $j=i+1$ and $i \in \lrB{\ell-1}$ is odd.
That is, $v_{i-1}=v_i$ means that the cross edges $u_{i-1}v_{i-1}$ and $v_iu_i$ are consecutive on the Hamiltonian cycle and share the common vertex $v_{i-1}=v_i$. Similarly, $u_i=u_{\lra{i+1}\bmod\ell}$ means that the cross edges $v_iu_i$ and $u_{\lra{i+1}\bmod\ell} \, v_{\lra{i+1}\bmod\ell}$ are consecutive on the Hamiltonian cycle and share the common vertex $u_i=u_{\lra{i+1}\bmod\ell}$.
Equivalently, the algorithm needs to find two collections of vertex-disjoint paths: 
\begin{align}
\B_1
&=
\lrC{P_1,\ldots,P_{\ell/2}}, \\%\quad \text{and} \quad
\B_2
&=\lrC{Q_1,\ldots,Q_{\ell/2}}
\end{align} 
such that 
\begin{itemize}
\item $P_i=\lrd{u_{2i-1},\ldots,u_{2i\bmod \ell}}$ and $Q_i=\lrd{v_{2i-2},\ldots,v_{2i-1}}$ for each $i\in\lrB{\ell/2}$;
%\item $\B_1$ is a set of vertex-disjoint paths in $\G\lrB{\V_1}$ covering all vertices in $\V_1$;
\item $\B_j$ is a set of vertex-disjoint paths in $\G\lrB{\V_j}$ covering all vertices in $\V_j$ for each $j \in \lrB{2}$.
\end{itemize}
This can be achieved by solving the \ac{TSPPK}. We first recall its definition.

\begin{definition}[\ac{TSPPK}]
Given a metric graph and $k$ vertex pairs $\lrA{s_1,t_1},\ldots,\lrA{s_k,t_k}$, the goal is to find, for each $i\in\lrB{k}$, a path between $s_i$ and $t_i$ such that the total cost of these paths is minimized and every vertex of the graph is covered by at least one path.
\end{definition}

Note that when $k=1$, the \ac{TSPPK} reduces to the \ac{TSPP}.
Recall that $\G\lrB{\V_1}$ is a metric graph. Moreover, the $\ell/2$ vertex pairs $\lrA{u_1,u_2},\ldots,\lrA{u_{\ell-1},u_{0}}$ are pairwise vertex-disjoint, since $u_i=u_{\lra{i+1}\bmod\ell}$ can occur only when $i \in \lrB{\ell-1}$ is odd, by definition. 
Hence, given $\G\lrB{\V_1}$ and these $\ell/2$ vertex pairs, we can apply the $\lrA{\alpha+\varepsilon}$-approximation algorithm~\cite{TV:2024:book} for the \ac{TSPPK} to obtain $\B_1$.

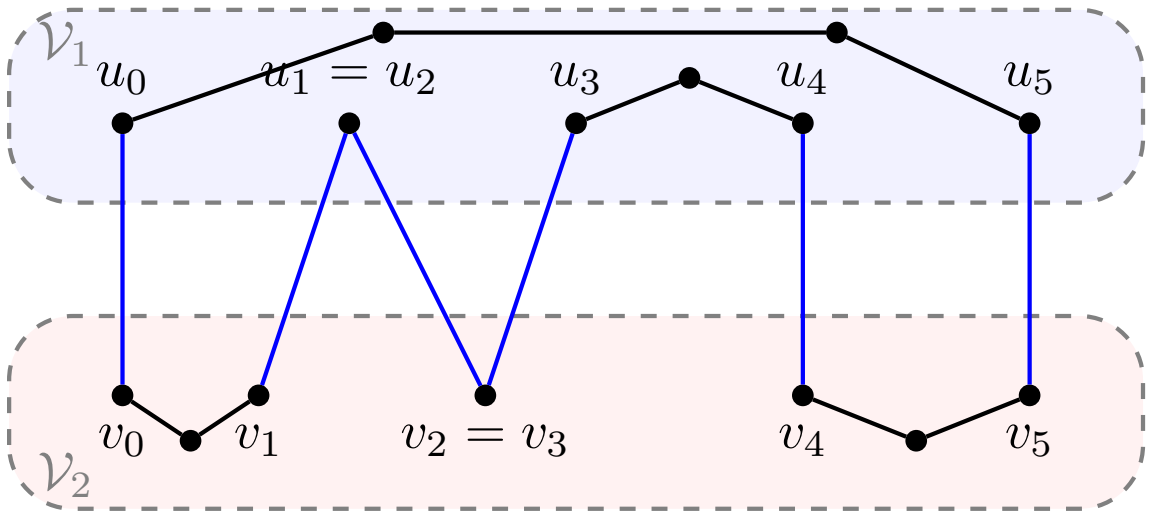
\begin{figure}[t!]
\centering
\begin{tikzpicture}[scale=0.75,
    vertex/.style={circle, draw=black, fill=black, inner sep=1.3pt},
    edge/.style={thick, black},
    crossedge/.style={thick, blue},
    region/.style={rounded corners=12pt, draw=gray, dashed, thick}
]

% ====================
% 1. 区域 V1, V2
% ====================
\filldraw[region, fill=blue!5] (-4.5, 1.5) rectangle (5.5, 3.2);
\node[text=gray] at (-4, 2.9) {$\V_1$};

\filldraw[region, fill=red!5] (-4.5, -3.2+2) rectangle (5.5, -1.5+
2);
\node[text=gray] at (-4, -2.9+2) {$\V_2$};

% ====================
% 2. 顶点
% ====================
% V1
\node[vertex, label=above:{$u_0$}] (u0) at (-3.5, 2.2) {};
\node[vertex, label=above:{$u_1 = u_2$}] (u12) at (-1.5, 2.2) {};
\node[vertex, label=above:{$u_3$}] (u3) at (0.5, 2.2) {};
\node[vertex, label=above:{$u_4$}] (u4) at (2.5, 2.2) {};
\node[vertex, label=above:{$u_5$}] (u5) at (4.5, 2.2) {};

% V2
\node[vertex, label=below:{$v_0$}] (v0) at (-3.5, -2.2+2) {};
\node[vertex, label=below:{$v_1$}] (v1) at (-2.3, -2.2+2) {};
\node[vertex, label=below:{$v_2 = v_3$}] (v23) at (-0.3, -2.2+2) {};
\node[vertex, label=below:{$v_4$}] (v4) at (2.5, -2.2+2) {};
\node[vertex, label=below:{$v_5$}] (v5) at (4.5, -2.2+2) {};

% ====================
% 3. 辅助点（路径内部）
% ====================
\node[vertex] (v01) at (-2.9, -2.6+2) {};
\node[vertex] (u34) at (1.5, 2.6) {};
\node[vertex] (v45) at (3.5, -2.6+2) {};
\node[vertex] (u50_1) at (2.8, 3.0) {};
\node[vertex] (u50_2) at (-1.2, 3.0) {};

% ====================
% 4. 边（全部无向）
% ====================

% 跨区域
\draw[crossedge] (u0) -- (v0);
\draw[crossedge] (v1) -- (u12);
\draw[crossedge] (u12) -- (v23);
\draw[crossedge] (v23) -- (u3);
\draw[crossedge] (u4) -- (v4);
\draw[crossedge] (v5) -- (u5);

% V2 内路径
\draw[edge] (v0) -- (v01) -- (v1);
\draw[edge] (v4) -- (v45) -- (v5);

% V1 内路径
\draw[edge] (u3) -- (u34) -- (u4);
\draw[edge] (u5) -- (u50_1) -- (u50_2) -- (u0);

\end{tikzpicture}
\caption{An illustration of Algorithm~\ref{algs}. 
The blue edges denote the enumerated ordered cross edges $\lra{u_0v_0,\ldots,u_{\ell-1}v_{\ell-1}}$, while the black edges depict the vertex-disjoint paths in $\G\lrb{\V_1}$ and $\G\lrb{\V_2}$ returned by the \ac{TSPPK} algorithm, which together form a Hamiltonian cycle respecting the prescribed order of the cross edges.}
\label{fig:hamiltonian_cycle}
\end{figure}

Note that in the definition of the \ac{TSPPK}, a vertex may be covered by two or more paths in $\B_1$. Since the $\ell/2$ vertex pairs are pairwise vertex-disjoint, any such vertex must be an internal (non-terminal) vertex of at least one path in $\B_1$. By the triangle inequality, it can be shortcut without increasing the cost. Therefore, after shortcutting, we can ensure that all paths in $\B_1$ are vertex-disjoint while the total cost does not increase. The same procedure yields $\B_2$.
The algorithm is summarized in Algorithm~\ref{algs}, and its illustration is provided in Fig.~\ref{fig:hamiltonian_cycle}.

\begin{algorithm}[t]
\caption{$\textsc{Small}\lrA{\ell}$}
\label{algs}
\begin{algorithmic}[1]
\REQUIRE The graph $\G=\lrA{\V,\E}$ and an even integer $\ell>0$.
\ENSURE A Hamiltonian cycle in $\G$.

\STATE Initialize $\W\coloneqq\emptyset$.

\FOR{all ordered sequences of $\ell$ cross edges\\ 
$\lrA{u_0v_0,\ldots, u_{\ell-1}v_{\ell-1}}$}\label{algsl2}

\IF{there exists $i<j$ with $\ell-1>j-i>1$ such that $v_{i-1}= v_{j-1}$ or $u_{i}= u_{j\bmod\ell}$}
\STATE \textbf{continue}
\ENDIF

\STATE Apply the $\lrA{\alpha+\varepsilon}$-approximation algorithm~\cite{TV:2024:book} for the \ac{TSPPK} to find a set $\B_1=\lrC{P_1,\ldots,P_{\ell/2}}$ of vertex-disjoint paths that cover all vertices in $\V_1$, where $P_i=\lrd{u_{2i-1},\ldots,u_{2i\bmod\ell}}$ for each $i\in\lrB{\ell/2}$. \label{algsl6}

\STATE Apply the $\lrA{\alpha+\varepsilon}$-approximation algorithm~\cite{TV:2024:book} for the \ac{TSPPK} to find a set $\B_2=\lrC{Q_1,\ldots,Q_{\ell/2}}$ of vertex-disjoint paths that covers all vertices in $\V_2$, where $Q_i=\lrd{v_{2i-2},\ldots,v_{2i-1}}$ for each $i\in\lrB{\ell/2}$.\label{algsl7}

\STATE $\W\coloneqq\W\cup\lrC{\sigma}$, where 
$\E\lrA{\sigma}=\E\lrA{\B_1}\cup \E\lrA{\B_2}\cup\lrc{u_iv_i}_{i=0}^{\ell-1}$.
\label{algsl8}
\ENDFOR
\RETURN $\arg\min_{\sigma\in\W}\w\lrA{\sigma}$.\label{algsl9}
\end{algorithmic}
\end{algorithm}

\begin{lemma}[$\textsc{Small}\lrA{\ell}$] \label{approxsmall}
For the \ac{BTSP} with $N_\c=\ell=O\lrA{1/\varepsilon}$, $\textsc{Small}\lrA{\ell}$ is an $\lrA{\alpha+\varepsilon}$-approximation algorithm with running time $N^{O\lrA{\ell}}f\lrA{N, k=O\lrA{1/\varepsilon}, \varepsilon}$, where $f\lrA{N, k, \varepsilon}$ denotes the running time of the $\lrA{\alpha+\varepsilon}$-approximation algorithm~\cite{TV:2024:book} for the \ac{TSPPK}.
\end{lemma}
\begin{IEEEproof}
Recall that $N_\c=\Size{\E_\c\lrA{\sigma^*}}$ denotes the number of cross edges in $\E\lrA{\sigma^*}$, which is even. 
When $N_\c=\ell$, we assume that 
\begin{align} \label{eq:sigma*}
\sigma^*= \lrd{u^*_0,v^*_0,\ldots,v^*_1,u^*_1, %\ldots,u^*_{\ell-2},v^*_{\ell-2},
\ldots,v^*_{\ell-1},u^*_{\ell-1}\ldots,u^*_0}.
\end{align}
where we may have $v^*_{i-1}=v^*_{j-1}$ or $u^*_i=u^*_{j\bmod\ell}$ if and only if $j=i+1$ and $i \in \lrB{\ell-1}$ is odd.

\paragraph*{Approximation Ratio}

We analyze the solution $\sigma$ computed by $\textsc{Small}\lrA{\ell}$ when the ordered cross edges enumerated in line~\ref{algsl2} of Algorithm~\ref{algs} are exactly those in $\sigma^*$, i.e., 
\begin{align}
\lrA{u_0v_0,\ldots, u_{\ell-1}v_{\ell-1}} = \lrA{u^*_0v^*_0,\ldots, u^*_{\ell-1}v^*_{\ell-1}}.
\end{align}
By line~\ref{algsl8}, we have 
\begin{equation}\label{eqs1}
\w\lrA{\sigma} = \w\lrA{\B_1,\B_2,\E_\c\lrA{\sigma^*}}.
\end{equation} 
From $\sigma^*$, we can extract $\ell/2$ vertex-disjoint paths $\B^*_1=\lrc{P^*_1,\ldots, P^*_{\ell/2}}$ in $\G\lrB{\V_1}$ that cover all vertices in $\V_1$, where $P^*_i=\lrd{u^*_{2i-1},\ldots,u^*_{2i\bmod\ell}}$ for each $i\in\lrB{\ell/2}$. 
Similarly, we extract $\ell/2$ vertex-disjoint paths $\B^*_2=\lrc{Q^*_1,\ldots, Q^*_{\ell/2}}$ in $\G\lrB{\V_2}$ that cover all vertices in $\V_2$, where $Q^*_i=\lrd{v^*_{2i-2},\ldots,v^*_{2i-1}}$ for each $i\in\lrB{\ell/2}$.
Note that $\B^*_1$ and $\B^*_2$ are feasible solutions to the \acp{TSPPK} considered in lines~\ref{algsl6} and \ref{algsl7}. 
Since $\B_1$ and $\B_2$ are obtained by applying the $\lrA{\alpha+\varepsilon}$-approximation algorithm~\cite{TV:2024:book} for the \ac{TSPPK}, we have 
\begin{equation}\label{eqs2}
\w\lrA{\B_1,\B_2}\leq \lrA{\alpha+\varepsilon} \, \w\lrA{\B^*_1,\B^*_2}.
\end{equation}

By the definition of $\sigma^*$, we have 
\begin{equation}\label{eqs3}
\w\lrA{\sigma^*} = \w\lrA{\B^*_1,\B^*_2,\E_\c\lrA{\sigma^*}}.
\end{equation}
Combining \eqref{eqs1}--\eqref{eqs3}, we obtain  
\begin{align}
\w\lrA{\sigma} \leq \lrA{\alpha+\varepsilon} \, \w\lrA{\sigma^*}.
\end{align}
Therefore, by line~\ref{algsl9}, the best solution returned by $\textsc{Small}\lrA{\ell}$ has cost at most $\lrA{\alpha+\varepsilon} \, \w\lrA{\sigma^*}$.

\paragraph*{Running Time}

Since the graph $\G$ contains at most $N^2$ cross edges, line~\ref{algsl2} enumerates at most $\binom{N^2}{\ell}\ell!= N^{O\lrA{\ell}}$ ordered sequences of $\ell$ cross edges.
For each enumeration, the algorithm computes a solution to the \ac{BTSP}, where the running time is dominated by two calls to the $\lrA{\alpha+\varepsilon}$-approximation algorithm~\cite{TV:2024:book} for the \ac{TSPPK}. Therefore, the overall running time of $\textsc{Small}\lrA{\ell}$ is $N^{O\lrA{\ell}}f\lrA{N, k=O\lrA{1/\varepsilon}, \varepsilon}$. %, where $f\lrA{N, \varepsilon}$ denotes the running time of the $\lrA{\alpha+\varepsilon}$-approximation algorithm for the \ac{TSPPK}.
\end{IEEEproof}

\subsection{The Algorithm \textsc{Large}}
The algorithm \textsc{Large} is a $\lrA{2\alpha+O\lrA{1/N_\c}}$-approximation algorithm for the \ac{BTSP}. %, where $\alpha$ denotes the best-known approximation ratio for the metric \ac{TSP}. 
Hence, when $N_\c=\Omega\lrA{1/\varepsilon}$, it achieves an approximation of $2\alpha+O\lrA{\varepsilon}$.
Recall that \textsc{Large} simply enumerates two independent cross edges $e_1=u_1v_1$ and $e_2=u_2v_2$, where $u_1,u_2\in \V_1$ and $v_1,v_2\in \V_2$. 
It then computes an approximate Hamiltonian path $P\lrA{u_1,u_2}$ between $u_1$ and $u_2$ in $\G\lrB{\V_1}$ and an approximate Hamiltonian path $P\lrA{v_1,v_2}$ between $v_1$ and $v_2$ in $\G\lrB{\V_2}$ by using the $\lrA{\alpha+\varepsilon/4}$-approximation algorithm for the \ac{TSPP}~\cite{TVZ:22:SICOMP}. 
Finally, it constructs a Hamiltonian cycle by combining these paths with the cross edges. The algorithm is summarized in Algorithm~\ref{algl}.

\begin{algorithm}[t]
\caption{\textsc{Large}}
\label{algl}
\begin{algorithmic}[1]
\REQUIRE The graph $\G=\lrA{\V,\E}$.
\ENSURE A Hamiltonian cycle in $\G$.

\STATE Initialize $\W\coloneqq\emptyset$.

\FOR{all possible pairs of independent cross edges\\ 
$e_1=u_1v_1$ and $e_2=u_2v_2$ with $u_1,u_2\in \V_1$ and\\
$v_1,v_2\in \V_2$}\label{algll2}

\STATE Find $\lrA{\alpha+{\varepsilon}/{4}}$-approximate Hamiltonian paths 
$P\lrA{u_1,u_2}$ and $P\lrA{v_1,v_2}$ in the graphs $\G\lrB{\V_1}$ and $\G\lrB{\V_2}$, respectively, by using the algorithm in~\cite{TVZ:22:SICOMP}.\label{algll3}

\STATE $\W\coloneqq\W\cup\lrC{\sigma}$, where 
$\E\lrA{\sigma}=\E\lrA{P\lrA{u_1,u_2}}\cup \E\lrA{P\lrA{v_1,v_2}}\cup\lrC{e_1,e_2}$.
\label{algll4}

\ENDFOR
\RETURN $\arg\min_{\sigma\in\W}\w\lrA{\sigma}$.\label{algll5}
\end{algorithmic}
\end{algorithm}

\begin{lemma}[\textsc{Large}]\label{approxlarge}
For the \ac{BTSP} with $N_\c \geq 4$ and any constant $0<\varepsilon<1$, \textsc{Large} is a $\lrA{2\alpha+{\varepsilon}/{2}+{16}/{N_\c}}$-approximation algorithm with running time $O\lra{N^4} \, g\lrA{N,\varepsilon}$, where $g\lrA{N,\varepsilon}$ denotes the running time of the $\lrA{\alpha+\varepsilon/4}$-approximation algorithm for the \ac{TSPP}.
\end{lemma}
\begin{IEEEproof}
As outlined in the idea sketch of Section~\ref{sec:LargeOverview}, the algorithm relies on the existence of a pair of independent cross edges and the cost bounds in \eqref{eq:L:1} and \eqref{eq:L:2}. The proof formalizes these bounds.

Let $\ell= N_\c$ and let the optimal Hamiltonian cycle $\sigma^*$ be as in \eqref{eq:sigma*}.
For convenience, define $u_{\ell}\coloneqq u_{0}$ and $v_{\ell}\coloneqq v_0$.
For any $i\in\lrB{\ell/2}$, define 
\begin{align}
P\lrA{v_{2i-2},u_{2i}} = \lrd{v_{2i-2},\ldots,v_{2i-1},u_{2i-1},\ldots, u_{2i}},
\end{align}
which is the sub-path between $v_{2i-2}$ and $u_{2i}$ from $\sigma^*$, and let
\begin{align}
\Lambda_i
\coloneqq
\w\lrA{u_{2i-2}v_{2i-2},P\lrA{v_{2i-2},u_{2i}},u_{2i}v_{2i}}
\end{align}
for brevity. 
Then, since each edge of $\sigma^*$ appears in at most two of $\Lambda_1, \ldots, \Lambda_{\ell/2}$, we have
%\begin{align}
%&\sum_{i=1}^{{\ell}/{2}}
%\lrB{\w\lrA{u_{2i-2},v_{2i-2}}+\w\lrA{P\lrA{v_{2i-2},u_{2i}}}+\w\lrA{u_{2i},v_{2i}}}
%\nonumber \\
%& \leq 2\w\lrA{\sigma^*}.
%\end{align} 
\begin{align}
\sum_{i=1}^{{\ell}/{2}}
\Lambda_i  \leq 2\w\lrA{\sigma^*}.
\end{align} 
Thus, there exists some $i^*\in \lrB{\ell/2}$ with
%\begin{align}\label{eql2}
%\w\lrA{u_{2i^*-2},v_{2i^*-2}}
%&+\w\lrA{P\lrA{v_{2i^*-2},u_{2i^*}}}
%\nonumber \\
%&+\w\lrA{u_{2i^*},v_{2i^*}}
%\leq \frac{4}{\ell}\w\lrA{\sigma^*}.
%\end{align}
\begin{align}\label{eql2}
\Lambda_{i^*}
\leq \frac{4}{\ell}\w\lrA{\sigma^*}.
\end{align}

We select $u_{2i^*-2}v_{2i^*-2}$ and $u_{2i^*}v_{2i^*}$ as our target pair of independent cross edges (corresponding to $e_1$ and $e_2$).
Next, we construct a Hamiltonian path between $u_{2i^*-2}$ and $u_{2i^*}$ in $\G\lrB{\V_1}$, and another between $v_{2i^*-2}$ and $v_{2i^*}$ in $\G\lrB{\V_2}$.

\paragraph*{Hamiltonian Path in $\G\lrB{\V_1}$}

By shortcutting all vertices in $\V_2$ from $\sigma^*$, we obtain a Hamiltonian cycle $\sigma'_1$ in $\G\lrB{\V_1}$ with 
\begin{align}
\sigma'_1 = \lrd{u_{2i^*-2},\ldots,u_{2i^*-1},\ldots,u_{2i^*-2}}.
\end{align}
Clearly, by the triangle inequality, we have 
\begin{equation}\label{eql3}
\w\lrA{\sigma'_1}\leq \w\lrA{\sigma^*}.
\end{equation}
Hence, by shortcutting the repeated vertices in $\V_1$ from the multigraph formed by the edges in $\E\lrA{\sigma'} \!\uplus \! \lrC{u_{2i^*-2}u_{2i^*}\!}$, we obtain a Hamiltonian path between $u_{2i^*-2}$ and $u_{2i^*}$ in $\G\lrB{\V_1}$ with cost at most $\w\lrA{\sigma'_1} + \w\lrA{u_{2i^*-2}u_{2i^*}}$ by the triangle inequality.
Therefore, the minimum-cost Hamiltonian path $P^*\lrA{u_{2i^*-2},u_{2i^*}}$ between $u_{2i^*-2}$ and $u_{2i^*}$ in $\G\lrB{\V_1}$ satisfies
\begin{align}\label{eql4}
&\w\lrA{P^*\lrA{u_{2i^*-2},u_{2i^*}}}
\nonumber \\
&\leq \w\lrA{\sigma'_1,u_{2i^*-2}u_{2i^*}}
\nonumber \\
&\leq \w\lrA{\sigma^*} + \Lambda_{i^*} - \w\lrA{u_{2i^*}v_{2i^*}}
\end{align}
where the second inequality follows from \eqref{eql3} and the triangle inequality.

\paragraph*{Hamiltonian Path in $\G\lrB{\V_2}$}

By shortcutting all vertices in $\V_1$ from $\sigma^*$, we obtain a Hamiltonian cycle $\sigma'_2$ in $\G\lrB{\V_2}$ with
\begin{align}
\sigma'_2 = \lrd{v_{2i^*-2},\ldots,v_{2i^*-1},\ldots,v_{2i^*-2}}.
\end{align}
By the triangle inequality, we have 
\begin{equation}\label{eql5}
\w\lrA{\sigma'_2}\leq \w\lrA{\sigma^*}.
\end{equation}
Hence, by shortcutting the repeated vertices in $\V_2$ from the multigraph formed by the edges in $\E\lrA{\sigma'}\uplus \lrC{v_{2i^*-2}v_{2i^*}}$, we obtain a Hamiltonian path between $v_{2i^*-2}$ and $v_{2i^*}$ in $\G\lrB{\V_2}$ with cost at most $\w\lrA{\sigma'_2} + \w\lrA{v_{2i^*-2}v_{2i^*}}$ by the triangle inequality.
Therefore, the minimum-cost Hamiltonian path $P^*\lrA{v_{2i^*-2},v_{2i^*}}$ between $v_{2i^*-2}$ and $v_{2i^*}$ in $\G\lrB{\V_2}$ satisfies
\begin{align}\label{eql6}
&\w\lrA{P^*\lrA{v_{2i^*-2},v_{2i^*}}}
\nonumber \\
&\leq \w\lrA{\sigma'_2,v_{2i^*-2}v_{2i^*}}
\nonumber \\
&\leq \w\lrA{\sigma^*} + \Lambda_{i^*} - \w\lrA{u_{2i^*-2}v_{2i^*-2}},
\end{align}
where the second inequality follows from \eqref{eql5} and the triangle inequality.

\paragraph*{Approximation Ratio}

For brevity, let 
\begin{align}
\Sigma_i
\coloneqq
&\w\lrA{P^*\lrA{u_{2i-2},u_{2i}},P^*\lrA{v_{2i-2},v_{2i}}}
\nonumber \\
&+\w\lrA{u_{2i-2}v_{2i-2},v_{2i}u_{2i}}.
\end{align}
Then, combining \eqref{eql2}, \eqref{eql4}, and \eqref{eql6}, we have
\begin{align} \label{eq:L:AR}
%&\w\lrA{P^*\lrA{u_{2i^*-2},u_{2i^*}}}+\w\lrA{P^*\lrA{v_{2i^*-2},v_{2i^*}}}
%\nonumber \\
%&+\w\lrA{u_{2i^*-2},v_{2i^*-2}}+\w\lrA{v_{2i^*},u_{2i^*}}
%\nonumber \\
%&\leq 2\w\lrA{\sigma^*}+2\w\lrA{u_{2i^*-2},v_{2i^*-2}}\\
%&\quad+2\w\lrA{P\lrA{v_{2i^*-2},u_{2i^*}}}+2\w\lrA{v_{2i^*},u_{2i^*}}\\
\Sigma_{i^*}
&\leq 2\w\lrA{\sigma^*}+2\Lambda_{i^*}
\nonumber \\
&\leq 2\w\lrA{\sigma^*}+\frac{8}{\ell}\w\lrA{\sigma^*}.
\end{align}
Finally, we bound the cost of the constructed Hamiltonian cycle in $\G$.
Because $u_{2i^*-2}v_{2i^*-2}$ and $u_{2i^*}v_{2i^*}$ are independent cross edges, they will be evaluated during the enumeration in line~\ref{algll2} of Algorithm~\ref{algl}. 
For this specific pair, the combined cost of the solution $\sigma$ generated by the algorithm is at most
\begin{align}
&\lrA{\alpha+\varepsilon/4} \w\lrA{P^*\lrA{u_{2i^*-2},u_{2i^*}},P^*\lrA{v_{2i^*-2},v_{2i^*}}}
\nonumber \\
&\quad + \w\lrA{u_{2i^*-2}v_{2i^*-2},u_{2i^*}v_{2i^*}}
\nonumber \\
&\leq \lrA{\alpha+\varepsilon/4} \Sigma_{i^*}
\nonumber \\
&\leq \lrA{2\alpha+\varepsilon/2}\w\lrA{\sigma^*} + \lrA{\alpha+\varepsilon/4}\frac{8}{\ell}\w\lrA{\sigma^*}
\nonumber \\
&\leq \lrA{2\alpha+\varepsilon/2+16/\ell}\w\lrA{\sigma^*},
\end{align}
where the first inequality follows from lines~\ref{algll3} and \ref{algll4}, the second inequality follows from \eqref{eq:L:AR}, and the last inequality follows from $\alpha<1.5$~\cite{KKG:23:IPCO} and $\varepsilon<1$.
Therefore, by line~\ref{algll5}, the best solution returned by \textsc{Large} has cost at most $\lrA{2\alpha+\varepsilon/2+16/\ell}\w\lrA{\sigma^*}$.

\paragraph*{Running Time}
%For the running time, 
Since $\G$ contains at most $N^2$ cross edges, line~\ref{algll2} enumerates at most $O\lra{N^4}$ pairs of independent cross edges.
For each enumeration, the algorithm computes a solution to the \ac{BTSP}, where the running time is dominated by two calls to the $\lrA{\alpha+\varepsilon/4}$-approximation algorithm~\cite{TVZ:22:SICOMP} for the \ac{TSPP}. 
Hence, the overall running time of \textsc{Large} is $O\lra{N^4}\,g\lrA{N,\varepsilon}$, where $g\lrA{N,\varepsilon}$ denotes the running time of the $\lrA{\alpha+\varepsilon/4}$-approximation algorithm for the \ac{TSPP}.
\end{IEEEproof}

\begin{theorem}[$\lrA{2\alpha+\varepsilon}$-Approximation in Polynomial Time]
For the \ac{BTSP} with any constant $\varepsilon>0$, Algorithm~\ref{algf} achieves an approximation ratio of $2\alpha+\varepsilon$ in polynomial time.
\end{theorem}
\begin{IEEEproof}
Recall that $N_\c=\Size{\E_\c\lrA{\sigma^*}}$ denotes the number of cross edges in $\E\lrA{\sigma^*}$. 
We consider the following cases.

\paragraph*{Case~1 ($N_\c>\Ceil{{32}/{\varepsilon}}$)}

By Lemma~\ref{approxlarge}, the solution $\sigma_0$ obtained by \textsc{Large} in line~\ref{algsf2} of Algorithm~\ref{algf} satisfies 
\begin{align}
\w\lrA{\sigma_0}\leq \lrA{2\alpha+\varepsilon/2+16/N_\c}\w\lrA{\sigma^*}\leq \lrA{2\alpha+\varepsilon}\w\lrA{\sigma^*}.
\end{align}

\paragraph*{Case~2 ($N_\c\leq\Ceil{{32}/{\varepsilon}}$)}

By Lemma~\ref{approxsmall}, the solution $\sigma_{N_\c}$ obtained by $\textsc{Small}\lrA{N_\c}$ in line~\ref{algsf4} of Algorithm~\ref{algf} satisfies 
\begin{align}
\w\lrA{\sigma_{N_\c}}\leq \lrA{\alpha+\varepsilon} \w\lrA{\sigma^*}\leq \lrA{2\alpha+\varepsilon}\w\lrA{\sigma^*}.
\end{align}

Hence, by line~\ref{algsf6}, the best solution returned by Algorithm~\ref{algf} has cost at most $\lrA{2\alpha+\varepsilon}\w\lrA{\sigma^*}$.

\paragraph*{Running Time}

Note that for any fixed constant $\varepsilon > 0$, we have $f\lrA{N, k=O\lrA{1/\varepsilon}, \varepsilon}=N^{O\lrA{1}}$~\cite{TV:2024:book} and $g\lrA{N,\varepsilon}=N^{O\lrA{1}}$~\cite{TVZ:22:SICOMP} (although their explicit polynomial bounds are not specified).
Therefore, by Lemmas~\ref{approxsmall} and \ref{approxlarge}, the running time of Algorithm~\ref{algf} is at most
\begin{align}
\sum_{i=1}^{\Ceil{32/\varepsilon}}N^{O\lrA{i}}f\lrA{N, k=O\lrA{1/\varepsilon}, \varepsilon} + O\lrA{N^4}g\lrA{N,\varepsilon},
\end{align}
which is polynomial for any constant $\varepsilon$, but scales as $N^{\Omega\lrA{1/\varepsilon}}$ when $\varepsilon$ becomes arbitrarily small. 
We use $\Omega$ in the exponent to emphasize that while the running time is polynomial for any constant $\varepsilon$, the degree of the polynomial is strictly bounded below by $\Omega\lrA{1/\varepsilon}$ due to the underlying subroutines.
\end{IEEEproof}

\begin{remark}[Breaking the $3$-Approximation Barrier]
The approximation ratio $2\alpha +\varepsilon < 3$ represents a substantial improvement over the previous $7/2$-approximation algorithm, breaking the $3$-approximation barrier for the \ac{BTSP}. The key technical insight is the structural connection between the \ac{BTSP} and the \ac{TSPPK}, established through cross-edge enumeration. This connection, combined with our structural patching technique that enumerates only two independent cross edges, demonstrates that the approximability of the \ac{BTSP} is fundamentally governed by the number of cross edges in an optimal Hamiltonian cycle. We expect that this structural insight extends naturally to multi-partition \ac{BTSP} variants, facilitating the design of constant-factor approximation algorithms in such generalized settings.
\end{remark}

\section{A Quadratic $4$-Approximation Algorithm}
\label{sec:5}

In this section, we propose a $4$-approximation algorithm for the \ac{BTSP}  with running time $O\lra{N^2}$, improving upon the previous $7/2$-approximation algorithm that runs in $O\lra{N^3}$ time~\cite{MRM:17:TCS}.

\subsection{Algorithm Overview}

Our algorithm distinguishes the special and general cases based on the structure of $\G$ and the \ac{MST} $\T$ of $\G$.

\subsubsection{Singleton Case ($\Size{\V_1}=1$ or $\Size{\V_2}=1$)}
We begin by considering a \emph{singleton} case where either the vertex set $\V_1$ or $\V_2$ consists only of a single vertex, i.e., $\Size{\V_1}=1$ or $\Size{\V_2}=1$.
For this specific scenario, it was previously shown in~\cite{MRM:17:TCS} that a $3/2$-approximation algorithm exists with running time $O\lra{N^3}$, dominated by the computation of a minimum-cost matching as in Christofides' algorithm~\cite{C:22:ORF}.
We instead present a $2$-approximation algorithm with running time $O\lra{N^2}$, which is crucial for obtaining our efficient algorithm. 

\subsubsection{General Case ($\Size{\V_1}\geq 2$ and $\Size{\V_2}\geq 2$)}
Our algorithm starts by computing an \ac{MST} $\T$ of $\G$ in $O\lra{N^2}$ time~\cite{CLRS:2022:book} and classifying $\T$ based on its cross-edge structure. 
Let $\E_\c\lrA{\T}$ denote the set of cross edges in $\E\lrA{\T}$.
Removing all cross edges in $\E_\c\lrA{\T}$ from $\T$ yields collections $\mathscr{X} =\lrC{\X_1,\ldots, \X_p}$ and $\mathscr{Y} = \lrC{\Y_1,\ldots, \Y_\ell}$ of acyclic components $\X_i$ and $\Y_j$, where $\V\lrA{\X_i} \subseteq \V_1$ and $\V\lrA{\Y_j} \subseteq \V_2$ for all $i\in \lrB{p}$ and $j\in\lrB{\ell}$. 
For any component $\B \in \mathscr{X}\cup \mathscr{Y}$, the subgraph induced by $\V\lrA{\B}$, denoted by $\T\lrB{\V\lrA{\B}}$, is a spanning tree of $\G\lrB{\V\lrA{\B}}$.
Moreover, for any vertex set $\V'\subseteq \V$, the subgraph induced by $\V'$, denoted by $\T\lrB{\V'}$, is a set of trees.
The spanning tree $\T$ is called a \emph{1-\ac{MST}} if $\E_\c\lrA{\T}$ contains at least one pair of independent cross edges, and a \emph{0-\ac{MST}} otherwise. 
If $\T$ is a 0-\ac{MST}, then all its cross edges are incident to a single vertex. 
Depending on the number of such cross edges, we further distinguish two subcases: \emph{mono-cross 0-\ac{MST}} in which there is exactly one cross edge, and \emph{poly-cross 0-\ac{MST}} in which there are multiple cross edges.
If $\T$ is a 1-MST, its component tree $\C$ has vertices corresponding to the components in $\mathscr{X}\cup \mathscr{Y}$ and edges corresponding to the cross edges in $\E_\c\lrA{\T}$. %(see Section~\ref{sec:1-MST} for the formal definition). 
Depending on the depth of $\C$, we further distinguish two subcases: \emph{mono-depth 1-\ac{MST}} in which $\depth\lrA{\C} = 1$, and \emph{poly-depth 1-\ac{MST}} in which $\depth\lrA{\C} \geq 2$.

We first recall a well-known algorithm for computing a Hamiltonian path in a metric graph (see \cite[Lemma~2]{MRM:17:TCS}).

\begin{lemma}[Hamiltonian Path via Spanning Tree] \label{path}
In a metric graph $\HH=\lrA{\V_\HH,\E_\HH}$ with edge-cost function $w_\HH$, given any spanning tree $\T$ of $\HH$ and two distinct vertices $x,y\in \V_\HH$, there exists an $O\lrA{\Size{\V_\HH}}$-time algorithm that computes a Hamiltonian path $P\lrA{x,y}$ between $x$ and $y$ in $\HH$ such that 
\begin{align}
w_\HH\lrA{P\lrA{x,y}}\leq 2w_\HH\lrA{\T}-w_\HH\lrA{P_\T\lrA{x,y}}.
\end{align}
\end{lemma}
\begin{IEEEproof}
By doubling all edges in $\E_\HH\lrA{\T}\setminus \E_\HH\lrA{P_\T\lrA{x,y}}$, we obtain a multigraph $\HH'=\lrA{\V_\HH, \E'_\HH}$ with $\E'_\HH=\E_\HH\lrA{2\T}\setminus \E_\HH\lrA{P_\T\lrA{x,y}}$. 
Since $\odd\lrA{\HH'}=\lrC{x,y}$, there exists an \emph{Eulerian walk} $\sigma=\lrd{x,\ldots,y}$ that traverses all edges in $\E'_\HH$, which can be found in $O\lrA{\Size{\V_\HH}+\Size{\E'_\HH}}$ time~\cite{CLRS:2022:book}.
By shortcutting repeated vertices in $\sigma$, we obtain a Hamiltonian path $P\lrA{x,y}$ between $x$ and $y$ in $\HH$.
By the triangle inequality, %we have
\begin{align}
w_\HH\lrA{P\lrA{x,y}}
\leq w_\HH\lrA{\sigma}
=
2w_\HH\lrA{\T}-w_\HH\lrA{P_\T\lrA{x,y}}.
\end{align}
Moreover, since $\Size{\E_\HH\lrA{\T}}=\Size{\V_\HH}-1$, $\Size{\E'_\HH}=O\lrA{\Size{\V_\HH}}$, and hence the running time is $O\lrA{\Size{\V_\HH}}$.
\end{IEEEproof}

\subsection{The Algorithm for the Singleton Case}

We now present an $O\lra{N^2}$-time $2$-approximation algorithm for the singleton case ($\Size{\V_1}=1$ or $\Size{\V_2}=1$).

\begin{algorithm}[t!]
\caption{The algorithm for the singleton case} % ($\Size{\V_1}=1$ and $\Size{\V_2}>1$)}
\label{alg1}
\begin{algorithmic}[1]
\REQUIRE The graph $\G=\lrA{\V,\E}$.
\ENSURE A Hamiltonian cycle in $\G$.

\STATE Compute an \ac{MST} $\T_2$ of $\G\lrB{\V_2}$~\cite{CLRS:2022:book}.

\STATE Find two distinct vertices $x,y\in \V_2$ such that $\w\lrA{ux,uy}$ is minimized.

\STATE Compute a Hamiltonian path $P\lrA{x,y}$ between $x$ and $y$ in $\G\lrB{\V_2}$ using the spanning tree $\T_2$ and the algorithm in Lemma~\ref{path}.

\RETURN A Hamiltonian cycle $\sigma$ in $\G$ with\\ 
$\E\lrA{\sigma}=\E\lrA{P\lrA{x,y}}\cup\lrC{ux,uy}$.

\end{algorithmic}
\end{algorithm}

\begin{lemma}[Singleton Case] \label{res1}
When either $\V_1$ or $\V_2$ consists only of a single vertex, there exists a $2$-approximation algorithm with running time $O\lra{N^2}$ for the \ac{BTSP}.
\end{lemma}
\begin{IEEEproof}
The problem is trivial if $\Size{\V_1}=\Size{\V_2}=1$.
Suppose $\V_1=\lrC{u}$ and $\Size{\V_2}\geq 2$. 
Our algorithm proceeds as follows.

First, we compute an \ac{MST} $\T_2$ of $\G\lrB{\V_2}$~\cite{CLRS:2022:book}, and find two distinct vertices $x,y\in \V_2$ such that $\w\lrA{ux,uy}$ is minimized. 
Next, we compute a Hamiltonian path $P\lrA{x,y}$ between $x$ and $y$ in $\G\lrB{\V_2}$ using the spanning tree $\T_2$ and the algorithm in Lemma~\ref{path}.
Finally, we construct a Hamiltonian cycle $\sigma$ in $\G$ by combining the path $P\lrA{x,y}$ with the edges $\lrC{ux,uy}$; that is 
\begin{align}
\E\lrA{\sigma}=\E\lrA{P\lrA{x,y}}\cup\lrC{ux,uy}.
\end{align}
The full procedure is detailed in Algorithm~\ref{alg1}.

The running time is $O\lra{N^2}$, dominated by computing the spanning tree $\T_2$ in $\G\lrB{\V_2}$ and selecting the edges $\lrC{ux,uy}$.
Since $\sigma^*$ must contain exactly two cross edges incident to $u$ and a Hamiltonian path in $\G\lrB{\V_2}$, we have 
\begin{align}
\w\lrA{ux,uy,\T_2}\leq \w\lrA{\sigma^*}
\end{align}
by our algorithm. 
Moreover, since $\w\lrA{P\lrA{x,y}}\leq 2\w\lrA{\T_2}$ by Lemma~\ref{path}, we have 
\begin{align}
\w\lrA{\sigma}&=\w\lrA{P\lrA{x,y},ux,uy}
\nonumber \\
&\leq 2\w\lrA{\T_2,ux,uy}
\nonumber \\
&\leq 2\w\lrA{\sigma^*},
\end{align}
which completes the proof.
\end{IEEEproof}

%\subsection{The Algorithm for the Mono-Cross 0-\ac{MST} Case}
\subsection{The Algorithm for the 0-\ac{MST} Case}

Now we assume that $\T$ is a 0-MST.
Previously, \cite{MRM:17:TCS} showed that a $3$-approximation algorithm exists whose running time is dominated by computing two independent cross edges of minimum total cost. We improve upon this result by addressing the two subcases---mono-cross and poly-cross 0-\acp{MST}---separately.

\subsubsection{Mono-Cross 0-\ac{MST} Case}

%In this subsection, 
We present an improved $5/2$-approximation algorithm with running time $O\lra{N^2}$.

Our algorithm first selects two independent cross edges $\lrC{u_1v_1,u_2v_2}$ based on the unique cross edge in $\E_\c\lrA{\T}$, where $u_1,u_2\in \V_1$ and $v_1,v_2\in \V_2$. 
It then computes a Hamiltonian path $P\lrA{u_1,u_2}$ in $\G\lrB{\V_1}$ and a Hamiltonian path $P\lrA{v_1,v_2}$ in $\G\lrB{\V_2}$.
Finally, it constructs a Hamiltonian cycle in $\G$ by combining the paths $P\lrA{u_1,u_2}$ and $P\lrA{v_1,v_2}$ with the edges $\lrC{u_1v_1,u_2v_2}$, as in line~\ref{algll4} of Algorithm~\ref{algl}.
The algorithm is summarized in Algorithm~\ref{alg2}.

\begin{lemma}[Mono-Cross 0-\ac{MST}]\label{res2}
When $\T$ is a 0-\ac{MST} with exactly one cross edge, there exists a $5/2$-approximation algorithm with running time $O\lra{N^2}$ for the \ac{BTSP}.
\end{lemma}
\begin{IEEEproof}
We first analyze the running time of Algorithm~\ref{alg2}.

\paragraph*{Running Time}

Since $\G$ contains at most $N^2$ cross edges, line~\ref{2l2} runs in $O\lra{N^2}$ time. 
For line~\ref{2l3}, the cross edges $e'_1$ and $e'_2$ can be found in $O\lra{N^2}$ time by enumerating the edges $uv'$ and $u'v$ for all vertices $u'\in \V_1\setminus\lrC{u}$ and $v'\in \V_2\setminus\lrC{v}$.
Moreover, line~\ref{2l5} runs in $O\lra{N^2}$ time by Lemma~\ref{path}.
Hence, Algorithm~\ref{alg2} runs in $O\lra{N^2}$ time.

\begin{algorithm}[t!]
\caption{The algorithm for the mono-cross 0-\ac{MST} case}
\label{alg2}
\begin{algorithmic}[1]
\REQUIRE The graph $\G=\lrA{\V,\E}$ and its spanning tree $\T$.
\ENSURE A Hamiltonian cycle in $\G$.

\STATE Let $e_1=uv$ be the unique cross edge in $\E_\c\lrA{\T}$, where $u\in \V_1$ and $v\in \V_2$, and let $\T_i=\T\lrB{\V_i}$ for each $i\in\lrB{2}$.
%Removing $e_1$ from $\T$ results in a spanning tree, denoted by $\T_i$, in $\G\lrB{\V_{i}}$ for $i\in\lrB{2}$.

\STATE\label{2l2} Find a minimum-cost cross edge $e_2$ that is vertex disjoint from $e_1$ (which exists since $\Size{\V_1} \ge 2$ and $\Size{\V_2} \ge 2$).

\STATE\label{2l3} Find a pair of vertex-disjoint cross edges $e'_1$ and $e'_2$ such that $u\in e'_1$, $v\in e'_2$, and $\w\lrA{e'_1,e'_2}$ is minimized.

\STATE\label{2l4} Let $\E_\T=\lrC{e_1,e_2}$ if $\w\lrA{e_1,e_2}\leq \w\lrA{e'_1,e'_2}$, and $\E_\T=\lrC{e'_1,e'_2}$ otherwise. Relabel $\E_\T=\lrC{u_1v_1,u_2v_2}$, where $u_1,u_2\in \V_1$ and $v_1,v_2\in \V_2$.

\STATE\label{2l5} Compute Hamiltonian paths $P\lrA{u_1,u_2}$ in $\G\lrB{\V_1}$ and $P\lrA{v_1,v_2}$ in $\G\lrB{\V_2}$ by applying the algorithm in Lemma~\ref{path} to $\T_1$ and $\T_2$, respectively.

\RETURN\label{2l6} A Hamiltonian cycle $\sigma$ in $\G$ with\\ 
$\E\lrA{\sigma}=\E\lrA{P\lrA{u_1,u_2}}\cup \E\lrA{P\lrA{v_1,v_2}}\cup \E_\T$.
\end{algorithmic}
\end{algorithm}

\paragraph*{Approximation Ratio}

It remains to show that the Hamiltonian cycle $\sigma$ computed by Algorithm~\ref{alg2} is a $5/2$-approximate solution.
By Algorithm~\ref{alg2}, we have
\begin{align}
\E\lrA{\sigma}
&=\E\lrA{P\lrA{u_1,u_2}}\cup \E\lrA{P\lrA{v_1,v_2}}\cup \E_\T ; \\
\E\lrA{\T}
&=\E\lrA{\T_1}\cup \E\lrA{\T_2}\cup \lrC{e_1}.
\end{align} 
By Lemma~\ref{path}, we have $\w\lrA{P\lrA{u_1,u_2}}
\leq 2\w\lrA{\T_1}$ and $\w\lrA{P\lrA{v_1,v_2}} \leq 2\w\lrA{\T_2}$. 
Hence, 
\begin{equation}\label{1e1}
\w\lrA{\sigma}\leq 2\w\lrA{\T_1,\T_2}+\w\lrA{\E_\T},    
\end{equation}
where, by line~\ref{2l4},
\begin{align} \label{eq:L5:1}
\w\lrA{\E_\T}=\min\lrC{\w\lrA{e_1,e_2},\w\lrA{e'_1,e'_2}}.
\end{align}
Recall that $\sigma^*$ denotes a minimum-cost Hamiltonian cycle in $\G$. Since $\G$ is bipartitioned, $\sigma^*$ must contain an even number of cross edges.
We consider the following cases.

\begin{itemize}

\item
\textbf{Case~1} ($\sigma^*$ contains exactly two cross edges $e^*_1$ and $e^*_2$):
Since $\T$ is an \ac{MST} in $\G$, the unique cross edge $e_1$ of $\T$ must be a minimum-cost cross edge. Hence,
\begin{equation}\label{1e2}
\w\lrA{e_1}\leq \min\lrC{\w\lrA{e^*_1}, \w\lrA{e^*_2}}.    
\end{equation}

Note that $\E\lrA{\sigma^*}\setminus\lrC{e^*_1,e^*_2}$ forms spanning trees in $\G\lrB{\V_1}$ and $\G\lrB{\V_2}$. By the optimality of $\T$, we have
\begin{equation}\label{1e3}
\w\lrA{\T_1,\T_2}\leq \w\lrA{\sigma^*}-\w\lrA{e^*_1,e^*_2}.
\end{equation}

\end{itemize}

We further consider two subcases.

\begin{itemize}

\item
\textbf{Subcase~1.1} (either $e^*_1$ or $e^*_2$ is vertex-disjoint from $e_1$):\\
Assume without loss of generality that $e^*_i$ is vertex disjoint from $e_1$. 
We have $\w\lrA{e_2}\leq \w\lrA{e^*_i}$ by line~\ref{2l2} and $\w\lrA{e_1}\leq \w\lrA{e^*_{3-i}}$ by \eqref{1e2}. 
Thus, $\w\lrA{e_1,e_2}\leq \w\lrA{e^*_1,e^*_2}$. Then, by \eqref{1e3},
\begin{align} \label{eq:L5:2}
\w\lrA{\T_1,\T_2,e_1,e_2}\leq \w\lrA{\sigma^*}.
\end{align}
%Therefore, we have
%\begin{align}
%\w\lrA{\sigma}&\leq 2\w\lrA{\T_1}+2\w\lrA{\T_2}+\w\lrA{\E_\T} 
%\nonumber \\
%&\leq 2\w\lrA{\T_1}+2\w\lrA{\T_2}+\w\lrA{e_1}+\w\lrA{e_2}
%\nonumber \\
%&\leq 2\w\lrA{\sigma^*},    
%\end{align}
%where the first inequality follows from \eqref{1e1}, and the second from $\w\lrA{\E_\T}\leq \w\lrA{e_1}+\w\lrA{e_2}$ by line~\ref{2l4}.

\item
\textbf{Subcase~1.2} ($e^*_1$ and $e^*_2$ are not vertex-disjoint from $e_1$):
In this case, we have 
\begin{align}  \label{eq:L5:0}
\w\lrA{e'_1,e'_2}\leq \w\lrA{e^*_1,e^*_2}
\end{align}
by line~\ref{2l3}. Combining with \eqref{1e3},
\begin{align} \label{eq:L5:3}
\w\lrA{\T_1,\T_2,e'_1,e'_2}\leq \w\lrA{\sigma^*}.
\end{align}

\end{itemize}
In both subcases, it follows from \eqref{1e1}, \eqref{eq:L5:1}, and \eqref{eq:L5:2} (or \eqref{eq:L5:3}, respectively) that
\begin{align}
\w\lrA{\sigma}
&\leq 2\w\lrA{\sigma^*}.    
\end{align}

\begin{itemize}

\item
\textbf{Case~2} ($\sigma^*$ contains at least four cross edges):
Select a set $\E^*=\lrC{e^*_1,e^*_2,e^*_3,e^*_4}\subseteq \E_\c\lrA{\sigma^*}$ of four cross edges.
We may assume that at most three edges in $\E^*$ are incident to $u$ or $v$.
Otherwise, suppose that $e^*_1$ and $e^*_2$ are incident to $u$, while $e^*_3$ and $e^*_4$ are incident to $v$. 
Since $\sigma^*$ is a Hamiltonian cycle in $\G$, $\sigma^*$ must contain at least two additional cross edges that are vertex-disjoint from both $u$ and $v$, say $e^*_5$ and $e^*_6$. 
Then, $\E^*\coloneqq \lrC{e^*_3,e^*_4,e^*_5,e^*_6}$ is a desired set of four cross edges in $\E\lrA{\sigma^*}$ such that at most three edges in $\E^*$ are incident to $u$ or $v$.

\end{itemize}

We further consider two subcases.

\begin{itemize}

\item
\textbf{Subcase~2.1} (at most two edges in $\E^*$ are incident to $u$ or $v$): 
Assume without loss of generality that $e^*_3$ and $e^*_4$ are vertex-disjoint from $e_1$.
By line~\ref{2l2}, we have $\w\lrA{e_2}\leq \min\lrC{\w\lrA{e^*_3},\w\lrA{e^*_4}}$. Since $e_1$ is a minimum-cost cross edge as shown in (\ref{1e2}), we have $\w\lrA{e_1}\leq \min\lrC{\w\lrA{e^*_1},\w\lrA{e^*_2}}$. Hence, 
\begin{align} \label{eq:L5:4}
\w\lrA{e_1,e_2}\leq \frac{1}{2}\w\lrA{\E^*}\leq\frac{1}{2}\w\lrA{\sigma^*}.
\end{align}

\item
\textbf{Subcase~2.2} (exactly three edges in $\E^*$ are incident to $u$ or $v$): 
Assume without loss of generality that $e^*_1=uv'\neq e_1$, $e^*_2=u'v\neq e_1$, $e^*_3$ is the third edge incident to $u$ or $v$, and $e^*_4$ is vertex-disjoint from $e_1$.
Since $e_1$ is a minimum-cost cross edge, we have $\w\lrA{e_1,e_2}\leq \w\lrA{e^*_3,e^*_4}$ by line~\ref{2l2}. Moreover, by line~\ref{2l3}, we have %$\w\lrA{e'_1}+\w\lrA{e'_2}\leq \w\lrA{e^*_1}+\w\lrA{e^*_2}$ 
\eqref{eq:L5:0}. Hence, 
\begin{align} \label{eq:L5:5}
\w\lrA{e_1,e_2,e'_1,e'_2} \leq \w\lrA{\E^*}\leq \w\lrA{\sigma^*}.
\end{align}

\end{itemize}
In both subcases, since 
\begin{align}  
\w\lrA{\E_\T}
\leq \min\lrC{\w\lrA{e_1,e_2},\frac{1}{2}\w\lrA{e_1,e_2,e'_1,e'_2}}
\end{align}
by line~\ref{2l4}, it follows from \eqref{eq:L5:4} (or \eqref{eq:L5:5}, respectively) that
\begin{align} \label{eq:L5:6}
\w\lrA{\E_\T}
\leq 
\frac{1}{2}\w\lrA{\sigma^*}.
\end{align}
Combining with \eqref{1e1} and $\w\lrA{\T_1,\T_2}\leq \w\lrA{\T}\leq \w\lrA{\sigma^*}$ by the definitions of $\T_1$, $\T_2$, and $\T$, we obtain
\begin{align}
\w\lrA{\sigma}
&\leq \frac{5}{2}\w\lrA{\sigma^*},    
\end{align}
which completes the proof.
\end{IEEEproof}

\subsubsection{Poly-Cross 0-\ac{MST} Case}

We present an improved $2$-approximation algorithm with running time $O\lra{N^2}$.

Since $\T$ is a 0-\ac{MST} with multiple cross edges, we have either $\Size{\mathscr{X}}=1$ and $\Size{\mathscr{Y}}>1$ or $\Size{\mathscr{X}}>1$ and $\Size{\mathscr{Y}}=1$.
We may assume without loss of generality that $\Size{\mathscr{X}}=1$ and $\Size{\mathscr{Y}}=\ell>1$.
Moreover, let $u$ be the vertex in $\V_1$ to which all cross edges in $\E_\c\lrA{\T}$ are incident.

\begin{algorithm}[t]
\caption{The algorithm for the poly-cross 0-\ac{MST} case}
\label{alg3}  
\begin{algorithmic}[1]
\REQUIRE The graph $\G=\lrA{\V,\E}$ and its spanning tree $\T$.
\ENSURE A Hamiltonian cycle in $\G$.

\STATE Determine whether $\Size{\mathscr{X}}=1$ or $\Size{\mathscr{Y}}=1$. Assume without loss of generality that $\Size{\mathscr{X}}=1$ and $\Size{\mathscr{Y}}=\ell>1$.
Let $u$ be the vertex in $\V_1$ to which all cross edges in $\E_\c\lrA{\T}$ are incident.

\STATE\label{3l2} Find a minimum-cost cross edge $e_{\ell+1}$ that is not incident to $u$ (which exists since $\Size{\V_1} \ge 2$).

\STATE\label{3l3} Relabel $\mathscr{Y}=\lrC{\Y_1, \ldots, \Y_\ell}$ and $\E_\c\lrA{\T}=\lrC{e_1,\ldots,e_\ell}$ such that $e_i=uv_i$ connects $\V_1$ to $\V\lrA{\Y_i}$ for each $i\in\lrB{\ell}$ and $e_{\ell+1}=u'v_{\ell+1}$ connects $\V_1$ to $\V\lrA{\Y_\ell}$. 

\STATE\label{3ltree} Obtain a spanning tree $\T_2$ of $\G\lrB{\V_2}$ by combining all edges in $\E\lrA{\T\lrB{\V_2}}\cup \lrC{v_iv_{i+1}\mid i\in\lrB{\ell-1}}$.
 
\STATE\label{3ltsp} Compute Hamiltonian paths $P\lrA{u,u'}$ in $\G\lrB{\V_1}$ and $P\lrA{v_1,v_{\ell+1}}$ in $\G\lrB{\V_2}$ by applying the algorithm in Lemma~\ref{path} to $\T\lrB{\V_1}$ and $\T_2$, respectively.

\RETURN\label{3le} A Hamiltonian cycle $\sigma$ in $\G$ with\\ 
$\E\lrA{\sigma}=\E\lrA{P\lrA{u,u'}}\cup \E\lrA{P\lrA{v_1,v_{\ell+1}}}\cup \lrC{e_1,e_{\ell+1}}$.
\end{algorithmic}
\end{algorithm}

Similar to Algorithm~\ref{alg2}, our algorithm first selects two independent cross edges, then constructs two Hamiltonian paths in $\G\lrB{\V_1}$ and $\G\lrB{\V_2}$, respectively, and finally forms a Hamiltonian cycle in $\G$ by combining the paths with the cross edges.
Specifically, it first finds a minimum-cost cross edge $e_{\ell+1}$ that is not incident to $u$.
Relabel $\mathscr{Y}=\lrC{\Y_1, \ldots, \Y_\ell}$ and $\E_\c\lrA{\T}=\lrC{e_1,\ldots,e_\ell}$ such that $e_i=uv_i$ connects $\V_1$ to $\V\lrA{\Y_i}$ for each $i\in\lrB{\ell}$, and $e_{\ell+1}=u'v_{\ell+1}$ connects $\V_1$ to $\V\lrA{\Y_\ell}$.
The edges $\lrC{uv_1,u'v_{\ell+1}}$ serve as two independent cross edges, and combining all edges in $\E\lrA{\T\lrB{\V_2}}\cup \lrC{v_iv_{i+1}\mid i\in\lrB{\ell-1}}$ forms a spanning tree $\T_2$ of $\G\lrB{\V_2}$.
Then, the algorithm computes Hamiltonian paths $P\lrA{u,u'}$ in $\G\lrB{\V_1}$ and $P\lrA{v_1,v_{\ell+1}}$ in $\G\lrB{\V_2}$ by applying the algorithm in Lemma~\ref{path} to spanning trees $\T\lrB{\V_1}$ in $\G\lrB{\V_1}$ and $\T_2$ in $\G\lrB{\V_2}$, respectively.
Finally, a Hamiltonian cycle $\sigma$ in $\G$ is obtained by combining the paths $P\lrA{u,u'}$ and $P\lrA{v_1,v_{\ell+1}}$ with the cross edges $\lrC{uv_1,u'v_{\ell+1}}$; that is, 
\begin{align}\label{eq:L6:0}
\E\lrA{\sigma}=\E\lrA{P\lrA{u,u'}}\cup \E\lrA{P\lrA{v_1,v_{\ell+1}}}\cup \lrC{e_1,e_{\ell+1}}.
\end{align}
The algorithm is summarized in Algorithm~\ref{alg3}.

\begin{lemma}[Poly-Cross 0-\ac{MST}]\label{res3}
When $\T$ is a 0-\ac{MST} with multiple cross edges, there exists a $2$-approximation algorithm with running time $O\lra{N^2}$ for the \ac{BTSP}.
\end{lemma}
\begin{IEEEproof}
We first analyze the running time of Algorithm~\ref{alg3}.

\paragraph*{Running Time}

Since $\G$ contains at most $N^2$ cross edges, line~\ref{3l2} runs in $O\lra{N^2}$ time. 
For lines~\ref{3l3}--\ref{3le}, similar to the analysis in Lemma~\ref{res2}, the algorithm also runs in $O\lra{N^2}$ time. Hence, Algorithm~\ref{alg3} runs in $O\lra{N^2}$ time.

\paragraph*{Approximation Ratio}

It remains to show that the Hamiltonian cycle $\sigma$ computed by Algorithm~\ref{alg3} is a 2-approximate solution.
By line~\ref{3ltsp} and Lemma~\ref{path}, we obtain
\begin{align}\label{eq:L6:1}
\w\lrA{P\lrA{u,u'}}&\leq 2\w\lrA{\T\lrB{\V_1}};\\
%\end{align}
%\begin{align}
\label{eq:L6:2}
\w\lrA{P\lrA{v_1,v_{\ell+1}}} &\leq 2\w\lrA{\T_2}-\w\lrA{P_{\T_2}\lrA{v_1,v_{\ell+1}}}.
\end{align} 
Moreover, we have
\begin{align}\label{eq:L6:3}
&2\w\lrA{\T_2}-\w\lrA{P_{\T_2}\lrA{v_1,v_{\ell+1}}}
\nonumber \\
&\leq2\w\lrA{\T_2}-%\sum\nolimits_{i=1}^{\ell-1}\w\lrA{v_iv_{i+1}}
\w\lrA{v_1v_2, v_2v_3, \ldots, v_{\ell-1}v_\ell}
\nonumber \\
&\leq 2\w\lrA{\T\lrB{\V_2}}+ %\sum\nolimits_{i=1}^{\ell-1}\w\lrA{v_iv_{i+1}}
\w\lrA{v_1v_2, v_2v_3, \ldots, v_{\ell-1}v_\ell}
\nonumber \\
&\leq 2\w\lrA{\T\lrB{\V_2}}+ \sum\nolimits_{i=1}^{\ell-1}w\lrA{e_i,e_{i+1}}
\nonumber \\
&=2\w\lrA{\T\lrB{\V_2},\E_\c\lrA{\T}}-\w\lrA{e_1,e_\ell},
\end{align}
where the first inequality follows from the definition of $\T_2$ by line~\ref{3ltree} and the triangle inequality, the second from the definition of $\T\lrB{\V_2}$ by line~\ref{3ltree}, the third from the triangle inequality, and the last equality from $\E_\c\lrA{\T}=\sum_{i=1}^{\ell}\w\lrA{e_i}$ by definition.
Combining \eqref{eq:L6:1}--\eqref{eq:L6:3}, we have 
\begin{align}\label{2e0}
&\w\lrA{P\lrA{u,u'},P\lrA{v_1,v_{\ell+1}}}
\nonumber \\
&\leq 2\w\lra{\T\lrb{\V_1},\T\lrb{\V_2},\E_\c\lra{\T}}-\w\lra{e_1,e_\ell}
\nonumber \\
&= 2\w\lrA{\T}-\w\lrA{e_1,e_\ell},
\end{align}
where the inequality follows from the above analysis, and the equality follows from $\E\lrA{\T}=\E\lrA{\T\lrB{\V_1}}\cup \E\lrA{\T\lrB{\V_2}} \cup \E_\c\lrA{\T}$ by definition.
By \eqref{eq:L6:0} and \eqref{2e0}, we have
\begin{align}\label{2e1}
\w\lrA{\sigma}&=\w\lra{P\lra{u,u'},P\lra{v_1,v_{\ell+1}},e_1,e_{\ell+1}}
\nonumber \\
&\leq 2\w\lrA{\T} +  \w\lrA{e_{\ell+1}}.
\end{align}
Recall that $\sigma^*$ denotes a minimum-cost Hamiltonian cycle in $\G$. Since $\Size{\V_1}>1$, it must contain a cross edge $e^*_i$ that is not incident to $u$.
Hence, $\w\lrA{e_{\ell+1}}\leq \w\lrA{e^*_i}$ by line~\ref{3l2}.
Note that $\E\lrA{\sigma^*}\setminus\lrC{e^*_i}$ forms a spanning tree of $\G$. By the optimality of $\T$, we have $\w\lrA{\T}\leq \w\lrA{\sigma^*}-\w\lrA{e^*_i}$. Hence, $\w\lrA{\T,e_{\ell+1}}\leq \w\lrA{\sigma^*}$, so $\w\lrA{\sigma}\leq 2\w\lrA{\sigma^*}$ by \eqref{2e1}.
\end{IEEEproof}

\subsection{The Algorithm for the 1-\ac{MST} Case}
\label{sec:1-MST}

Now we consider that $\T$ is a 1-\ac{MST}.
With respect to the tree $\T$ of $\G$, we define the \emph{component tree} $\C=\lrA{\V_\C,\E_\C}$, where $\V_\C=\mathscr{X}\cup\mathscr{Y}$, and
\begin{align}
\E_\C = \{ \X_i\Y_j \mid\;& i \in \lrB{p},  j \in \lrB{\ell}, \exists\ e \in \E_\c\lrA{\T} \notag\\
&\text{s.t. } e \text{ connects } \V\lrA{\X_i} \text{ and } \V\lrA{\Y_j} \}.
\end{align}
Each vertex of $\C$ corresponds to a component in $\mathscr{X}\cup\mathscr{Y}$. 
Since $\T$ is a spanning tree of $\G$, $\C$ is also a tree.
%We distinguish two cases depending on whether $\depth\lrA{\C}=1$ or $\depth\lrA{\C}\geq 2$.
Previously, \cite{MRM:17:TCS} showed that a 3-approximation algorithm exists for the \emph{mono-depth} case ($\depth\lrA{\C}=1$), and that 
a $7/2$-approximation algorithm exists in $O\lra{N^3}$ time for the \emph{poly-depth} case ($\depth\lrA{\C}\geq 2$) \cite[Lemmas 12--13]{MRM:17:TCS}, where the cubic complexity stems from a Hamiltonian path subroutine \cite[Lemma~3]{MRM:17:TCS}. We improve upon this result by addressing the two sub-cases---mono-depth and poly-depth 1-\acp{MST}---separately.

\subsubsection{Mono-Depth 1-\ac{MST} Case}

We present an improved $8/3$-approximation algorithm with running time $O\lra{N^2}$.

Since $\depth\lrA{\C}=1$, we have either $\Size{\mathscr{X}}=1$ or $\Size{\mathscr{Y}}=1$. We assume without loss of generality that $\mathscr{X}=\lrC{\X_1}$ and $\mathscr{Y}=\lrC{\Y_1, \ldots, \Y_\ell}$. In this case, $\T\lrB{\V_1}$ coincides with the component $\X_1$, and each component $\Y_i \in \mathscr{Y}$ is connected to $\X_1$ by exactly one cross edge $e_i = u_i v_i \in \E_\c\lrA{\T}$, where $u_i \in \V_1$ and $v_i \in \V\lrA{\Y_i}$. 
Let $\M \subseteq \V_1$ be the set of vertices in $\V_1$ that are incident to at least one cross edge, and refer to the vertices in $\M$ as \emph{marked vertices}. For each $m \in \M$, let $\mathscr{Y}\lrA{m} \subseteq \mathscr{Y}$ represent the set of $\V_2$-components connected to $m$. Since $\T$ is a 1-\ac{MST}, it contains at least two independent cross edges, implying that $\Size{\M} \ge 2$.
Let $\T_{\M}$ denote the minimal subtree of $\T\lrB{\V_1}$ spanning $\M$. 
Note that every leaf of $\T_{\M}$ is a marked vertex.
Our approach systematically pre-processes the $\V_2$-components into $\Size{\M}$ local tours anchored at marked vertices, and then performs pairwise merging on $\T_\M$. 
The algorithm is summarized in Algorithm~\ref{alg:depth1}, where the tilde symbol denotes dynamically updated objects: $\tilde{\T}_\M$ is the running tree initialized to $\T_\M$, while $\tilde{\E}_{\V_2}$ and $\tilde{\E}_{\T_\M}$ are the accumulated edge multisets initialized to $\emptyset$.

\begin{algorithm}[t!]
\caption{The algorithm for the mono-depth 1-\ac{MST} case}
\label{alg:depth1}
\begin{algorithmic}[1]
\REQUIRE The graph $\G=\lrA{\V,\E}$ and its spanning tree $\T$.
\ENSURE A Hamiltonian cycle in $\G$.
\STATE Let $\M = \lrC{u \in \V_1 \mid \exists\ v \in \V_2, uv \in \E_\c\lrA{\T}}$.
\STATE Let $\T_\M$ be the minimal subtree of $\T\lrB{\V_1}$ spanning $\M$.

\STATE \textbf{Pre-processing:} 
\STATE For each $m \in \M$, locally merge all $\V_2$-components in $\mathscr{Y}\lrA{m}$ to form a single tour $\pi_m$ anchored at $m$, covering all vertices in $\mathscr{Y}\lrA{m}$ and the vertex $m$.

\STATE \textbf{Iterative Pairwise Merging:} 
\STATE Initialize $\tilde{\T}_\M\coloneqq \T_\M$, $\tilde{\E}_{\T_\M} \coloneqq \emptyset$, and $\tilde{\E}_{\V_2} \coloneqq \emptyset$.
\STATE Set $k \coloneqq 0$ if $\Size{\M}\bmod 2=0$, and $k \coloneqq 3$ otherwise. \label{set_threshold}
\WHILE{the number of marked vertices in $\tilde{\T}_\M$ exceeds~$k$, and there exist two marked vertices $u, v \in \tilde{\T}_\M$ such that removing $\E\lra{P_{\tilde{\T}_\M}\lrA{u,v}} \cup \lrC{u,v}$ from $\tilde{\T}_\M$ still leaves all remaining marked vertices in the same connected component}\label{mergeconditions}
    \STATE Update $\tilde{\T}_\M$ by removing $\E\lra{P_{\tilde{\T}_\M}\lrA{u,v}}$ and $\lrC{u,v}$.
    \STATE Merge the tours $\pi_u$ and $\pi_v$ into a path $P_{uv}=\lrd{u,\ldots,v}$ with $\V\lrA{P_{uv}}=\V\lrA{\pi_u}\cup \V\lrA{\pi_v}$.
    \STATE\label{alg7l11} Accumulate edge budgets:\\ $\tilde{\E}_{\V_2} \coloneqq \tilde{\E}_{\V_2} \uplus \E\lrA{P_{uv}}$ and $\tilde{\E}_{\T_\M} \coloneqq \tilde{\E}_{\T_\M} \uplus \E\lra{P_{\tilde{\T}_\M}\lrA{u,v}}$.
\ENDWHILE

\IF{$\Size{\M}\bmod 2=0$}
    \STATE\label{Hgraph1} Form a multigraph $\HH$ with $\E\lrA{\HH} = \lra{\E\lrA{2\T\lrB{\V_1}} \setminus \tilde{\E}_{\T_\M}} \uplus \tilde{\E}_{\V_2}$.
\ELSE
    \STATE Let the last three unmerged tours be $\pi_1, \pi_2, \pi_3$ with anchors $x_1, x_2, x_3 \in \M$, respectively.
    
    \IF{$x_1, x_2, x_3$ are on a path in $\tilde{\T}_\M$ (without loss of generality in the order $x_1, x_2, x_3$)}\label{alg7l17}
        \STATE Merge $\pi_1$ and $\pi_2$ into a path $P_{x_1x_2}=\lrd{x_1,\ldots,x_2}$ with $\V\lrA{P_{x_1x_2}}=\V\lrA{\pi_1}\cup \V\lrA{\pi_2}$, and then merge $P_{x_1x_2}$ with $\pi_3$ into a path $P_{x_1x_3}=\lrd{x_1,\ldots,x_3}$ with $\V\lrA{P_{x_1x_3}}=\lrC{x_1,x_3}\cup \V_2\lrA{\pi_1}\cup \V_2\lrA{\pi_2} \cup \V_2\lrA{\pi_3}$.
        \STATE\label{Hgraph2} Form a multigraph $\HH$ with $\E\lrA{\HH} = \lrb{\E\lrA{2\T\lrB{\V_1}} \setminus \lrc{\tilde{\E}_{\T_\M} \uplus \E\lra{P_{\tilde{\T}_\M}\lra{x_1,x_3}}}} \uplus \lra{\tilde{\E}_{\V_2} \uplus \E\lrA{P_{x_1x_3}}}$.
    \ELSE
        \STATE\label{alg7l21} Let $a \in \V\lra{\tilde{\T}_\M}$ be their common branching vertex.
        \STATE Select $x \in \lrC{x_1, x_2, x_3}$ minimizing $\w\lra{P_{\tilde{\T}_\M}\lra{a, x}}$; without loss of generality $x = x_1$. \label{assumex1}
        \STATE Merge $\pi_1, \pi_2, \pi_3$ into $P_{x_2x_3}=\lrd{x_2,\ldots,x_3}$ with $\V\lrA{P_{x_2x_3}}=\lrC{x_2, x_3} \cup \V_2\lrA{\pi_1}\cup \V_2\lrA{\pi_2}\cup \V_2\lrA{\pi_3}$.
        \STATE\label{alg7l24} Form a multigraph $\HH$ with $\E\lrA{\HH} = \lrb{\E\lrA{2\T\lrB{\V_1}} \setminus \lrc{\tilde{\E}_{\T_\M} \uplus \E\lra{P_{\tilde{\T}_\M}\lrA{a, x_2}} \uplus \E\lra{P_{\tilde{\T}_\M}\lrA{a, x_3}} } } \uplus \lra{\tilde{\E}_{\V_2} \uplus \E\lrA{P_{x_2 x_3}} }$.
    \ENDIF
\ENDIF

\STATE Extract a Hamiltonian cycle $\sigma$ by shortcutting the Eulerian tour of $\HH$ such that it visits all vertices in each merged $\V_2$-path consecutively.
\RETURN $\sigma$.
\end{algorithmic}
\end{algorithm}

To bound the cost of the Hamiltonian cycle $\sigma$, we formalize the operations in Algorithm~\ref{alg:depth1}. 
The shortcutting relies on the triangle inequality: for any $x, y$ on the same side of the partition, the direct edge cost $\w\lrA{xy}$ is at most the cost of any path between them in $\G$, as mentioned in Section~\ref{sec:3.2}.

\paragraph*{Phase~1 (Pre-processing or Tour Formation)} 

\begin{figure}[t!]
\centering
\subfigure[$\Size{\mathscr{Y}\lrA{x_i}}=1$]{
\begin{tikzpicture}[scale=0.75, baseline=(xi.base),
    vertex/.style={circle, draw=black, fill=black, inner sep=1.2pt},
    subvtx/.style={circle, draw=black, fill=black, inner sep=1.0pt},
    crossedge/.style={thick, blue, ->, >=stealth},
    pathline/.style={thick, black, ->, >=stealth},
    msttree/.style={thick, gray!60, dashed}
]

% 统一 bounding box（关键对齐）
\path[use as bounding box] (-2.6,-1.7) rectangle (2.6,1.4);

% 节点
\node[vertex, label=right:{\small $x_i$}] (xi) at (0, 1.5) {};
\node[vertex, label=above left:{\small $y_i = y_i'$}] (yi) at (0, 0.2) {}; 

\node[subvtx] (v1) at (-0.8, -0.8) {};
\node[subvtx] (v2) at (0.8, -0.8) {};
\node[subvtx] (v3) at (0, -1.3) {};

% 组件边界
\draw[lightgray, thick] (0, -0.6) ellipse (1.2cm and 1.1cm);
% \node[text=gray] at (1.2, -1.6) {\small $\mathscr{Y}\lrA{x_i}$};

% \ac{MST}
\draw[msttree] (xi) -- (yi);
% \node[text=gray] at (0.7, 1.1) {\scriptsize $e_i \in T$};
\draw[msttree] (yi) -- (v1);
\draw[msttree] (yi) -- (v2);
\draw[msttree] (yi) -- (v3);

% Tour
\draw[pathline] (yi) to[bend right=20] (v1);
\draw[pathline] (v1) to[bend right=20] (v3);
\draw[pathline] (v3) to[bend right=20] (v2);
\draw[pathline] (v2) to[bend right=20] (yi);

% cross edge
\draw[crossedge] (xi) to[bend right=25] (yi);
\draw[crossedge] (yi) to[bend right=25] (xi);

\end{tikzpicture}
}
\subfigure[$\Size{\mathscr{Y}\lrA{x_i}}=2$]{
\begin{tikzpicture}[scale=0.75, baseline=(xi.base),
    vertex/.style={circle, draw=black, fill=black, inner sep=1.2pt},
    subvtx/.style={circle, draw=black, fill=black, inner sep=1.0pt},
    crossedge/.style={thick, blue, ->, >=stealth},
    pathline/.style={thick, black, ->, >=stealth},
    msttree/.style={thick, gray!60, dashed}
]

% 统一 bounding box（关键对齐）
\path[use as bounding box] (-2.6,-1.7) rectangle (2.6,1.4);

% 节点
\node[vertex, label=right:{\small $x_i$}] (xi) at (0, 1.5) {};

% 左组件
\node[vertex, label=above left:{\small $y_i$}] (yi) at (-1.4, 0.2) {};
\node[subvtx] (v1) at (-2.0, -0.8) {};
\node[subvtx] (v2) at (-0.8, -1.3) {};

% 右组件
\node[vertex, label=above right:{\small $y_i'$}] (yip) at (1.4, 0.2) {};
\node[subvtx] (v3) at (0.8, -1.3) {};
\node[subvtx] (v4) at (2.0, -0.8) {};

% 组件边界
\draw[lightgray, thick] (-1.4, -0.6) ellipse (1.0cm and 1.1cm);
\draw[lightgray, thick] (1.4, -0.6) ellipse (1.0cm and 1.1cm);
% \node[text=gray] at (0, -0.4) {\small $\mathscr{Y}\lrA{x_i}$};

% \ac{MST}
\draw[msttree] (xi) -- (yi);
\draw[msttree] (xi) -- (yip);
\draw[msttree] (yi) -- (v1);
\draw[msttree] (yi) -- (v2);
\draw[msttree] (yip) -- (v3);
\draw[msttree] (yip) -- (v4);

% Tour
\draw[pathline] (yi) to[bend right=20] (v1);
\draw[pathline] (v1) to[bend right=20] (v2);
\draw[pathline] (v2) to[bend right=15] (v3);
\draw[pathline] (v3) to[bend right=20] (v4);
\draw[pathline] (v4) to[bend right=20] (yip);

% cross edge
\draw[crossedge] (xi) to[bend right=15] (yi);
\draw[crossedge] (yip) to[bend right=15] (xi);

\end{tikzpicture}
}
\caption{An illustration of Phase~1 (pre-processing). The thick dashed gray lines represent the edges in $\T$. Blue arrows indicate the cross edges, and black arrows represent the local $\V_2$-paths.}
\label{fig:preprocessing}
\end{figure}
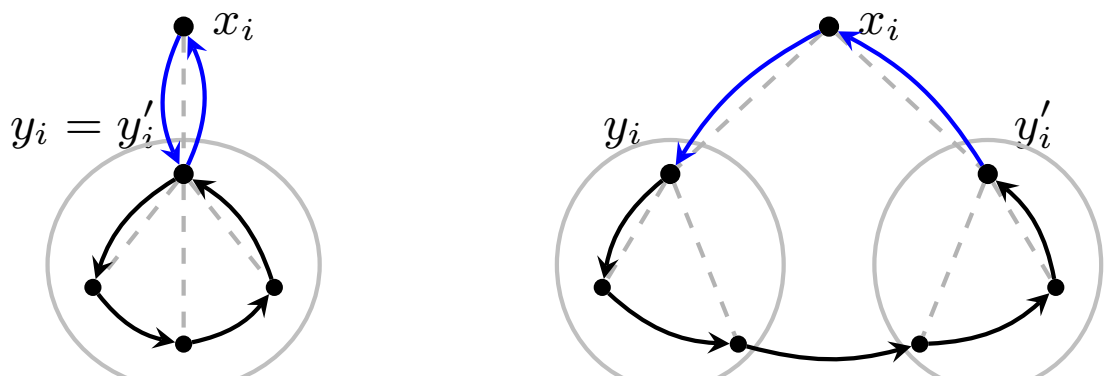

For each $x_i \in \M$, we consider the connected $\V_2$-components $\mathscr{Y}\lrA{x_i}$. By doubling the spanning trees $\T\lrB{\mathscr{Y}\lrA{x_i}}$ and the associated cross edges $\E_\c\lrA{x_i}$, we form an Eulerian subgraph. Shortcutting this subgraph entirely within $\V_2$ yields a local $\V_2$-path with two endpoints $y_{i}, y_{i}' \in \V_2$ connected to $x_i$, where $y_{i}=y_{i}'$ when $\Size{\mathscr{Y}\lrA{x_i}}=1$.
This $\V_2$-path forms a local tour $\pi_i = \lrd{x_i,y_i,\ldots, y_i',x_i}$ anchored at $x_i$. If $\Size{\mathscr{Y}\lrA{x_i}}>1$, this walk is in fact a cycle, as illustrated in Fig.~\ref{fig:preprocessing}.
The total cost of these walks is at most
\begin{align}\label{eqf1}
\sum_{x_i\in \M} \w\lrA{\pi_i} &\leq \sum_{x_i\in \M} 2\w\lrA{\T\lrB{\mathscr{Y}\lrA{x_i}},\E_\c\lrA{x_i}}
\nonumber \\
&= 2\w\lrA{\T} - 2\w\lrA{\T\lrB{\V_1}}.
\end{align}

\begin{figure}[t]
\centering
\begin{tikzpicture}[scale=0.75,
    vertex/.style={circle, draw=black, fill=black, inner sep=1.2pt},
    subvtx/.style={circle, draw=black, fill=black, inner sep=1.0pt},
    crossedge/.style={thick, blue, ->, >=stealth},
    pathline/.style={thick, black, ->, >=stealth},
    msttree/.style={thick, gray!60, dashed},
    jumpedge/.style={thick, red, ->, >=stealth}, % 新增：红色跳跃边
    dropedge/.style={thick, gray!60, dotted, ->, >=stealth} % 新增：被丢弃的边
]

% 统一 bounding box 保证居中
% \path[use as bounding box] (-4.5,-2.2) rectangle (4.5,2.6);

% ====================
% 1. 绘制 V1 和 V2 的区域背景
% ====================
% \filldraw[rounded corners=10pt, draw=gray, dashed, thick, fill=blue!5] (-4.3, 1.0) rectangle (4.3, 2.5);
% \node[text=gray, font=\large] at (-3.8, 1.7) {$\V_1$};

% \filldraw[rounded corners=10pt, draw=gray, dashed, thick, fill=red!5] (-4.3, -2.0) rectangle (4.3, 0.6);
% \node[text=gray, font=\large] at (-3.8, -0.7) {$\V_2$};

% ====================
% 2. 放置节点
% ====================
% V1 节点 (x_a 和 x_b) 及路径中间点
\node[vertex, label=above:{\small $x_a$}] (xa) at (-2.5, 1.3) {};
\node[vertex, label=above:{\small $x_b$}] (xb) at (2.5, 1.3) {};
\node[subvtx] (xm) at (0, 1.8) {}; % P_\T_\M 路径的中间示意节点

% 左侧 V2 组件节点 (\pi_a)
\node[vertex, label=above left:{\small $y_a'$}] (yap) at (-3.5, 0.2) {};
\node[vertex, label=above right:{\small $y_a$}] (ya) at (-1.5, 0.2) {};
\node[subvtx] (va1) at (-3.5, -1.0) {};
\node[subvtx] (va2) at (-1.5, -1.0) {};

% 右侧 V2 组件节点 (\pi_b)
\node[vertex, label=above left:{\small $y_b$}] (yb) at (1.5, 0.2) {};
\node[vertex, label=above right:{\small $y_b'$}] (ybp) at (3.5, 0.2) {};
\node[subvtx] (vb1) at (1.5, -1.0) {};
\node[subvtx] (vb2) at (3.5, -1.0) {};

% 标示 V2 组件的边界
\draw[lightgray, thick] (-2.5, -0.6) ellipse (1.6cm and 1.2cm);
\node[text=gray] at (-2.5, -1.5) {\small $\pi_a$};

\draw[lightgray, thick] (2.5, -0.6) ellipse (1.6cm and 1.2cm);
\node[text=gray] at (2.5, -1.5) {\small $\pi_b$};

% ====================
% 3. 绘制路径与连线
% ====================
% 树路径 P_{\tilde{\T}_\M}(x_a, x_b)
\draw[msttree] (xa) to (xm) to (xb);
\node[above, text=gray] at (0, 0.9) {\small $P_{\tilde{\T}_\M}\lrA{x_a, x_b}$};

% V2 内部的 Tour 路径 (黑色箭头)
\draw[pathline] (yap) to[bend right=20] (va1);
\draw[pathline] (va1) to[bend right=20] (va2);
\draw[pathline] (va2) to[bend right=20] (ya);

\draw[pathline] (yb) to[bend right=20] (vb1);
\draw[pathline] (vb1) to[bend right=20] (vb2);
\draw[pathline] (vb2) to[bend right=20] (ybp);

% 跨区域边 - 保留的部分 (蓝色粗箭头)
\draw[crossedge] (xa) to[bend right=15] (yap);
\draw[crossedge] (ybp) to[bend right=15] (xb);

% 跨区域边 - 被 Drop 的部分 (灰色虚线 + "drop" 标签)
\draw[dropedge] (ya) to[bend right=15] node[midway, right, text=gray] {} (xa);
\draw[dropedge] (xb) to[bend right=15] node[midway, left, text=gray] {} (yb);

% 合并引入的 V2-jump 边 (红色醒目箭头)
\draw[jumpedge] (ya) to[bend right=15] node[midway, below] {\small $\V_2$-jump} (yb);

\end{tikzpicture}
\caption{An illustration of Phase~2 (iterative pairwise merging). To merge tours $\pi_a$ and $\pi_b$, the cross edges $x_a y_a$ and $x_b y_b$ (gray dotted lines) are removed, and a $\V_2$-jump edge (red arrow) is introduced. This $\V_2$-jump connects them into a single path $P_{x_a x_b}$ from $x_a$ to $x_b$.}
\label{fig:merging}
\end{figure}
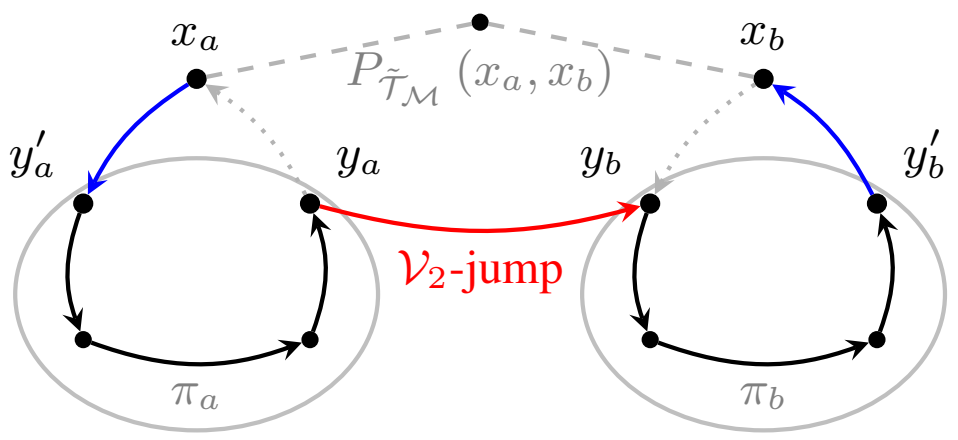

\paragraph*{Phase~2 (Iterative Pairwise Merging)} We define a merge operation between two tours $\pi_a$ and $\pi_b$, anchored at $x_a$ and $x_b$ respectively, where the tours connect to their anchors via the nodes $y_a, y_a'$ and $y_b, y_b'$ in $\V_2$ (note that $y_a=y_a'$ and/or $y_b=y_b'$ may hold).
To merge them into $P_{x_ax_b}=\lrd{x_a,\ldots,x_b}$, we drop the cross edges $x_ay_a$ and $x_by_b$, and introduce the $\V_2$-jump edge $y_ay_b$, 
as illustrated in Fig.~\ref{fig:merging}.
By the triangle inequality, this jump has cost at most
\begin{equation} \label{eqf2}
\w\lrA{y_ay_b} \le \w\lrA{y_ax_a,P_{\tilde{\T}_\M}\lrA{x_a, x_b}, x_by_b}.
\end{equation}
Compared to the previous cost $\w\lrA{\pi_a,\pi_b}$, this operation of constructing $P_{x_ax_b}$ additionally consumes exactly one copy of every edge in $\E\lra{P_{\tilde{\T}_\M}\lrA{x_a, x_b}}$. 
The total cost of these paths is $\w\lra{\tilde{\E}_{\V_2}}$ by line~\ref{alg7l11}. 
Hence, 
\begin{align}\label{eqf3}
\w\lra{\tilde{\E}_{\V_2}} &\leq \sum_{x_a,x_b: \, \text{paired}} \w\lrA{\pi_a,\pi_b, P_{\tilde{\T}_\M}\lrA{x_a, x_b}}
\nonumber \\
& = \w\lra{\tilde{\E}_{\T_\M}} +\sum_{x_a,x_b: \, \text{paired}} \w\lrA{\pi_a,\pi_b}.
\end{align}
The merge condition in line~\ref{mergeconditions} ensures that %the paths 
$P_{\tilde{\T}_\M}\lrA{x_a, x_b}$ are edge-disjoint across merge operations. Thus,  by line~\ref{alg7l11},
\begin{equation} \label{eqf4}
\w\lra{\tilde{\E}_{\T_\M}} \le \w\lrA{\T\lrB{\V_1}}.
\end{equation}
The termination of this iterative merging is classified into three cases, formalized in the following lemmas.

%We classify the termination of this iterative merging into three lemmas.

\begin{lemma}[Even Termination] \label{lem:perfect_pairing}
When $\Size{\M}\bmod 2=0$, Algorithm~\ref{alg:depth1} achieves an approximation ratio of $2$.
\end{lemma}
\begin{IEEEproof}
All local tours are merged in Phase~2 (iterative pairwise merging) by line~\ref{mergeconditions}, yielding
\begin{align}
\sum_{x_a,x_b: \, \text{paired}}\w\lrA{\pi_a,\pi_b} = \sum_{x_i\in \M} \w\lrA{\pi_i}.
\end{align}
Then, by \eqref{eqf1} and \eqref{eqf3}, we have
\begin{equation} \label{eqf5}
\w\lra{\tilde{\E}_{\V_2}}
\leq 2\w\lrA{\T} - 2\w\lrA{\T\lrB{\V_1}} + \w\lra{\tilde{\E}_{\T_\M}}.
\end{equation}
By line~\ref{Hgraph1}, 
\begin{align}
\E\lrA{\HH} = \lra{\E\lrA{2\T\lrB{\V_1}} \setminus \tilde{\E}_{\T_\M} } \uplus \tilde{\E}_{\V_2}.
\end{align}
Since the paired paths $P_{\tilde{\T}_\M}\lrA{x_a, x_b}$ in $\tilde{\T}_\M$ are edge-disjoint, by \eqref{eqf5}, we obtain
\begin{equation} \label{eqf6}
\w\lrA{\HH} = 2\w\lrA{\T\lrB{\V_1}} - \w\lra{\tilde{\E}_{\T_\M}} + \w\lra{\tilde{\E}_{\V_2}} \leq 2\w\lrA{\T}.
\end{equation}

To extract a Hamiltonian cycle $\sigma$, observe that $\HH$ is connected---as $\E\lrA{2\T\lrB{\V_1}} \setminus \tilde{\E}_{\T_\M}$ retains at least one full copy of $\T\lrB{\V_1}$---and Eulerian---as replacing any path $P_{\tilde{\T}_\M}\lrA{u,v}$ with the corresponding merged path $P_{uv}$ exactly preserves the parity of degrees at $u$ and $v$. 
We extract an Eulerian tour $\sigma_\HH$ of $\HH$, assuming without loss of generality that $\sigma_\HH$ starts and ends at a vertex in $\V_1$.
By definition, all merged paths $P_{x_ax_b}$ are vertex-disjoint and together cover all vertices in $\V_2$. Hence, each vertex in $\V_2$ has a degree of exactly $2$ in $\HH$ and appears exactly once in $\sigma_\HH$. Furthermore, each marked vertex $x_i \in \M$ is incident to at most one edge with an endpoint in $\V_2$ in $\HH$. Thus, in the tour $\sigma_\HH$, at most one appearance of $x_i$ is incident to a vertex in $\V_2$.
Consequently, any vertex appearing more than once in $\sigma_\HH$ must lie in $\V_1$. Moreover, except for at most one appearance (which connects to $\V_2$), all other appearances of this vertex are incident to two vertices in $\V_1$. Since the triangle inequality holds within $\V_1$, all such internal $\V_1$ appearances can be shortcut without increasing the total cost.

By shortcutting all redundant appearances in $\V_1$, we obtain a valid Hamiltonian cycle $\sigma$ where every vertex is visited exactly once. Therefore, by \eqref{eqf6}, we obtain
\begin{align}
\w\lrA{\sigma} \le \w\lrA{\sigma_\HH} = \w\lrA{\HH}\leq 2\w\lrA{\T} \le 2\w\lrA{\sigma^*},
\end{align}
which completes the proof.
\end{IEEEproof}

\begin{lemma}[Path Termination] \label{lem:3_path}
If the last three tours are on a path in $\tilde{\T}_\M$, Algorithm~\ref{alg:depth1} achieves an~approximation~ratio~of~$2$.
\end{lemma}
\begin{IEEEproof}
By lines~\ref{alg7l17} and~\ref{Hgraph2}, the algorithm merges $\pi_1$ and $\pi_2$ into $P_{x_1x_2}$, and subsequently with $\pi_3$ into $P_{x_1x_3}$. 
The sum of the two $\V_2$-jumps is bounded by the intermediate tree paths $P_{\tilde{\T}_\M}\lrA{x_1, x_2}$ and $P_{\tilde{\T}_\M}\lrA{x_2, x_3}$, which partition $P_{\tilde{\T}_\M}\lrA{x_1, x_3}$. Thus, we obtain
\begin{equation} \label{eq2f1}
\w\lra{P_{x_1x_3}} \leq \w\lra{\pi_1,\pi_2,\pi_3,P_{\tilde{\T}_\M}\lra{x_1,x_3}}.
\end{equation}
In this case, we also have
\begin{align} \label{eq:L8}
\w\lrA{\pi_1,\pi_2,\pi_3}+\sum_{x_a,x_b: \, \text{paired}}\w\lrA{\pi_a,\pi_b} 
= \sum_{x_i\in \M} \w\lrA{\pi_i}.
\end{align}
By \eqref{eqf1}, \eqref{eqf3}, and \eqref{eq2f1}, and the construction of $\HH$ in line~\ref{Hgraph2},  
\begin{align}
\w\lrA{\HH} &= 2\w\lra{\T\lrB{\V_1}} \! - \!\w\lra{\tilde{\E}_{\T_\M},\!P_{\tilde{\T}_\M}\lrA{x_1,x_3}}
\!+\! \w\lra{\tilde{\E}_{\V_2},\! P_{x_1x_3}}
\nonumber \\
&\le 2\w\lrA{\T\lrB{\V_1}} - \w\lra{\tilde{\E}_{\T_\M}} + \w\lra{\tilde{\E}_{\V_2},\pi_1,\pi_2,\pi_3}
\nonumber \\
&\le 2\w\lrA{\T}.
\end{align}
Since $\HH$ remains Eulerian and connected on $\V$, the Hamiltonian cycle $\sigma$ can be extracted exactly as in Lemma~\ref{lem:perfect_pairing}, yielding $\w\lrA{\sigma} \le 2\w\lrA{\sigma^*}$.
\end{IEEEproof}

\begin{lemma}[Star Termination] \label{lem:3_star}
If the pairwise merging stalls at a 3-star bottleneck, Algorithm~\ref{alg:depth1} achieves an approximation ratio of $8/3$.
\end{lemma}
\begin{IEEEproof}
Let $x_1, x_2, x_3$ be the anchors of the last three tours $\pi_1, \pi_2, \pi_3$ branching from a vertex $a$.
Assume that $\pi_i=\lrd{x_i, y_i, \ldots, y'_i, x_i}$, and that $x_1$ minimizes $\w\lra{P_{\tilde{\T}_\M}\lrA{a, x}}$ over all $x \in \lrC{x_1, x_2, x_3}$ by line~\ref{assumex1}.
By lines~\ref{alg7l21}--\ref{alg7l24}, the algorithm merges them into a path $P_{x_2 x_3}=\lrd{x_2,y'_2,\ldots,y'_3,x_3}$. 
This merging requires two $\V_2$-jumps: one from $\pi_2$ to $\pi_1$ via $y_2y_1$, and one from $\pi_1$ to $\pi_3$ via $y_1'y_3$, as illustrated in Fig.~\ref{fig:3star_merge}.

\begin{figure}[t]
\centering
\begin{tikzpicture}[scale=0.75,
    vertex/.style={circle, draw=black, fill=black, inner sep=1.2pt},
    subvtx/.style={circle, draw=black, fill=black, inner sep=1.0pt},
    crossedge/.style={thick, blue, ->, >=stealth},
    pathline/.style={thick, black, ->, >=stealth},
    msttree/.style={thick, gray!60, dashed},
    jumpedge/.style={thick, red, ->, >=stealth}, 
    dropedge/.style={thick, gray!60, dotted, ->, >=stealth}
]

% ====================
% 1. 顶部汇合点 a 及 V1 路径
% ====================
\node[vertex, label=left:{\small $a$}] (a) at (0, 3) {};

% x 坐标设定：x2 在左，x1 在中，x3 在右
\node[vertex, label=left:{\small $x_2$}] (x2) at (-4, 1.5) {};
\node[vertex, label=left:{\small $x_1$}] (x1) at (0, 1.5) {};
\node[vertex, label=right:{\small $x_3$}] (x3) at (4, 1.5) {};

% 路径上的中间点 m
\node[subvtx] (m2) at ($(a)!0.5!(x2)$) {};
\node[subvtx] (m1) at ($(a)!0.5!(x1)$) {};
\node[subvtx] (m3) at ($(a)!0.5!(x3)$) {};

% 绘制 P_{\T_\M} 路径 (灰色虚线)
\draw[msttree] (a) -- (m2) -- (x2)
    node[midway, above left, text=gray, font=\scriptsize]
    {$P_{\tilde{\T}_\M}(a,x_2)$};

\draw[msttree] (a) -- (m1) -- (x1)
    node[midway, left, text=gray, font=\scriptsize]
    {$P_{\tilde{\T}_\M}(a,x_1)$};

\draw[msttree] (a) -- (m3) -- (x3)
    node[midway, above right, text=gray, font=\scriptsize]
    {$P_{\tilde{\T}_\M}(a,x_3)$};
% \node[text=gray, font=\scriptsize] at (-2.3, 3.1) {$P_{\tilde{\T}_\M}(a,x_2)$};
% \node[text=gray, font=\scriptsize] at (0, 3.1) {$P_{\tilde{\T}_\M}(a,x_1)$};
% \node[text=gray, font=\scriptsize] at (2.3, 3.1) {$P_{\tilde{\T}_\M}(a,x_3)$};

% ====================
% 2. 三个 Tour 组件 (C2, C1, C3)
% ====================

% --- C2 (Left) ---
\node[vertex, label=above left:{\small $y_2'$}] (y2p) at (-5, 0.2) {};
\node[vertex, label=above right:{\small $y_2$}] (y2) at (-3, 0.2) {};
\node[subvtx] (v2a) at (-4, -0.8) {};
\draw[lightgray, thick] (-4, -0.33) ellipse (1.4cm and 1cm);
\node[text=gray] at (-4, -1.1) {\small $\pi_2$};
\draw[pathline] (y2p) to[bend right=30] (v2a);
\draw[pathline] (v2a) to[bend right=30] (y2);

% --- C1 (Middle) ---
\node[vertex, label=above left:{\small $y_1$}] (y1) at (-1, 0.2) {};
\node[vertex, label=above right:{\small $y_1'$}] (y1p) at (1, 0.2) {};
\node[subvtx] (v1a) at (0, -0.8) {};
\draw[lightgray, thick] (0, -0.33) ellipse (1.4cm and 1cm);
\node[text=gray] at (0, -1.1) {\small $\pi_1$};
\draw[pathline] (y1) to[bend right=30] (v1a);
\draw[pathline] (v1a) to[bend right=30] (y1p);

% --- C3 (Right) ---
\node[vertex, label=above left:{\small $y_3$}] (y3) at (3, 0.2) {};
\node[vertex, label=above right:{\small $y_3'$}] (y3p) at (5, 0.2) {};
\node[subvtx] (v3a) at (4, -0.8) {};
\draw[lightgray, thick] (4, -0.33) ellipse (1.4cm and 1cm);
\node[text=gray] at (4, -1.1) {\small $\pi_3$};
\draw[pathline] (y3) to[bend right=30] (v3a);
\draw[pathline] (v3a) to[bend right=30] (y3p);

% ====================
% 3. 合并操作：Drop, Keep, and Jumps
% ====================

% Keep (端点连边)
\draw[crossedge] (x2) to[bend right=15] (y2p);
\draw[crossedge] (y3p) to[bend right=15] (x3);

% Drop (被丢弃的 cross edges)
\draw[dropedge] (y2) to[bend right=15] (x2);
\draw[dropedge] (x1) to[bend right=15] (y1);
\draw[dropedge] (y1p) to[bend right=15] (x1);
\draw[dropedge] (x3) to[bend right=15] (y3);

% V2-Jumps (红色合并边)
\draw[jumpedge] (y2) to[bend left=20] node[midway, below] {\scriptsize $\V_2\text{-jump}$} (y1);
\draw[jumpedge] (y1p) to[bend left=20] node[midway, below] {\scriptsize $\V_2\text{-jump}$} (y3);

\end{tikzpicture}
\caption{An illustration of merging three tours in the 3-star bottleneck case. To form the path $P_{x_2 x_3}$, we remove four cross edges (dotted) and introduce two $\V_2$-jump edges (red).}
\label{fig:3star_merge}
\end{figure}
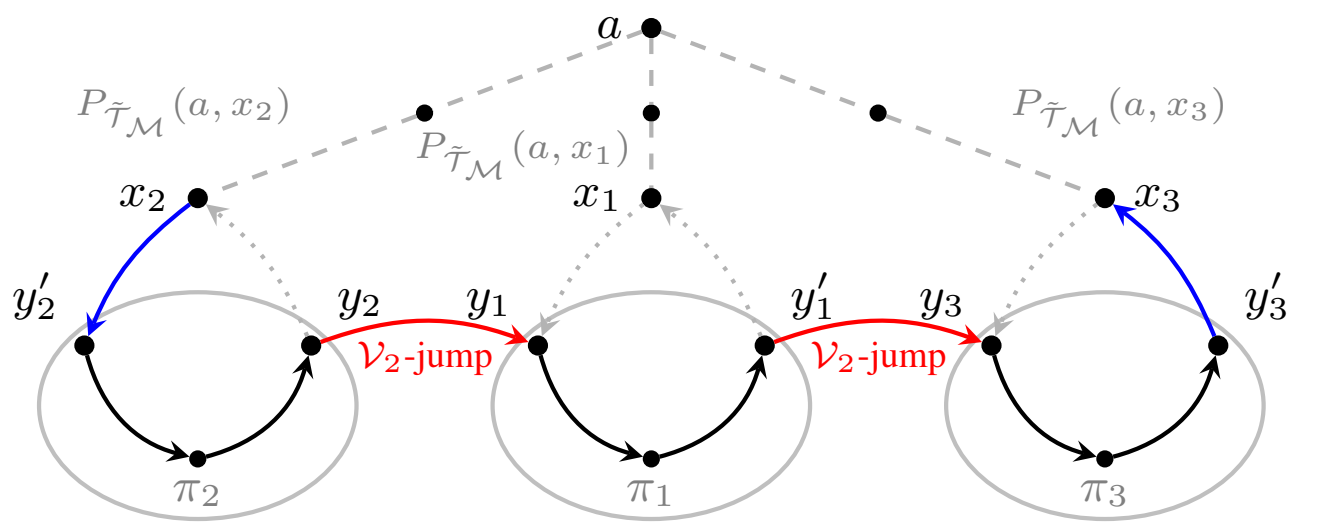

By the triangle inequality, the costs of these jumps are bounded by their corresponding paths through $a$ and $x_1$. In particular, we have
\begin{align}
\w\lrA{y_2y_1}&\le \w\lrA{y_2x_2, P_{\tilde{\T}_\M}\lra{x_2, a},P_{\tilde{\T}_\M}\lra{a, x_1}, x_1y_1}; \\
\w\lrA{y_1'y_3} &\le \w\lrA{y_1'x_1,P_{\tilde{\T}_\M}\lra{x_1, a},P_{\tilde{\T}_\M}\lra{a, x_3},x_3y_3}.
\end{align}
By summing these bounds, we obtain
\begin{align} \label{eq3f1}
\w\lrA{P_{x_2 x_3}} &\le \w\lra{\pi_1,\pi_2,\pi_3,P_{\tilde{\T}_\M}\lrA{a, x_2},P_{\tilde{\T}_\M}\lrA{a, x_3}} 
\nonumber \\
&\quad + 2\w\lra{P_{\tilde{\T}_\M}\lrA{a, x_1}}.
\end{align}
By \eqref{eqf1}, \eqref{eqf3}, \eqref{eq:L8}, and \eqref{eq3f1}, and the construction of $\HH$ in line~\ref{alg7l24}, we have
\begin{align}
\w\lrA{\HH} &= 2\w\lrA{\T\lrB{\V_1}} - \w\lra{\tilde{\E}_{\T_\M}, P_{\tilde{\T}_\M}\lrA{a, x_2},P_{\tilde{\T}_\M}\lrA{a, x_3}}
\nonumber \\
&\quad + \w\lra{\tilde{\E}_{\V_2},P_{x_2 x_3}}
\nonumber \\
&\le 2\w\lrA{\T\lrB{\V_1}} - \w\lra{\tilde{\E}_{\T_\M}} + \w\lra{\tilde{\E}_{\V_2},\pi_1,\pi_2,\pi_3}
\nonumber \\
&\quad + 2\w\lra{P_{\tilde{\T}_\M}\lrA{a, x_1}}
\nonumber \\
&\le 2\w\lra{\T,P_{\tilde{\T}_\M}\lrA{a, x_1}}.
\end{align}
Since there are at least 3 edge-disjoint branches meeting at $a$ in $\tilde{\T}_\M$, an averaging argument guarantees that selecting the branch $x_1$ minimizing $\w\lra{P_{\tilde{\T}_\M}\lrA{a, x}}$ yields 
\begin{align}
2\w\lra{P_{\tilde{\T}_\M}\lrA{a, x_1}} \le \frac{2}{3} \w\lra{\tilde{\T}_\M} \le \frac{2}{3} \w\lrA{\sigma^*}.
\end{align}
Thus, 
%\begin{align}
$\w\lrA{\HH} \le \frac{8}{3}\w\lrA{\sigma^*}$.
%\end{align}

For the topological validity, removing the paths $P_{\tilde{\T}_\M}\lrA{a, x_2}$ and $P_{\tilde{\T}_\M}\lrA{a, x_3}$ from $2\T\lrB{\V_1}$ reduces the degree of $a$ by $2$ (preserving its even parity) and reduces the degrees of $x_2$ and $x_3$ by $1$. The inclusion of the new path $P_{x_2x_3}$ adds degree 1 to both $x_2$ and $x_3$, restoring them to even parity. Furthermore, since $P_{\tilde{\T}_\M}\lrA{a, x_1}$ is not subtracted from $2\T\lrB{\V_1}$ and $x_1$ is not an endpoint of the new path $P_{x_2 x_3}$, $x_1$ trivially retains its even parity.
Thus, $\HH$ remains entirely Eulerian. 
The Hamiltonian cycle $\sigma$ can also be extracted exactly as in Lemma~\ref{lem:perfect_pairing} with 
\begin{align}
\w\lrA{\sigma} \le \frac{8}{3}\w\lrA{\sigma^*},
\end{align}
which completes the proof.
\end{IEEEproof}

%\begin{lemma}[{\fontsize{9.5pt}{10pt}\selectfont Mono-Depth 1-\ac{MST}}]\label{res4}
%\scalebox{0.95}[1]{When $\T$ is a 1-\ac{MST} whose} component tree has depth $1$, 
\begin{lemma}[Mono-Depth 1-\ac{MST}]\label{res4}
When $\T$ is a 1-\ac{MST} with $\depth\lrA{\C} = 1$,
there exists an $8/3$-approximation algorithm with running time $O\lra{N^2}$ for the \ac{BTSP}.
\end{lemma}
\begin{IEEEproof}
By Lemmas~\ref{lem:perfect_pairing}--\ref{lem:3_star}, Algorithm~\ref{alg:depth1} terminates in one of three cases: even termination, path termination, or 3-star bottleneck. 
%reduces the problem to either a perfect parity state or a 3-star bottleneck. 
In the worst-case (the 3-star), the approximation ratio is $8/3$. We now justify the algorithmic liveness and the $O\lra{N^2}$ time complexity.

To guarantee the liveness of the \texttt{while} loop in line~\ref{mergeconditions}, we show that as long as the number of remaining marked vertices $\Size{\M_r}$ exceeds $k$, a valid pair $\lrA{u, v}$ can be found. Let $\T_{\M_r}$ be the minimal subtree of $\T\lrB{\V_1}$ that spans the current set of marked vertices $\M_r$. By the property of minimal spanning trees, every leaf of $\T_{\M_r}$ must be a vertex in $\M_r$. We select such a leaf $u$ and let $v$ be its closest neighbor in $\M_r$ along the structure of $\T_{\M_r}$. The path $P_{\tilde{\T}_\M}\lrA{u, v}$ contains no other marked vertices internally. Removing the edges of this path and the vertices $u, v$ is equivalent to pruning a pendant branch of the spanning structure $\T_{\M_r}$. Since $u$ is a leaf and no other marked vertices reside on the path to $v$, this operation cannot disconnect the remaining vertices in $\M_r \setminus \lrC{u, v}$. Thus, connectivity is preserved, and the loop can proceed until $\Size{\M_r} \le k$.

In terms of time complexity, the algorithm is dominated by the pairwise merging phase. Phase~1 (pre-processing) requires a standard depth-first-search to identify components and form local tours, taking $O\lrA{N}$ time. In Phase~2 (iterative pairwise merging), each iteration of the \texttt{while} loop involves finding a pair $\lrA{u, v}$ and updating the tree. By maintaining the current spanning structure and a list of its leaves, identifying a pair takes $O\lrA{N}$ time. With at most $N/2$ iterations, the total time for merging is $O\lra{N^2}$. Finally, the construction of the Eulerian multigraph $\HH$, the extraction of the Eulerian tour, and the metric shortcutting are all standard procedures running in $O\lrA{N}$ time~\cite{CLRS:2022:book}. Therefore, the overall complexity is $O\lra{N^2}$.
\end{IEEEproof}

% By Lemmas~\ref{lem:perfect_pairing}--\ref{lem:3_star}, any tree $\T_\M$ can be iteratively reduced to either a perfect parity state or an odd-parity 3-star bottleneck. 
% Taking the worst-case bound, the approximation ratio is at most $8/3$. The tour formation, multiset edge replacement, and shortcutting can be fully implemented via DFS in $O\lra{N^2}$ time.

\subsubsection{Poly-Depth 1-\ac{MST} Case}

We present an improved $4$-approximation algorithm in $O\lra{N^2}$ time.

\begin{lemma}[Poly-Depth 1-\ac{MST}] \label{res5}
When $\T$ is a 1-\ac{MST} with $\depth\lrA{\C}\geq 2$, there exists a $4$-approximation algorithm with running time $O\lra{N^2}$ for the \ac{BTSP}.
\end{lemma}

\begin{IEEEproof}
The $7/2$-approximation algorithm of \cite[Lemmas~12 and 13]{MRM:17:TCS} runs in $O\lra{N^3}$ time. The cubic complexity stems from the following Hamiltonian path subroutine \cite[Lemma~3]{MRM:17:TCS}:
given a metric graph $\HH=\lrA{\V_\HH,\E_\HH}$ with edge-cost function $w_\HH$ and any two distinct vertices $x,y\in \V_\HH$, an $O\lra{\Size{\V_\HH}^3}$-time procedure computes a Hamiltonian path $P\lrA{x,y}$ such that $w_\HH\lrA{P\lrA{x,y}}\leq 1.5 \, w_\HH\lrA{\sigma^*_\HH}$, where $\sigma^*_\HH$ is a minimum-cost Hamiltonian cycle in $\HH$.

We replace this subroutine with a faster alternative that trades a tighter approximation ratio for quadratic running time. 
Specifically, we first compute an \ac{MST} $\T^*_\HH$ of $\HH$, and then apply the algorithm in Lemma~\ref{path} to obtain a Hamiltonian path between $x$ and $y$ in $\HH$ with cost at most $2w_\HH\lrA{\T^*_\HH}$. Since $w_\HH\lrA{\T^*_\HH}\leq w_\HH\lrA{\sigma^*_\HH}$, the resulting path satisfies 
\begin{align}
w_\HH\lrA{P\lrA{x,y}}\leq 2 \, w_\HH\lrA{\sigma^*_\HH}
\end{align}
and is computed in $O\lra{\Size{\V_\HH}^2}$ time. Substituting this $O\lra{\Size{\V_\HH}^2}$-time subroutine into the algorithm of \cite{MRM:17:TCS} inflates the approximation ratio from $7/2$ to $4$ while reducing the overall running time from $O\lra{N^3}$ to $O\lra{N^2}$, which completes the proof.
\end{IEEEproof}

\begin{theorem}[$4$-Approximation in Quadratic Time] \label{thm:2}
For the \ac{BTSP}, there exists a $4$-approximation algorithm with running time $O\lra{N^2}$.
\end{theorem}
\begin{IEEEproof}
The result follows from Lemmas~\ref{res1}--\ref{res5}.
\end{IEEEproof}

\begin{remark}[Breaking the Cubic-Time Barrier]
Theorem~\ref{thm:2} reduces the running time from $O\lra{N^3}$ to $O\lra{N^2}$---a full polynomial factor improvement---breaking the cubic-time barrier for the \ac{BTSP} while maintaining the approximation ratio at a constant value. This complexity is achieved by replacing the previous cubic-time matching subroutines with our $O\lra{N^2}$-time Hamiltonian path subroutine, complemented by structural insights into \ac{MST} topology. Beyond the overall $4$-approximation guarantee, we obtain strictly tighter subcase ratios of $5/2$, $2$, and $8/3$ for mono-cross 0-\acp{MST}, poly-cross 0-\acp{MST}, and mono-depth 1-\acp{MST}, respectively. The $8/3$-approximation for mono-depth 1-\acp{MST} merits particular emphasis, as it is achieved through a tailored iterative pairwise merging strategy that systematically resolves topological bottlenecks arising in three-tour merges. We expect that the refined cost-control arguments developed here extend to further algorithmic refinements, facilitating the design of $O\lra{N^2}$-time algorithms that break the $7/2$-approximation barrier.
\end{remark}

\section{Conclusion}
\label{sec:6}

In this paper, we have significantly advanced the state of the art for the \ac{BTSP}, a natural generalization of the metric \ac{TSP} that models scenarios where the triangle inequality is partially violated across different vertex partitions. We have designed two new approximation algorithms that offer distinct, highly favorable trade-offs between approximation guarantees and computational efficiency.
First, we have presented a new algorithmic framework that yields an approximation ratio of $2\alpha + \varepsilon$ in polynomial time for any constant $\varepsilon > 0$. By establishing a structural relationship between cross edges of the optimal Hamiltonian cycle and the \ac{TSPPK}, this algorithm substantially improves upon the previous $7/2$ ratio. 
Second, we have developed a highly efficient $O\lra{N^2}$-time algorithm that guarantees a $4$-approximation. By systematically classifying the topology of the \ac{MST} and replacing cubic-time matching bottlenecks with an efficient $O\lra{N^2}$ Hamiltonian path subroutine, we have achieved both faster execution and substantially tighter subcase ratios (down to $5/2$, $2$, and $8/3$) for distinct \ac{MST} topologies. These results open several promising directions for future research. One interesting direction is to generalize the structural insights from our first algorithm to \ac{BTSP} instances where the vertex set is partitioned into three or more subsets. Furthermore, the edge-budgeting and iterative pairwise merging techniques introduced in our quadratic-time algorithm provide a solid analytical foundation. 
Together, these contributions enrich both the algorithmic theory and the practical applicability of the \ac{BTSP} in networked systems.

\section*{Acknowledgment}
AI tools were used only for language polishing, grammatical refinement, and figure generation. All algorithmic and theoretical results were developed independently by the authors without using AI tools.

\balance

% Generated by IEEEtran.bst, version: 1.14 (2015/08/26)


\begin{thebibliography}{10}
\providecommand{\url}[1]{#1}
\csname url@samestyle\endcsname
\providecommand{\newblock}{\relax}
\providecommand{\bibinfo}[2]{#2}
\providecommand{\BIBentrySTDinterwordspacing}{\spaceskip=0pt\relax}
\providecommand{\BIBentryALTinterwordstretchfactor}{4}
\providecommand{\BIBentryALTinterwordspacing}{\spaceskip=\fontdimen2\font plus
\BIBentryALTinterwordstretchfactor\fontdimen3\font minus
  \fontdimen4\font\relax}
\providecommand{\BIBforeignlanguage}[2]{{%
\expandafter\ifx\csname l@#1\endcsname\relax
\typeout{** WARNING: IEEEtran.bst: No hyphenation pattern has been}%
\typeout{** loaded for the language `#1'. Using the pattern for}%
\typeout{** the default language instead.}%
\else
\language=\csname l@#1\endcsname
\fi
#2}}
\providecommand{\BIBdecl}{\relax}
\BIBdecl

\bibitem{TV:2024:book}
V.~Traub and J.~Vygen, \emph{Approximation {A}lgorithms for {T}raveling
  {S}alesman {P}roblems}.\hskip 1em plus 0.5em minus 0.4em\relax Cambridge
  University Press, 2024.

\bibitem{ZKPZZS:26:IEEE_O_CSTO}
J.~Zhao, U.~Khalid, S.~N. Paing, C.~Zhang, Z.-H. Zhan, and H.~Shin,
  ``Approximation algorithms for vehicle routing problems: {A} survey,''
  \emph{{IEEE} Commun. Surveys Tuts.}, vol.~28, pp. 7391--7440, 2026.

\bibitem{TLLY:25:IEEE_J_EVC}
H.~Tong, M.~Li, J.~Liu, and X.~Yao, ``How do dynamic events change the fitness
  landscape of traveling salesman problems?'' \emph{{IEEE} Trans. Evol.
  Comput.}, vol.~29, no.~5, pp. 1463--1474, 2025.

\bibitem{GMZTCD:25:IEEE_J_EVC}
T.~Guo, Y.~Mei, M.~Zhang, K.~Tang, K.~Cai, and W.~Du, ``Enhanced evolution of
  parallel algorithm portfolio for vehicle routing problem via transfer
  optimization,'' \emph{{IEEE} Trans. Evol. Comput.}, {E}arly Access.

\bibitem{BMMD:24:IEEE_J_ITS}
A.~Bogyrbayeva, M.~Meraliyev, T.~Mustakhov, and B.~Dauletbayev, ``Machine
  learning to solve vehicle routing problems: {A} survey,'' \emph{{IEEE} Trans.
  Intell. Transp. Syst.}, vol.~25, no.~6, pp. 4754--4772, 2024.

\bibitem{LIY:24:IEEE_J_ITS}
Z.~Lyu, M.~Z. Islam, and A.~J. Yu, ``A scalable and adaptable supervised
  learning approach for solving the traveling salesman problems,'' \emph{{IEEE}
  Trans. Intell. Transp. Syst.}, vol.~25, no.~11, pp. 17\,092--17\,104, 2024.

\bibitem{LXHZYI:23:IEEE_J_EVC}
Y.~Liu, L.~Xu, Y.~Han, X.~Zeng, G.~G. Yen, and H.~Ishibuchi, ``Evolutionary
  multimodal multiobjective optimization for traveling salesman problems,''
  \emph{{IEEE} Trans. Evol. Comput.}, vol.~28, no.~2, pp. 516--530, 2024.

\bibitem{JH:22:IEEE_J_EVC}
S.-J. Jian and S.-Y. Hsieh, ``A niching regression adaptive memetic algorithm
  for multimodal optimization of the {E}uclidean traveling salesman problem,''
  \emph{{IEEE} Trans. Evol. Comput.}, vol.~27, no.~5, pp. 1413--1426, 2023.

\bibitem{ZZCSTZWD:2022:IEEE_J_ITS}
R.~Zhang, C.~Zhang, Z.~Cao, W.~Song, P.~S. Tan, J.~Zhang, B.~Wen, and
  J.~Dauwels, ``Learning to solve multiple-{TSP} with time window and
  rejections via deep reinforcement learning,'' \emph{{IEEE} Trans. Intell.
  Transp. Syst.}, vol.~24, no.~1, pp. 1325--1336, 2022.

\bibitem{XLZ:21:IEEE_J_ITS}
X.~Xu, J.~Li, and M.~Zhou, ``Bi-objective colored traveling salesman
  problems,'' \emph{{IEEE} Trans. Intell. Transp. Syst.}, vol.~23, no.~7, pp.
  6326--6336, 2021.

\bibitem{MLZD:21:IEEE_J_ITS}
X.~Meng, J.~Li, M.~Zhou, and X.~Dai, ``A dynamic colored traveling salesman
  problem with varying edge weights,'' \emph{{IEEE} Trans. Intell. Transp.
  Syst.}, vol.~23, no.~8, pp. 13\,549--13\,558, 2021.

\bibitem{YHHW:19:IEEE_J_EVC}
X.~Yan, H.~Huang, Z.~Hao, and J.~Wang, ``A graph-based fuzzy evolutionary
  algorithm for solving two-echelon vehicle routing problems,'' \emph{{IEEE}
  Trans. Evol. Comput.}, vol.~24, no.~1, pp. 129--141, 2019.

\bibitem{WCLY:16:IEEE_J_ITS}
X.~Wang, T.-M. Choi, H.~Liu, and X.~Yue, ``Novel ant colony optimization
  methods for simplifying solution construction in vehicle routing problems,''
  \emph{{IEEE} Trans. Intell. Transp. Syst.}, vol.~17, no.~11, pp. 3132--3141,
  2016.

\bibitem{ABC:2006:book}
D.~L. Applegate, R.~E. Bixby, V.~Chv{\'a}tal, and W.~J. Cook, \emph{The
  {T}raveling {S}alesman {P}roblem: {A} {C}omputational {S}tudy}.\hskip 1em
  plus 0.5em minus 0.4em\relax Princeton University Press, 2006.

\bibitem{ZVCZ:26:IEEE_J_ITS}
N.~Zhang, F.~Viadero-Monasterio, J.~Chen, and H.~Zhang, ``On-line distributed
  model predictive scheduling for multi-vehicle routing problems with
  lane-change and platoon maneuvers,'' \emph{{IEEE} Trans. Intell. Transp.
  Syst.}, vol.~27, no.~5, pp. 5988--6001, 2026.

\bibitem{JZWMCL:25:IEEE_J_ITS}
Y.-H. Jia, Q.~Zheng, Y.~Wang, Y.~Mei, W.-N. Chen, and Z.~Lin, ``A neural solver
  with traversal-based feature representation and adjacent attention for
  capacitated arc routing problem,'' \emph{{IEEE} Trans. Intell. Transp.
  Syst.}, vol.~26, no.~12, pp. 22\,329--22\,343, 2025.

\bibitem{ZLZBYMZ:25:IEEE_J_ITS}
B.~Zuo, W.~Li, J.~Zhao, T.~Bai, L.~Yang, Z.~Ma, and Y.~Zhu, ``{CMIP}:
  {C}ombining constructive model with improvement policy for large-scale
  min-max multiple traveling salesman problem,'' \emph{{IEEE} Trans. Intell.
  Transp. Syst.}, vol.~27, no.~1, pp. 1550--1564, 2025.

\bibitem{CLXLEC:24:IEEE_J_ITS}
H.~Cha, C.~Lee, C.~Xie, Q.-C. Lu, J.~Eun, and T.~Cheong, ``An exact {A}*-based
  tree search algorithm for {TSP} with sequence-and-load dependent risk,''
  \emph{{IEEE} Trans. Intell. Transp. Syst.}, vol.~25, no.~9, pp.
  10\,817--10\,834, 2024.

\bibitem{LDNXW:24:IEEE_J_ITS}
J.~Li, B.~T. Dai, Y.~Niu, J.~Xiao, and Y.~Wu, ``Multi-type attention for
  solving multi-depot vehicle routing problems,'' \emph{{IEEE} Trans. Intell.
  Transp. Syst.}, vol.~25, no.~11, pp. 17\,831--17\,840, 2024.

\bibitem{LZC:23:IEEE_J_ITS}
Z.~Ling, Y.~Zhang, and X.~Chen, ``A deep reinforcement learning based real-time
  solution policy for the traveling salesman problem,'' \emph{{IEEE} Trans.
  Intell. Transp. Syst.}, vol.~24, no.~6, pp. 5871--5882, 2023.

\bibitem{KJKJJ:23:IEEE_J_ITS}
G.~Kim, H.~Jin, M.~Kim, K.~Jang, and I.~G. Jang, ``Loop-wise route
  representation and its optimization formulation for symmetric traveling
  salesman problems,'' \emph{{IEEE} Trans. Intell. Transp. Syst.}, vol.~24,
  no.~9, pp. 9490--9500, 2023.

\bibitem{JMZ:22:IEEE_J_EVC}
Y.-H. Jia, Y.~Mei, and M.~Zhang, ``Confidence-based ant colony optimization for
  capacitated electric vehicle routing problem with comparison of different
  encoding schemes,'' \emph{{IEEE} Trans. Evol. Comput.}, vol.~26, no.~6, pp.
  1394--1408, 2022.

\bibitem{ZXFZ:22:IEEE_J_ITS}
Y.~Zhou, W.~Xu, Z.-H. Fu, and M.~Zhou, ``Multi-neighborhood simulated
  annealing-based iterated local search for colored traveling salesman
  problems,'' \emph{{IEEE} Trans. Intell. Transp. Syst.}, vol.~23, no.~9, pp.
  16\,072--16\,082, 2022.

\bibitem{LGN:21:IEEE_J_ITS}
B.~Lin, B.~Ghaddar, and J.~Nathwani, ``Deep reinforcement learning for the
  electric vehicle routing problem with time windows,'' \emph{{IEEE} Trans.
  Intell. Transp. Syst.}, vol.~23, no.~8, pp. 11\,528--11\,538, 2021.

\bibitem{ZMZZ:20:IEEE_J_ITS}
J.~Zhao, M.~Mao, X.~Zhao, and J.~Zou, ``A hybrid of deep reinforcement learning
  and local search for the vehicle routing problems,'' \emph{{IEEE} Trans.
  Intell. Transp. Syst.}, vol.~22, no.~11, pp. 7208--7218, 2020.

\bibitem{WST:20:IEEE_J_ITS}
X.~Wang, S.~Shao, and J.~Tang, ``Iterative local-search heuristic for weighted
  vehicle routing problem,'' \emph{{IEEE} Trans. Intell. Transp. Syst.},
  vol.~22, no.~6, pp. 3444--3454, 2020.

\bibitem{XLZ:20:IEEE_J_ITS}
X.~Xu, J.~Li, and M.~Zhou, ``Delaunay-triangulation-based variable neighborhood
  search to solve large-scale general colored traveling salesman problems,''
  \emph{{IEEE} Trans. Intell. Transp. Syst.}, vol.~22, no.~3, pp. 1583--1593,
  2020.

\bibitem{FHTGTTO:20:IEEE_J_ITS}
L.~Feng, Y.~Huang, I.~W. Tsang, A.~Gupta, K.~Tang, K.~C. Tan, and Y.-S. Ong,
  ``Towards faster vehicle routing by transferring knowledge from customer
  representation,'' \emph{{IEEE} Trans. Intell. Transp. Syst.}, vol.~23, no.~2,
  pp. 952--965, 2020.

\bibitem{ZQL:19:IEEE_J_ITS}
Z.~Zhang, H.~Qin, and Y.~Li, ``Multi-objective optimization for the vehicle
  routing problem with outsourcing and profit balancing,'' \emph{{IEEE} Trans.
  Intell. Transp. Syst.}, vol.~21, no.~5, pp. 1987--2001, 2019.

\bibitem{AA:18:IEEE_J_ITS}
S.~Ait~Bouziaren and B.~Aghezzaf, ``An improved augmented
  $\varepsilon$-constraint and branch-and-cut method to solve the {TSP} with
  profits,'' \emph{{IEEE} Trans. Intell. Transp. Syst.}, vol.~20, no.~1, pp.
  195--204, 2018.

\bibitem{MLDD:17:IEEE_J_ITS}
X.~Meng, J.~Li, X.~Dai, and J.~Dou, ``Variable neighborhood search for a
  colored traveling salesman problem,'' \emph{{IEEE} Trans. Intell. Transp.
  Syst.}, vol.~19, no.~4, pp. 1018--1026, 2017.

\bibitem{KOHTZ:15:IEEE_J_ITS}
G.~Kim, Y.-S. Ong, C.~K. Heng, P.~S. Tan, and N.~A. Zhang, ``City vehicle
  routing problem ({City VRP}): {A} review,'' \emph{{IEEE} Trans. Intell.
  Transp. Syst.}, vol.~16, no.~4, pp. 1654--1666, 2015.

\bibitem{CYL:2025:nQI}
M.~Cattelan, S.~Yarkoni, and W.~Lechner, ``Parallel circuit implementation of
  variational quantum algorithms,'' \emph{npj Quantum Inf.}, vol.~11, no.~1,
  p.~27, 2025.

\bibitem{NRSTD:25:IEEE_J_ITS}
C.~Niu, P.~Rastogi, J.~Soman, K.~Tamuli, and V.~Dixit, ``Applying quantum
  computing to solve multicommodity network flow problem,'' \emph{{IEEE} Trans.
  Intell. Transp. Syst.}, vol.~26, no.~4, pp. 5094--5101, 2025.

\bibitem{SCYB:2023:nQI}
A.~Skolik, M.~Cattelan, S.~Yarkoni, T.~B{\"a}ck, and V.~Dunjko, ``Equivariant
  quantum circuits for learning on weighted graphs,'' \emph{npj Quantum Inf.},
  vol.~9, no.~1, p.~47, 2023.

\bibitem{GA:23:IEEE_J_ITS}
K.~Gautam and C.~W. Ahn, ``Quantum path integral approach for vehicle routing
  optimization with limited qubit,'' \emph{{IEEE} Trans. Intell. Transp.
  Syst.}, vol.~25, no.~5, pp. 3244--3258, 2023.

\bibitem{ABAPF:22:IEEE_J_ITS}
U.~Azad, B.~K. Behera, E.~A. Ahmed, P.~K. Panigrahi, and A.~Farouk, ``Solving
  vehicle routing problem using quantum approximate optimization algorithm,''
  \emph{{IEEE} Trans. Intell. Transp. Syst.}, vol.~24, no.~7, pp. 7564--7573,
  2022.

\bibitem{GKZ:2022:nQI}
A.~Glos, A.~Krawiec, and Z.~Zimbor{\'a}s, ``Space-efficient binary optimization
  for variational quantum computing,'' \emph{npj Quantum Inf.}, vol.~8, no.~1,
  p.~39, 2022.

\bibitem{GJ:1979:book}
M.~R. Garey and D.~S. Johnson, \emph{Computers and {I}ntractability}.\hskip 1em
  plus 0.5em minus 0.4em\relax Freeman San Francisco, 1979, vol. 174.

\bibitem{SG:76:JACM}
S.~Sahni and T.~Gonzalez, ``${P}$-complete approximation problems,'' \emph{J.
  ACM}, vol.~23, no.~3, pp. 555--565, 1976.

\bibitem{WS:2011:book}
D.~P. Williamson and D.~B. Shmoys, \emph{The {D}esign of {A}pproximation
  {A}lgorithms}.\hskip 1em plus 0.5em minus 0.4em\relax Cambridge University
  Press, 2011.

\bibitem{C:22:ORF}
N.~Christofides, ``Worst-case analysis of a new heuristic for the travelling
  salesman problem,'' \emph{Oper. Res. Forum}, vol.~3, no.~1, 2022.

\bibitem{KKG:21:STOC}
A.~R. Karlin, N.~Klein, and S.~O. Gharan, ``A (slightly) improved approximation
  algorithm for metric {TSP},'' in \emph{Proc. ACM Symp. Theory Comput.
  (STOC)}, 2021, pp. 32--45.

\bibitem{KKG:23:IPCO}
------, ``A deterministic better-than-$3/2$ approximation algorithm for metric
  {TSP},'' in \emph{Proc. Conf. Integer Program. Comb. Optim. (IPCO)}, 2023,
  pp. 261--274.

\bibitem{AB:95:SIAMDM}
T.~Andreae and H.-J. Bandelt, ``Performance guarantees for approximation
  algorithms depending on parametrized triangle inequalities,'' \emph{SIAM J.
  Discrete Math.}, vol.~8, no.~1, pp. 1--16, 1995.

\bibitem{GGMSTZ:10:IEEE_J_EPM}
J.~S. Gyorfi, D.~R. Gamota, S.~M. Mok, J.~B. Szczech, M.~Toloo, and J.~Zhang,
  ``Evolutionary path planning with subpath constraints,'' \emph{{IEEE} Trans.
  Electron. Packag. Manuf.}, vol.~33, no.~2, pp. 143--151, 2010.

\bibitem{SHES:16:IJCAI}
M.~Safilian, S.~M. Hashemi, S.~Eghbali, and A.~Safilian, ``An approximation
  algorithm for the subpath planning problem,'' in \emph{Proc. Int. Joint Conf.
  Artif. Intell. (IJCAI)}, 2016, pp. 669--675.

\bibitem{SYKK:17:IJCAI}
H.~Sumita, Y.~Yonebayashi, N.~Kakimura, and K.~Kawarabayashi, ``An improved
  approximation algorithm for the subpath planning problem and its
  generalization,'' in \emph{Proc. Int. Joint Conf. Artif. Intell. (IJCAI)},
  2017, pp. 4412--4418.

\bibitem{ZLG:22:ISAAC}
J.~Zhou, P.~Li, and J.~Guo, ``Parameterized approximation algorithms for
  {TSP},'' in \emph{Proc. Int. Symp. Algorithms Comput. (ISAAC)}, 2022, pp.
  50:1--50:16.

\bibitem{ZZG:25:COCOON}
J.~Zhou, Z.~Zhang, and J.~Guo, ``An {FPT} factor-$11$ approximation algorithm
  for {TSP},'' in \emph{Proc. Computing and Combinatorics Conf. (COCOON)},
  2025, pp. 283--296.

\bibitem{ZXPX:25:AAAI}
J.~Zhao, M.~Xiao, J.~Peng, and Z.~Xiong, ``Improved approximation algorithms
  for clustered {TSP} and subgroup planning,'' in \emph{Proc. AAAI Conf. Artif.
  Intell. (AAAI)}, 2025, pp. 26\,742--26\,749.

\bibitem{ZSX:26:AAAI}
J.~Zhao, Z.~Sheng, and M.~Xiao, ``{FPT} approximation algorithms for {TSP} on
  non-metric graphs,'' in \emph{Proc. AAAI Conf. Artif. Intell. (AAAI)}, 2026,
  pp. 37\,195--37\,202.

\bibitem{MRM:17:TCS}
U.~Mohan, S.~Ramani, and S.~Mishra, ``Constant factor approximation algorithm
  for {TSP} satisfying a biased triangle inequality,'' \emph{Theor. Comput.
  Sci.}, vol. 657, pp. 111--126, 2017.

\bibitem{BBV:2008:incollection}
R.~Baldacci, M.~Battarra, and D.~Vigo, ``Routing a heterogeneous fleet of
  vehicles,'' in \emph{The {V}ehicle {R}outing {P}roblem: {L}atest {A}dvances
  and {N}ew {C}hallenges}.\hskip 1em plus 0.5em minus 0.4em\relax Springer,
  2008, pp. 3--27.

\bibitem{BHS:25:IEEE_J_JSAC}
T.~T. Bui, D.~V. Huynh, V.~Sharma, K.~Singh, O.~A. Dobre, H.~Shin, and T.~Q.
  Duong, ``{UAV}-assisted physical layer security for space–air–ground
  integrated networks {(SAGIN)} with multiple eavesdroppers,'' \emph{{IEEE} J.
  Sel. Areas Commun.}, vol.~44, pp. 1888--1902, 2026.

\bibitem{KPN:25:IEEE_M_WC}
U.~Khalid, U.~I. Paracha, Z.~Naveed, T.~Q. Duong, M.~Z. Win, and H.~Shin,
  ``Quantum fusion intelligence for integrated satellite-ground remote
  sensing,'' \emph{{IEEE} Wireless Commun.}, vol.~32, no.~3, pp. 46--55, Jun.
  2025.

\bibitem{KBR:25:IEEE_M_NET}
A.~Khan, T.~T. Bui, J.~ur~Rehman, S.~Chatzinotas, S.~L. Cotton, O.~A. Dobre,
  T.~Q. Duong, and H.~Shin, ``Integrated non-terrestrial and terrestrial
  quantum anonymous networks,'' \emph{{IEEE} Netw.}, vol.~39, no.~3, pp.
  196--206, May 2025.

\bibitem{KUW:24:IEEE_M_COM}
U.~Khalid, M.~S. Ulum, M.~Z. Win, and H.~Shin, ``Integrated satellite-ground
  variational quantum sensing networks,'' \emph{{IEEE} Commun. Mag.}, vol.~62,
  no.~10, pp. 20--27, Oct. 2024.

\bibitem{ZPFSW:23:IEEE_J_JSAC}
F.~Zaman, S.~N. Paing, A.~Farooq, H.~Shin, and M.~Z. Win, ``Concealed quantum
  telecomputation for anonymous {6G URLLC} networks,'' \emph{{IEEE} J. Sel.
  Areas Commun.}, vol.~41, no.~7, pp. 2278--2296, Jul. 2023.

\bibitem{ZFUJSW:23:IEEE_M_WC}
F.~Zaman, A.~Farooq, M.~A. Ullah, H.~Jung, H.~Shin, and M.~Z. Win, ``Quantum
  machine intelligence for {6G} {URLLC},'' \emph{{IEEE} Wireless Commun.},
  vol.~30, no.~2, pp. 22--30, Apr. 2023.

\bibitem{H:91:ORL}
J.~A. Hoogeveen, ``Analysis of christofides' heuristic: {S}ome paths are more
  difficult than cycles,'' \emph{Oper. Res. Lett.}, vol.~10, no.~5, pp.
  291--295, 1991.

\bibitem{AKS:15:JACM}
H.-C. An, R.~Kleinberg, and D.~B. Shmoys, ``Improving christofides' algorithm
  for the s-t path {TSP},'' \emph{J. ACM}, vol.~62, no.~5, pp. 1--28, 2015.

\bibitem{S:13:IPCO}
A.~Seb{\H{o}}, ``Eight-fifth approximation for the path {TSP},'' in \emph{Proc.
  Conf. Integer Program. Comb. Optim. (IPCO)}, 2013, pp. 362--374.

\bibitem{V:16:SIAMDM}
J.~Vygen, ``Reassembling trees for the traveling salesman,'' \emph{SIAM J.
  Discrete Math.}, vol.~30, no.~2, pp. 875--894, 2016.

\bibitem{GV:18:MP}
C.~Gottschalk and J.~Vygen, ``Better s-t-tours by {G}ao trees,'' \emph{Math.
  Program.}, vol. 172, no.~1, pp. 191--207, 2018.

\bibitem{S:16:FOCS}
A.~Sebo and A.~Van~Zuylen, ``The salesman's improved paths: {A} $3/2+ 1/34$
  approximation,'' in \emph{Proc. {IEEE} Symp. Found. Comput. Sci. (FOCS)},
  2016, pp. 118--127.

\bibitem{TV:19:JACM}
V.~Traub and J.~Vygen, ``Approaching $3/2$ for the s-t-path {TSP},'' \emph{J.
  ACM}, vol.~66, no.~2, pp. 1--17, 2019.

\bibitem{Z:19:SODA}
R.~Zenklusen, ``A $1.5$-approximation for path {TSP},'' in \emph{Proc. ACM-SIAM
  Symp. Discrete Algorithms (SODA)}, 2019, pp. 1539--1549.

\bibitem{TVZ:22:SICOMP}
V.~Traub, J.~Vygen, and R.~Zenklusen, ``Reducing path {TSP} to {TSP},''
  \emph{SIAM J. Comput.}, vol.~51, no.~3, pp. 20--24, 2022.

\bibitem{B:25:SODA}
M.~B{\"{o}}hm, Z.~Friggstad, T.~M{\"{o}}mke, and J.~Spoerhase, ``Approximating
  traveling salesman problems using a bridge lemma,'' in \emph{Proc. ACM-SIAM
  Symp. Discrete Algorithms (SODA)}, 2025, pp. 1166--1177.

\bibitem{A:01:NETWORKS}
T.~Andreae, ``On the traveling salesman problem restricted to inputs satisfying
  a relaxed triangle inequality,'' \emph{Networks}, vol.~38, no.~2, pp. 59--67,
  2001.

\bibitem{B:02:TCS}
H.-J. B{\"o}ckenhauer, J.~Hromkovi{\v{c}}, R.~Klasing, S.~Seibert, and
  W.~Unger, ``Towards the notion of stability of approximation for hard
  optimization tasks and the traveling salesman problem,'' \emph{Theor. Comput.
  Sci.}, vol. 285, no.~1, pp. 3--24, 2002.

\bibitem{BC:00:IPL}
M.~A. Bender and C.~Chekuri, ``Performance guarantees for the {TSP} with a
  parameterized triangle inequality,'' \emph{Inf. Process. Lett.}, vol.~73, no.
  1-2, pp. 17--21, 2000.

\bibitem{M:15:IPL}
T.~M{\"o}mke, ``An improved approximation algorithm for the traveling salesman
  problem with relaxed triangle inequality,'' \emph{Inf. Process. Lett.}, vol.
  115, no.~11, pp. 866--871, 2015.

\bibitem{FHPS:06:ALGOR}
L.~Forlizzi, J.~Hromkovi, G.~Proietti, and S.~Seibert, ``On the stability of
  approximation for hamiltonian path problems,'' \emph{Algorithmic Oper. Res.},
  vol.~1, no.~1, pp. 31--45, 2006.

\bibitem{B:08:JDALG}
H.-J. B{\"o}ckenhauer, D.~Bongartz, J.~Hromkovi{\v{c}}, R.~Klasing,
  G.~Proietti, S.~Seibert, and W.~Unger, ``On $k$-connectivity problems with
  sharpened triangle inequality,'' \emph{J. Discrete Algorithms}, vol.~6,
  no.~4, pp. 605--617, 2008.

\bibitem{CHHK:22:ALGCA}
L.-H. Chen, S.-Y. Hsieh, L.-J. Hung, and R.~Klasing, ``On the approximability
  of the single allocation $p$-hub center problem with parameterized triangle
  inequality,'' \emph{Algorithmica}, vol.~84, no.~7, pp. 1993--2027, 2022.

\bibitem{CCHKK:20:ICS}
S.-C. Chang, L.-H. Chen, L.-J. Hung, S.-S. Kao, and R.~Klasing, ``The hardness
  and approximation of the densest $k$-subgraph problem in parameterized metric
  graphs,'' in \emph{Proc. Int. Comput. Symp. (ICS)}, 2020, pp. 126--130.

\bibitem{KM:2018:incollection}
R.~Klasing and T.~M{\"o}mke, ``A modern view on stability of approximation,''
  in \emph{Adventures Between Lower Bounds and Higher Altitudes: {E}ssays
  Dedicated to Juraj Hromkovi{\v{c}} on the Occasion of His 60th
  Birthday}.\hskip 1em plus 0.5em minus 0.4em\relax Springer, 2018, pp.
  393--408.

\bibitem{TDZ:04:ICRA}
G.~Tong-ying, Q.~Dao-kui, and D.~Zai-li, ``Research of path planning for
  polishing robot based on improved genetic algorithm,'' in \emph{Proc. IEEE
  Int. Conf. Robotics Biomimetics (ICRA)}, 2004, pp. 334--338.

\bibitem{BEX:24:arXiv}
E.~Bampis, B.~Escoffier, and M.~Xefteris, ``Improved {FPT} approximation for
  non-metric {TSP},'' \emph{arXiv Preprint}, vol. abs/2407.08392, 2024.

\bibitem{ZZWG:25:COCOON}
J.~Zhou, Z.~Zhang, Y.~Wen, and J.~Guo, ``From metric to general graphs: {FPT}
  constant-factor approximation algorithms for three location problems,'' in
  \emph{Proc. Computing and Combinatorics Conf. (COCOON)}, 2025, pp. 297--310.

\bibitem{CLRS:2022:book}
T.~H. Cormen, C.~E. Leiserson, R.~L. Rivest, and C.~Stein, \emph{Introduction
  to {A}lgorithms}.\hskip 1em plus 0.5em minus 0.4em\relax MIT Press, 2022.

\end{thebibliography}
\end{document}